\documentclass[12pt,a4paper]{amsart}

\allowdisplaybreaks

\usepackage{amsmath}
\usepackage{amstext}
\usepackage{amssymb}
\usepackage{amsbsy}
\usepackage{array}
\usepackage{appendix}
\usepackage{amsfonts}

\usepackage{pifont}

\usepackage{youngtab}
\usepackage{ytableau}

\usepackage{braket}
\usepackage{bbm}

\usepackage{color}
\usepackage{caption}
\usepackage{epic}
\usepackage{eepic}
\usepackage{empheq}
\usepackage{eurosym}
\usepackage{epsfig}

\usepackage{float}
\usepackage{framed}

\usepackage{graphicx}

\usepackage{indentfirst}
\usepackage[latin9]{inputenc}

\usepackage{latexsym}
\usepackage{longtable}
\usepackage{lscape}

\usepackage{marginnote}

\usepackage[numbers,square,comma, compress]{natbib}

\usepackage{pgf}
\usepackage[parfill]{parskip}
\usepackage{pxfonts}
\usepackage{pdfpages}
\usepackage{pstricks}

\usepackage{rotating}

\usepackage{subcaption}
\usepackage{simplewick}

\usepackage{tcolorbox}
\usepackage{tikz}
\usepackage{tabularx}
\usepackage{txfonts}

\usepackage{varioref}

\usepackage{wasysym}

\def\be{\begin{equation}}
\def\ee{\end{equation}}
\def\ba{\begin{array}}
\def\ea{\end{array}}

\def\bQ{\bar{Q}}
\def\bV{\bar{V}}
\def\bW{\bar{W}}

\def\ll {\left\lgroup}
\def\rr{\right\rgroup}

\def\leq{\leqslant}
\def\geq{\geqslant}

\def\ba{\pmb{a}}
\def\bb{\pmb{b}}
\def\bq{\pmb{q}}
\def\bx{\bar{x}}
\def\by{\bar{y}}

\def\bu{\bar{u}}

\def\balpha{\bar{\alpha}} 
\def\bemptyset{\pmb{\emptyset}} 

\def\i{\iota}

\def\x{x^{(1)}}

\def\bref\textbf{\ref}

\def\ll { \left\lgroup}
\def\rr{\right\rgroup}

\def\union{\mathop{\bigcup}}

\def\det{\operatorname{det}}

\def\ba{ \pmb{a}}
\def\bb{ \pmb{b}}
\def\bc{ \pmb{c}}

\def\bk{ \pmb{k}}

\def\bs{ \pmb{s}}
\def\bu{ \pmb{u}}

\def\bx{ \pmb{x}}
\def\by{ \pmb{y}}

\def\bA{\pmb{A}}
\def\bB{\pmb{B}}
\def\bC{\pmb{C}}
\def\bD{\pmb{D}}
\def\bP{\pmb{P}}
\def\bN{\pmb{N}}
\def\bQ{\pmb{Q}}

\def\bV{\pmb{V}}
\def\bW{\pmb{W}}
\def\bY{\pmb{Y}}

\def\balpha{\pmb{\alpha}} 
\def\bDelta{\pmb{\Delta}} 
\def\bi{\pmb{\iota }}
\def\bj{\pmb{\jmath}}

\def\i{\iota}

\let\emptyset\varnothing

\newcommand{\cmark}{\text{\ding{51}}}
\newcommand{\xmark}{\text{\ding{55}}}

\newcommand{\cA}{\mathcal{A}}

\newcommand{\cC}{\mathcal{C}}

\newcommand{\cF}{\mathcal{F}}

\newcommand{\cH}{\mathcal{H}}

\newcommand{\cL}{\mathcal{L}}
\newcommand{\cM}{\mathcal{M}}
\newcommand{\cN}{\mathcal{N}}

\newcommand{\cS}{\mathcal{S}}

\newcommand{\cW}{\mathcal{W}}

\newcommand{\cZ}{\mathcal{Z}}

\newcommand{\CC}{\mathbb{C}}

\newcommand{\ZZ}{\mathbb{Z}}

\newcommand{\eo}{\epsilon_1}
\newcommand{\et}{\epsilon_2}

\newcommand{\1}{\textbf{1.}}
\newcommand{\2}{\textbf{2.}}
\newcommand{\3}{\textbf{3.}}
\newcommand{\4}{\textbf{4.}}
\newcommand{\5}{\textbf{5.}}
\newcommand{\6}{\textbf{6.}}

\newcommand{\wsquare}{\square}
\newcommand{\bsquare}{\blacksquare}

\renewcommand{\mod}{\textup{mod}\,}

\usepackage[scale= 0 .825]{geometry}

\makeatletter

\numberwithin{equation}{section}
\numberwithin{figure}{section}

\numberwithin{equation}{section}
\numberwithin{figure}{section}

\restylefloat{figure}

\begin{document} 

\title[]{
Coloured refined topological vertices and 
\\ 
parafermion conformal field theories
}

\author[]{Wee Chaimanowong and Omar Foda} 

\address{
School of Mathematics and Statistics, 
University of Melbourne, 
Royal Parade, Parkville, Victoria 3010, Australia
} 

\email{
n.chaimanowong@student.unimelb.edu.au, 
omar.foda@unimelb.edu.au
} 


\dedicatory{To Professor Mikio Sato on his 90th birthday}

\begin{abstract} 
We extend the definition of the refined topological vertex $\cC$ to an $n$-coloured 
refined topological vertex $\cC_n$ that depends on $n$ free bosons, and compute the 
5D strip partition function made of $N$ pairs of $\cC_n$ vertices and conjugate 
$\cC^{\, \star}_n$ vertices. 
Using geometric engineering and the AGT correspondence, the 4D limit of this strip 
partition function is identified with a (normalized) matrix element of a (primary 
state) vertex operator that intertwines two (arbitrary descendant) states in 
a (generically non-rational) 2D conformal field theory with $\ZZ_{\, n}$ parafermion 
primary states. 
\end{abstract} 

\maketitle

\section{Introduction} 
\label{section.01}

\subsection{Background}
In a 2D conformal field theory, a correlation function is a sum of 
(or an integral over) products of holomorphic and anti-holomorphic 
conformal blocks. Methods to compute the conformal block include 
\1 making use of a null state that flows in an internal channel in 
the block to derive and solve a differential equation for the block
\cite{belavin.polyakov.zamolodchikov},
\2 representing the conformal block in terms of a Coulomb gas 
of charges with a background charge and screening charges, then 
evaluating the conformal block as an integral over the positions 
of screening charges 
\cite{nienhuis, dotsenko.fateev.01, dotsenko.fateev.02}, and 
\3 using the analytic properties of the conformal block to derive 
and solve a recursion relation that can be solved for the block 
\cite{zamolodchikov.01,zamolodchikov.02}.
These methods are powerful and lead to deep insights into the analytic structure 
of the conformal blocks, but they are also non-algorithmic in the sense that the 
answer cannot (in general) be written explicitly and directly, and become complicated to apply in 
the presence of vertex operators of a sufficiently-large highest-weight charge
\footnote{\,
Vertex operators that require multiple screening charges in the Coulomb gas 
approach.
}, 
and for 5- and higher-point conformal blocks
\footnote{\,
The more powerful (elliptic) version of the recursion relation is available 
only for 4-point conformal blocks \cite{zamolodchikov.02, poghossian}. 
}. 

\subsection{Conformal blocks as products of normalized matrix elements}
An algorithmic approach to computing the conformal blocks is to regard them 
as products of matrix elements 
$\cM^{\, 2D}$ of primary-state vertex operators between arbitrary descendant states,
these being normalized by Shapovalov matrix elements.
However, no closed-form expressions for these matrix elements are known 
\footnote{\, 
In $\cW_2$ conformal field theories, 
the Virasoro algebra $\cW_2$ provides sufficiently-many constraints to 
allow us to compute any Virasoro matrix element on an element-by-element 
basis, but in $\cW_N$ theories, $N= 3, 4, \cdots$, more conditions than 
those provided by the $\cW_N$ algebra are required \cite{kanno.matsuo.zhang}, 
and the computation of the matrix elements, while still on an element-by-element 
basis, is more complicated than in the $\cW_2$ case.
}.

\subsection{From 2D conformal blocks to 4D instanton partition functions} 
An approach to compute the normalized matrix elements 
$\cM^{\, 2D}$ in closed form is the AGT correspondence 
\cite{alday.gaiotto.tachikawa, mironov.morozov, wyllard.01}, which applies 
to  $\cW_N \times \cH$ conformal field theories, where the $\cW_N$ algebra, 
generated by chiral spin-2, spin-3, $\cdots$, spin-$N$ currents, is augmented
by a Heisenberg algebra $\cH$ generated by a chiral spin-1 current
\footnote{\,
The contributions of the $\cH$ algebra factor out in the result 
of conformal block computations in $\cW_N \times \cH$ theories, and 
one obtains conformal blocks in $\cW_N$ theories.
}.
The AGT correspondence identifies matrix element $\cM^{\, 2D}$ 
in 2D $\cW_N \times \cH$ conformal field theories and instanton 
partition functions $\cZ^{\, 4D}_{\, N}$, in 4D $\cN = 2$ supersymmetric 
Yang-Mills theories with matter in bifundemental 
$SU\ll N \rr$ representations. 

\subsection{From 4D instanton partition functions to 5D strip partition 
functions} The 4D instanton partition function $\cZ^{\, 4D}_{\, N}$ is 
engineered by taking the 4D limit of a 5D topological string strip partition 
function $\cS^{\, 5D}_{\, N}$ \cite{katz.klemm.vafa, katz.mayr.vafa},
obtained by gluing topological vertices 
\cite{
iqbal, aganagic.klemm.marino.vafa, 
awata.kanno.01, awata.kanno.02, 
iqbal.kozcaz.vafa, iqbal.kashani-poor}.
In this sense, a topological vertex is the most fundamental building 
block of the correlation functions in a 2D conformal field theory.

\subsection{2D parafermion conformal field theories}
In \cite{
alfimov.belavin.tarnopolsky,
alfimov.tarnopolsky, 
belavin.belavin.bershtein, 
belavin.bershtein.feigin.litvinov.tarnopolsky,
belavin.bershtein.tarnopolsky, 
belavin.feigin, 
belavin.mukhametzhanov, 
belavin.wyllard,
bonelli.maruyoshi.tanzini.01, 
bonelli.maruyoshi.tanzini.02,
itoyama.oota.yoshioka,
nishioka.tachikawa, 
spodyneiko, 
wyllard.02}, 
and other works, 4D instanton partition functions on $\CC^{\, 2} / \ZZ_{\, n}$ 
were related \footnote{\,
The action of $\ZZ_{\, n}$ on $\CC^{\, 2} = \CC_{\, 1} \times \CC_{\, 2}$ is 
$\ll z_{\, 1}, \, z_{\, 2} \rr 
\mapsto
\ll \omega \, z_{\, 1}, \, \omega^{\, -1} \, z_{\, 2} \rr$,
where
$z_{\, 1} \in \CC_{\, 1},
\, 
z_{\, 2} \in \CC_{\, 2},
\, 
\omega^{\, n} \, = \, 1$ 
\cite{belavin.bershtein.feigin.litvinov.tarnopolsky}
}, 
using an extension of the AGT correspondence, to matrix elements in non-rational 
2D conformal field theories based on the algebra 
$
\cA \ll N, n \rr = 
\ll 
\widehat{sl} \ll N \rr_n 
\times 
\widehat{sl} \ll N \rr_p
\, / \, 
\widehat{sl} \ll N \rr_{n + p}
\rr 
\times
\widehat{sl} \ll n \rr_N 
\times 
\cH
$
\footnote{\,
Our notation for $\cA \ll N, n \rr$ is adapted to that used in the present work 
and as such it differs, in an obvious way, from that used in the original papers. 
Further, we stress that the discussion of the conformal field theories in the original 
papers on the subject, and definitely in the present work, is restricted to non-rational 
conformal field theories. The space of states in rational theories based on 
$\cA \ll N, n \rr$ contains degenerate representations with null states that 
require special treatment.
}. 
However, no connection with 5D topological string partition functions and topological 
vertices was made.

\subsection{In this work}
\1 We propose an extension of the refined topological vertex $\cC$ of 
\cite{iqbal.kozcaz.vafa}, 
constructed using a single free boson, to a refined topological 
vertex $\cC_n$ constructed using $n$ free bosons, and a conjugate 
vertex $\cC^{\, \star}_n$.  
\2 We compute $\cS^{\, 5D}_{\, N, \, n}$, the 5D topological string strip 
partition function that consists of $N$ pairs of vertices where
each pair consists of a single $\cC_n$ vertex and a single $\cC^{\, \star}_n$.  
\3 We take the 4D limit of 
$\cS^{\, 5D}_{\, N,  \,  n}$ of to obtain 
$\cS^{\, 4D}_{\, N,  \,  n}$, and identify the result, using geometric 
engineering \cite{katz.klemm.vafa, katz.mayr.vafa}, with 
$\cZ^{\, 4D}_{\, N, \, n}$, the 4D instanton 
partition function of matter in a bifundamental representation of 
$\textit{SU} \ll N \rr$ in $\CC^{\, 4} / \ZZ_{\, n}$. 
\4 We use the AGT correspondence, as defined in 
\cite{
alfimov.belavin.tarnopolsky, 
alfimov.tarnopolsky, 
belavin.belavin.bershtein, 
belavin.bershtein.feigin.litvinov.tarnopolsky,
belavin.bershtein.tarnopolsky, 
belavin.feigin, 
belavin.mukhametzhanov, 
belavin.wyllard,
bonelli.maruyoshi.tanzini.01, 
bonelli.maruyoshi.tanzini.02, 
spodyneiko, 
wyllard.02}, 
to identify 
$\cZ^{\, 4D}_{\, N, \, n}$ with 
a matrix element 
$\cM^{\, 2D}_{
\, \cA \ll  N, \, n \rr
} 
\ll \balpha_{\, L}, \, \balpha, \, \balpha_{\, R} \rr$ 
of a primary vertex operator that carries a highest-weight charge 
$\balpha$ between left and right states, 
$\langle \, \balpha_{\, L} \, |$ and $| \, \balpha_{\, R} \, \rangle$, 
where
$\langle \, \balpha_{\, L} \, |$ and $| \, \balpha_{\, R} \, \rangle$ 
are arbitrary descendant states.
\5 We obtain the linear relation between the K\"ahler parameters 
of the 5D instanton partition function and the parameters of the 
corresponding $\cA \ll N, \, n \rr$ matrix element,
\6 We discuss in detail the differences in normalizations of these 
objects, and show that these differences cancel out when gluing 
matrix elements to compute conformal blocks.

\subsection{Outline of contents}
In section \textbf{2}, we recall the combinatorics of partitions and
Young diagrams that is used in the sequel, 
and then in Section \textbf{3}, we do the same for 
symmetric functions in infinitely-many 
variables, with emphasis on the Schur
functions, Heisenberg 
algebras, and correspondences between 
them.
In \textbf{4}, we introduce the $n$-coloured refined topological vertex 
$\cC_{\, n}$, and the conjugate vertex $\cC^{\, \star}_{\, n}$. 
In \textbf{5}, we compute $\cS^{\, 5D}_{\, N, \, n}$, 
the 5D \textit{SU(N)} topological string strip partition function 
made of $N$ pairs of vertices. 
In \textbf{6}, we compute $\cS^{\, 4D}_{\, N, \, n}$, the 4D limit of 
$\cS^{\, 5D}_{\, N, \, n}$, and identify it with the 4D instanton partition 
function $\cZ^{\, 4D}_{\, N, \, n}$. 
In \textbf{7}, we use the 5D strips to compute 5D web diagrams, and then  
in \textbf{8}, we take the 4D limit of the 5D web diagrams. 
In \textbf{9}, we reproduce a 4-point conformal block computed in 
\cite{alfimov.tarnopolsky}, and in \textbf{10}, we make a number of comments.

\section{Young and Maya diagrams}
\label{section.02.refined.topological.vertex}

\subsection{Young diagrams}
\label{partitions.young.diagrams} 
A partition $Y = \ll y_1, y_2, \cdots \rr$, of a non-negative integer 
$|\,  Y \,|$, is a set of non-negative, non-increasing integers
$y_i \geq y_{i+1} \geq 0$, $\sum_{i = 1} y_i= | \, Y \, |$, and can 
be represented as a Young diagram, that consists of rows such that 
row $i$ has $y_i$ cells (see Figure \ref{young.diagram.01}).
We use $y_i$ for the $i$-th row as well as for the number of cells in 
that row, and $Y^{\intercal}$ for the transpose of $Y$. 
In this work, a Young diagram has infinitely-many rows.
By a finite Young diagram, we mean a Young diagram that
has fintely many non-null rows.
The null Young diagram 
$Y = \emptyset$ is such that all rows are null.
Given a set of $n$ Young diagrams $\bY = \ll Y_1, \cdots, Y_n \rr$,
define

\begin{equation}
| \, \bY \, | = 
\sum_{i = 1}^n
| \, Y_i \, |, 
\quad 
\bY^\intercal = 
\ll 
Y^{\, \intercal}_1, \cdots, 
Y^{\, \intercal}_n 
\rr
\end{equation}

\subsubsection{The (infinite) borderline of a (finite) Young diagram}
\begin{figure}
\begin{tikzpicture}[scale=.6]

\node at (- 1.0,   0.5) {$i = 1$};
\node at (- 1.0, - 0.5) {$i = 2$};
\node at (- 1.0, - 1.5) {$i = 3$};
\node at (- 1.0, - 2.5) {$i = 4$};
\node at (- 1.0, - 3.5) {$i = 5$};

\node at ( -0.2,   1.4) {$j =  $};
\node at (  0.5,   1.5) {$    1$};
\node at (  1.5,   1.5) {$    2$};
\node at (  2.5,   1.5) {$    3$};
\node at (  3.5,   1.5) {$    4$};
\node at (  4.5,   1.5) {$    5$};
\node at (  5.5,   1.5) {$    6$};
\node at (  6.5,   1.5) {$    7$};

\draw [thick] (0, 0) rectangle (1,1);
\draw [very thick] (5, 1.00)--(7, 1.00);
\draw [very thick] (5, 0.95)--(7, 0.95);
\draw [very thick] (5.00, 1)--(5.00, 0);
\draw [very thick] (5.05, 1)--(5.05, 0);
\draw [very thick] (4,  0.00)--(5.05,  0.00);
\draw [very thick] (4, -0.05)--(5.05, -0.05);
\draw [very thick] (4.00, 0)--(4.00, -1);
\draw [very thick] (4.05, 0)--(4.05, -1);
\draw [very thick] (2, -1.00)--(4.05, -1.00);
\draw [very thick] (2, -1.05)--(4.05, -1.05);
\draw [very thick] (2.00, -1)--(2.00, -2);
\draw [very thick] (2.05, -1)--(2.05, -2);
\draw [very thick] (0, -2.00)--(2.05, -2.00);
\draw [very thick] (0, -2.05)--(2.05, -2.05);
\draw [very thick] (0.00, -2)--(0.00, -4);
\draw [very thick] (0.05, -2)--(0.05, -4);

\draw [thick] (1, 0) rectangle (1,1);
\draw [thick] (2, 0) rectangle (1,1);
\draw [thick] (3, 0) rectangle (1,1);
\draw [thick] (4, 0) rectangle (1,1);
\draw [thick] (5, 0) rectangle (1,1);
\draw [thick] (0,-1) rectangle (1,1);
\draw [thick] (1,-1) rectangle (1,1);
\draw [thick] (2,-1) rectangle (1,1);
\draw [thick] (3,-1) rectangle (1,1);
\draw [thick] (4,-1) rectangle (1,1);
\draw [thick] (0,-2) rectangle (1,1);
\draw [thick] (1,-2) rectangle (1,1);
\draw [thick] (2,-2) rectangle (1,1);

\node at (1.5,-0.5) {$\cmark$};
\node at (4.5,-1.5) {$\xmark$};

\draw [thick] (0,-3)--(0.2,-3);
\draw [thick] (6,1)--(6,0.8);

\end{tikzpicture}
\caption{
{\it
The Young diagram $Y = \ll 5, 4, 2 \rr$, and the corresponding 
transpose Young diagram 
$Y^{\, \intercal} = \ll 3, 3, 2, 2, 1 \rr$.
The infinite borderline is shown in thick
lines.  
The rows are numbered from top to bottom, 
the columns are from left to right, with
null-rows and null-columns included. 
The arm-length and leg-length of the box with $\cmark$ are $A_{\cmark, Y} = 2, A^+_{\cmark, Y} = 5/2, A^{++}_{\cmark, Y} = 3, L_{\cmark, Y} = 1, L^+_{\cmark, Y} = 3/2, L^{++}_{\cmark, Y} = 2$. Similarly for the box with $\xmark$, we have $A_{\xmark, Y} = -3, A^+_{\xmark, Y} = -5/2, A^{++}_{\xmark, Y} = -2, L_{\xmark, Y} = -2, L^+_{\xmark, Y} = -3/2, L^{++}_{\xmark, Y} = -1$.
}
}
\label{young.diagram.01}
\end{figure}
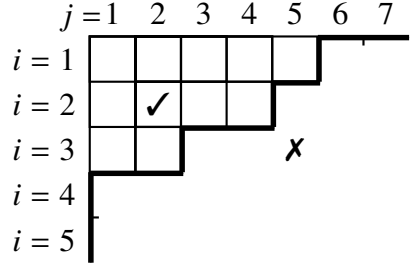

The union of the positive $x$-axis and the negative $y$-axis, that 
is the borderline of the south-east quadrant, is the borderline of 
the null Young diagram $Y = \emptyset$.
The (infinite) borderline of a (finite) Young diagram is the union 
of 
the right vertical   boundaries of the right-most cells of each row, 
the lower horizontal boundaries of the bottom     cells of each column,
the semi-infinite segment of the positive $x$-axis to the right, and 
the semi-infinite segment of the negative $y$-axis below the Young 
diagram, as in Figure \ref{young.diagram.01}.

\subsubsection{Cells}

We use $\wsquare$ for a cell (a square) in the south-east quadrant
of the plane, and refer to the coordinates of $\wsquare$ as $\ll i, j \rr$.
If $\wsquare \in Y$, then 
$i$ is the $Y$-row-number,    counted from top to bottom, and 
$j$ is the $Y$-column-number, counted from left to right, 
that $\wsquare$ lies in. 
If $\wsquare \notin Y$, 
we still regard $i$ ($j$) as a $Y$-row-number ($Y$-column-number) 
albeit that row (column) is null. 
In other words, the coordinates 
$\ll i, j \rr$ of a cell are measured with respect to the (original) 
boundaries of the south-east quadrant, rather than with respect to 
the borderline of any specific Young diagram.

\subsubsection{Arms, legs, and hooks}
\label{arms.legs} Consider a cell $\wsquare$ with coordinates 
$\ll i, j \rr$. We define the lengths of 
the           arm $A^{  }_{\wsquare, Y}$, 
half-extended arm $A^{+ }_{\wsquare, Y}$, 
     extended arm $A^{++}_{\wsquare, Y}$, 
the           leg $L^{  }_{\wsquare, Y}$,
half-extended leg $L^{+ }_{\wsquare, Y}$, 
     extended leg $L^{++}_{\wsquare, Y}$, of $\wsquare$  
with respect to the Young diagram $Y$, 

\begin{eqnarray}
A^{  }_{\wsquare, Y} & = & y_i               - \, j, \quad
A^{+ }_{\wsquare, Y}   =   A^{  }_{\wsquare, Y} + \frac12,            \quad 
A^{++}_{\wsquare, Y}   =   A^{  }_{\wsquare, Y} +      1,  
\\
L^{  }_{\wsquare, Y} & = & y_j^{\intercal} - \, i,   \quad
L^{+ }_{\wsquare, Y}   =   L^{  }_{\wsquare, Y} + \frac12,            \quad        
L^{++}_{\wsquare, Y}   =   L^{  }_{\wsquare, Y} +      1
\end{eqnarray}

Note that $A_{\wsquare,Y}$ and $L_{\wsquare,Y}$ are negative when 
$\wsquare \notin Y$. The hook of a cell $\wsquare$, with respect 
to the borderline of a Young diagram $Y$ is,

\begin{equation}
H_{\wsquare} = A_{\wsquare, Y} + L_{\wsquare, Y} + 1
\end{equation}

\subsubsection{Charged Young diagrams}
We define a charged Young diagram as a pair $\ll Y, c_{\, Y} \rr$, where $Y$ is a finite Young 
diagram and $c_{\, Y} \in \ZZ$ is the charge. 

\subsection{Maya diagrams}\label{Maya.diagram.construction}
A Maya diagram is an infinite 1-dimensional lattice with a black stone 
or a white stone on each segment, the segments are labelled by a position 
coordinate $c_{\, p}$, such that 
sufficiently far to the left, all stones are black, and 
sufficiently far to the right, all stones are white \cite{miwa.jimbo.date.book}.

\subsubsection{The ground-state, 0-charge Maya diagram}
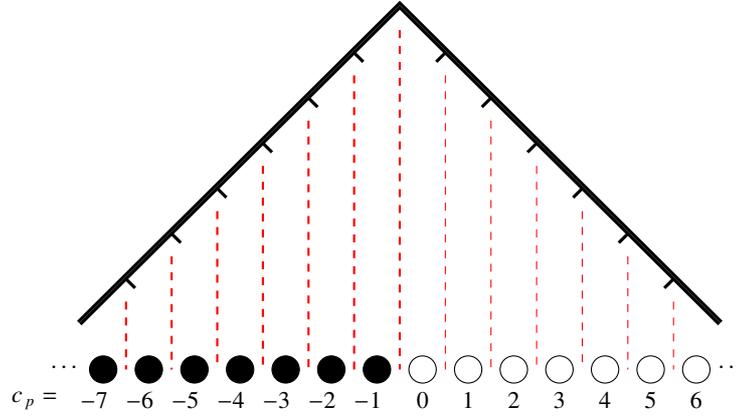
\begin{figure}
\begin{center}
\begin{scriptsize}
\begin{tikzpicture}[scale = 0.6]

\draw [very thick] (-1.5,1.0)--(5.5,8.0)--(12.5,1.0);
\draw [very thick] (-1.55,1.05)--(5.5,8.1)--(12.55,1.05);

\draw [very thick] ( 6.3,6.8)--(6.5,7.0);
\draw [very thick] ( 7.3,5.8)--(7.5,6.0);
\draw [very thick] ( 8.3,4.8)--(8.5,5.0);
\draw [very thick] ( 9.3,3.8)--(9.5,4.0);
\draw [very thick] ( 10.3,2.8)--(10.5,3.0);
\draw [very thick] ( 11.3,1.8)--(11.5,2.0);
\draw [very thick] ( -0.5,2.0)--(-0.3,1.8);
\draw [very thick] ( 0.5,3.0)--(0.7,2.8);
\draw [very thick] ( 1.5,4.0)--(1.7,3.8);
\draw [very thick] ( 2.5,5.0)--(2.7,4.8);
\draw [very thick] ( 3.5,6.0)--(3.7,5.8);
\draw [very thick] ( 4.5,7.0)--(4.7,6.8);

\foreach \i in {0,...,6} 
{
\draw [thick, dashed, red] (\i - .5, \i + 1.5)--(\i - .5, 0);
}
\foreach \j in {7,...,12} 
{
\draw [thin, dashed, red](\j -.5, 13.5 -\j)--(\j - .5, 0);
}
\node [left]  at (-1.3,0) {$\cdots$};
\node [right] at (12.3,0) {$\cdots$};
\foreach \x in {0,...,2}
{
\draw [fill=black!100] (\x- 1,0) circle (0.3);
\draw (\x+10,0) circle (0.3);
}

\draw [fill=black!100] (2,0) circle (0.3);
\draw [fill=black!100] (3,0) circle (0.3);
\draw [fill=black!100] (4,0) circle (0.3);
\draw [fill=black!100] (5,0) circle (0.3);
\draw (6,0) circle (0.3);
\draw (7,0) circle (0.3);
\draw (8,0) circle (0.3);
\draw (9,0) circle (0.3);

\foreach \a in {7, 6, ..., 1} 
{
\node [below] at (- \a + 5.8, -.3) {$- \a$};
}
\foreach \a in {0, 1, ..., 6} 
{
\node [below] at (  \a + 6.0, -.3) {$  \a$};
}			
\node [below] at (-2.5, -.3) {$c_{\, p} = $};
\end{tikzpicture}
\end{scriptsize}
\end{center}
\caption{ \it The ground-state, 0-charge Maya diagram.
That no stones are shuffled corresponds to 
introducing a null Young diagram. 
That the charge is zero corresponds to 
positioning the apex between $\ll c_{\, p} = -1 \rr$, and 
$\ll c_{\, p} = 0 \rr$.}
\label{young.maya.01}
\end{figure}

The simplest Maya diagram is the ground-state,  0-charge Maya diagram 
$M_{\, 0}$, where 
all segments from 
position $\ll c_{\, p} = - \infty \rr$, to 
position $\ll c_{\, p} = - 1 \rr$, 
inclusive, carry black stones, 
and 
all segments from 
position $\ll c_{\, p} = 0 \rr$, to 
position $\ll c_{\, p} = \infty \rr$, inclusive, 
carry white stones, see Figure \ref{young.maya.01}.
From $M_{\, 0}$, we can generate all other diagrams by applying 
a charged Young diagram $\ll Y, c_{\, Y} \rr$
to $M_{\, 0}$, such that, $Y$ shuffles the black and the white stones 
at finite distances from the origin, and   
$c_{\, Y}$ shifts 
the positions of \textit{all} stones to the right by the same distance 
$c_{\, Y}$.
The result is an excited state, charge-$c_{\, Y}$ Maya diagram. 

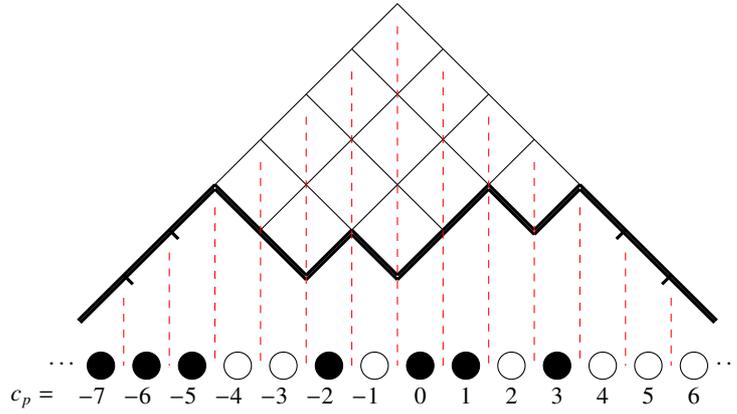
\begin{figure}
\begin{center}
\begin{scriptsize}
\begin{tikzpicture}[scale = 0.6]

\draw [thin] (-1.5,1.0)--(5.5,8.0)--(12.5,1.0);
\draw [thin] ( 2.5,3.0)--(6.5,7.0);
\draw [thin] ( 3.5,2.0)--(7.5,6.0);
\draw [thin] ( 5.5,2.0)--(8.5,5.0);
\draw [thin] ( 8.5,3.0)--(9.5,4.0);
\draw [thin] ( 1.5,4.0)--(3.5,2.0);
\draw [thin] ( 2.5,5.0)--(5.5,2.0);
\draw [thin] ( 3.5,6.0)--(6.5,3.0);
\draw [thin] ( 4.5,7.0)--(8.5,3.0);

\draw [very thick] (-1.5, 1.0)--(1.5, 4.0);
\draw [very thick] (-1.45,0.95)--(1.5,3.9);

\draw [very thick] ( 1.5,4.0)--(3.5,2.0);
\draw [very thick] ( 1.5,3.9)--(3.5,1.9);

\draw [very thick] ( 3.5,2.0)--(4.5,3.0);
\draw [very thick] ( 3.5,1.9)--(4.5,2.9);

\draw [very thick] ( 4.5,3.0)--(5.5,2.0);
\draw [very thick] ( 4.5,2.9)--(5.5,1.9);

\draw [very thick] ( 5.5,2.0)--(7.5,4.0);
\draw [very thick] ( 5.5,1.9)--(7.5,3.9);

\draw [very thick] ( 7.5,4.0)--(8.5,3.0);
\draw [very thick] ( 7.5,3.9)--(8.5,2.9);

\draw [very thick] ( 8.5,3.0)--(9.5,4.0);
\draw [very thick] ( 8.5,2.9)--(9.5,3.9);

\draw [very thick] ( 9.5, 4.0)--(12.50, 1.00);
\draw [very thick] ( 9.5, 3.9)--(12.45, 0.95);

\draw [very thick] ( 10.3,2.8)--(10.5,3.0);
\draw [very thick] ( 11.3,1.8)--(11.5,2.0);
\draw [very thick] ( -0.5,2.0)--(-0.3,1.8);
\draw [very thick] ( 0.5,3.0)--(0.7,2.8);

\foreach \i in {0,...,6} 
{
\draw [thin, dashed, red] (\i - .5, \i + 1.5)--(\i - .5, 0);
}
\foreach \j in {7,...,12} 
{
\draw [thin, dashed, red](\j -.5, 13.5 -\j)--(\j - .5, 0);
}
\node [left]  at (-1.3,0) {$\cdots$};
\node [right] at (12.3,0) {$\cdots$};
\foreach \x in {0,...,2}
{
\draw [fill=black!100] (\x- 1,0) circle (0.3);
\draw                 (\x+10,0) circle (0.3);
}

\draw                  (2,0) circle (0.3);
\draw                  (3,0) circle (0.3);
\draw [fill=black!100]  (4,0) circle (0.3);
\draw                  (5,0) circle (0.3);
\draw [fill=black!100]  (6,0) circle (0.3);
\draw [fill=black!100]  (7,0) circle (0.3);
\draw                  (8,0) circle (0.3);
\draw [fill=black!100]  (9,0) circle (0.3);

\foreach \a in {7, 6, ..., 1} 
{
\node [below] at (- \a + 5.8, -.3) {$- \a$};
}
\foreach \a in {0, 1, ..., 6} 
{
\node [below] at (  \a + 6.0, -.3) {$  \a$};
}				

\node [below] at (-2.5, -.35) {$c_{p} = $};

\end{tikzpicture}
\end{scriptsize}
\end{center}
\caption{ \it An excited-state, 0-charge Maya diagram with a shuffling 
that corresponds to $Y = \ll 4,3,3,2 \rr$.
The apex of the Young diagram is between 
position $\ll c_{\, p} = -1 \rr$, and 
position $\ll c_{\, p} = 0 \rr$.
}
\label{young.maya.02}
\end{figure}

\subsubsection{Introducing a Young diagram}
Starting from the ground-state, 0-charge Maya diagram $M_{\, 0}$, we can 
use a finite Young diagram $Y$ to shuffle the black and white stones 
at finite distances from the origin and produce an excited-state Maya diagram 
as follows. 
\1 We position the diagram $Y$ as in Figure \ref{young.maya.02}, so that 
the apex of $Y$ projects on the point between positions $\ll -1 \rr$
and $\ll 0 \rr$ on the Maya diagram.
The infinite borderline of $Y$ then consists of upward and downward segments
$\ll \diagup, \diagdown \rr$, where
all segments are $\diagup$   sufficiently far to the left, and 
all segments are $\diagdown$ sufficiently far to the right.
\2 We map the infinite borderline to a Maya diagram according to,  

\begin{equation}
\diagup   \rightleftharpoons \CIRCLE, 
\quad
\quad
\diagdown \rightleftharpoons \Circle
\end{equation}

The result is an excited-state, 0-charge Maya diagram, where the configuration
of the black and white stones is 
in bijection with a finite Young diagram $Y$.
Note that the position coordinates 
$c_{\, p}$ are the same as in the ground-state Maya diagram.

\subsubsection{Introducing a charge}
\label{charged.Maya}

\begin{figure}
\begin{center}
\begin{scriptsize}
\begin{tikzpicture}[scale = 0.6]

\draw [thin] (-1.5,1.0)--(5.5,8.0)--(12.5,1.0);
\draw [thin] ( 2.5,3.0)--(6.5,7.0);
\draw [thin] ( 3.5,2.0)--(7.5,6.0);
\draw [thin] ( 5.5,2.0)--(8.5,5.0);
\draw [thin] ( 8.5,3.0)--(9.5,4.0);
\draw [thin] ( 1.5,4.0)--(3.5,2.0);
\draw [thin] ( 2.5,5.0)--(5.5,2.0);
\draw [thin] ( 3.5,6.0)--(6.5,3.0);
\draw [thin] ( 4.5,7.0)--(8.5,3.0);

\draw [very thick] (-1.50, 1.00)--(1.5, 4.0);
\draw [very thick] (-1.45, 0.95)--(1.5, 3.9);

\draw [very thick] ( 1.5,4.0)--(3.5,2.0);
\draw [very thick] ( 1.5,3.9)--(3.5,1.9);

\draw [very thick] ( 3.5,2.0)--(4.5,3.0);
\draw [very thick] ( 3.5,1.9)--(4.5,2.9);

\draw [very thick] ( 4.5,3.0)--(5.5,2.0);
\draw [very thick] ( 4.5,2.9)--(5.5,1.9);

\draw [very thick] ( 5.5,2.0)--(7.5,4.0);
\draw [very thick] ( 5.5,1.9)--(7.5,3.9);

\draw [very thick] ( 7.5,4.0)--(8.5,3.0);
\draw [very thick] ( 7.5,3.9)--(8.5,2.9);

\draw [very thick] ( 8.5,3.0)--(9.5,4.0);
\draw [very thick] ( 8.5,2.9)--(9.5,3.9);

\draw [very thick] ( 9.5, 4.00)--(12.50, 1.00);
\draw [very thick] ( 9.5, 3.90)--(12.45, 0.95);

\draw [very thick] ( 10.3,2.8)--(10.5,3.0);
\draw [very thick] ( 11.3,1.8)--(11.5,2.0);
\draw [very thick] ( -0.5,2.0)--(-0.3,1.8);
\draw [very thick] ( 0.5,3.0)--(0.7,2.8);

\foreach \i in {0,...,6} 
{
\draw [thin, dashed, red] (\i - .5, \i + 1.5)--(\i - .5, 0);
}
\foreach \j in {7,...,12} 
{
\draw [thin, dashed, red](\j -.5, 13.5 -\j)--(\j - .5, 0);
}
\node [left]  at (-1.3,0) {$\cdots$};
\node [right] at (12.3,0) {$\cdots$};
\foreach \x in {0,...,2}
{
\draw [fill=black!100] (\x- 1,0) circle (0.3);
\draw                 (\x+10,0) circle (0.3);
}

\draw                  (2,0) circle (0.3);
\draw                  (3,0) circle (0.3);
\draw [fill=black!100]  (4,0) circle (0.3);
\draw                  (5,0) circle (0.3);
\draw [fill=black!100]  (6,0) circle (0.3);
\draw [fill=black!100]  (7,0) circle (0.3);
\draw                  (8,0) circle (0.3);
\draw [fill=black!100]  (9,0) circle (0.3);

\foreach \a in {5, 4, ..., 1} 
{
\node [below] at (- \a + 3.8, -.3) {$- \a$};
}
\foreach \a in {0, 1, ..., 8} 
{
\node [below] at (  \a + 4.0, -.3) {$  \a$};
}						

\node [below] at (-3, -.35) {$c_{p} + c_Y = $};

\end{tikzpicture}
\end{scriptsize}
\end{center}
\caption{ \it An excited-state charge-2 Maya diagram, with a shuffling 
that corresponds to $Y = \ll 4, 3, 3, 2 \rr$.
The apex of the Young diagram is between 
position $\ll c_{\, p} = 1 \rr$, and 
position $\ll c_{\, p} = 2 \rr$. 
}
\label{young.maya.04}
\end{figure}
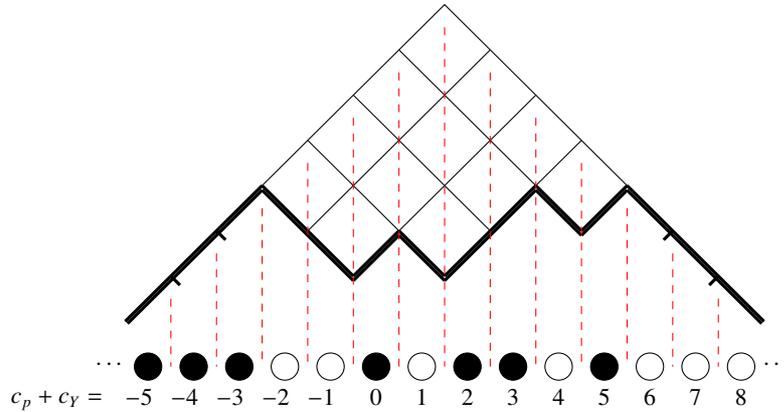

We can introduce a charge $c_{\, Y}$ into any 0-charge Maya diagram 
by shifting all stones globally by $c_{\, Y}$ segments to the right, as 
in Figure \ref{young.maya.04}.
A stone that is at position $c_{p}$ before introducing the charge is 
shifted to position $\ll c_{\, p} + c_{\, Y} \rr$.

\section{Symmetric functions and Heisenberg algebras}
\label{section.03}
\textit{We recall basic definitions related to symmetric functions, 
Heisenberg algebras and relations between them.
}

\subsection{Exponentiated sequences}
Given any two sequences of integers $\ba = \ll a_1,a_2,\cdots \rr, \bb = \ll b_1,b_2,\cdots \rr$ and two variables $x, y$ we define

\begin{equation}
    x^{\, \ba} \, y^{\, \bb} = 
\ll x^{a_1} y^{b_1}, x^{a_2} y^{b_2}, \cdots \rr,  
\end{equation}

In particular, given a Young diagram
$Y = \ll y_1,y_2,\cdots \rr$,
and an infinite sequence of integers $\bi = \ll 1,2,\cdots \rr$, we have 

\begin{equation}
x^{\, \bi}\, y^{\, \pm Y}  = \ll x  \, y^{\, \pm y_1}, x^2\, y^{\, \pm y_2} \cdots \rr, 
\quad
x^{\, \bi - 1}\, y^{\, \pm Y}  = \ll y^{\, \pm y_1}, x\, y^{\, \pm y_2} \cdots \rr, 
\quad
\cdots 
\end{equation}

For the purposes of section \ref{section.04},
we define the sub-sequence,   

\begin{equation}
\label{subsequence.symbol}
[x^{\, \ba} y^{\, \bb}]_c = 
\ll x^{\, a_i} y^{\, b_i} \, | \, i =
1, 2, \cdots, 
a_i + b_i = c \, \mod{\, n} \rr
\end{equation}

For example, for $n = 3$ and 
$Y = \ll 4, 3, 3, 2 \rr$,  
$x^{\, \bi} \, y^{\, -Y} = 
\ll x y^{-4}, x^2 y^{-3}, x^3 y^{-3}, x^4 y^{-2}, x^5, x^6 \cdots \rr$, 
we have 
$[x^{\, \bi} \, y^{\, -Y}]_0 = 
\ll 
x\, y^{-4}, x^3 \, y^{-3}, x^6, \cdots 
\rr$.

\subsection{The power-sum symmetric functions}
Given 
$\bx = \ll x_1, x_2, \cdots \rr$, 
the power-sum symmetric function
$p_n \ll \bx \rr$,
$n \in \ll 0, 1, \cdots \rr$,
is
\footnote{\,
Ch. I, p. 23, in \cite{macdonald.book}
}, 
	
\begin{equation}
p_{\, 0} \ll \bx \rr = 1,
\quad  
p_{\, n} \ll \bx \rr = \sum_{\, i} x_i^{\, n}
,
\quad 
n = 1, 2, \cdots,  
\end{equation}
	
and $p_{\, Y} \ll \bx \rr$, indexed by a Young diagram $Y = \ll y_1, y_2, \cdots \rr$, is
\footnote{\,
Ch. I, p. 24, in \cite{macdonald.book}
}, 
	
\begin{equation}
p_{\, Y}   \ll \bx \rr \, = \, 
p_{\, y_1} \ll \bx \rr\, 
p_{\, y_2} \ll \bx \rr\, 
\cdots
\end{equation}

\subsection{The inner product of power-sum functions}
\label{power.sum.inner.product.schur.basis}
Consider the ring of symmetric functions in a set of variables $\bx = \ll x_1, x_2 \cdots \rr$, 
with constant coefficients. In this case, the power-sum functions are
defined to be orthogonal with inner product
\cite{macdonald.book}
	
\begin{equation}
\langle\, p_{\, Y_1} \ll \bx \rr\, |\, p_{\, Y_2} \ll \bx \rr \rangle_{\, q \, t} = 
z_{\, Y_1}\, \delta_{Y_1 Y_2},
\quad
z_Y 
=
1^{n_1} \ll n_1 ! \rr
2^{n_2} \ll n_2 ! \rr 
\cdots, 
\label{young.diagram.power.sum.inner.product}
\end{equation} 
 
\subsection{The Schur function} 
Given a Young diagram 
$Y = \ll y_1, y_2, \cdots \rr$,
$Y^{\, \intercal} = 
\ll 
y^{\, \intercal}_1, 
y^{\, \intercal}_2, \cdots \rr$,
and a set of variables 
$\bx = \ll x_1, x_2, \cdots, x_n \rr$,
such that 
$| \, \bx \, | \geq y^{\, \intercal}_1$, 
the Schur function
$s_{\, Y_1} \ll\, \bx \, \rr$
is 
\footnote{\,
Ch. I, p. 40, in \cite{macdonald.book}
}, 

\begin{equation}
    s_Y\ll \bx \rr = \frac{\det\ll x_i^{y_j + n - j} \rr_{1\leq i,j \leq n}}{\prod_{1\leq i,j \leq n}\ll x_i - x_j \rr}
\end{equation}

\subsection{The skew Schur function}
For two partitions $Y_1$ and $Y_2$, the product of the Schur 
functions,

\begin{equation}
s_{\, Y_1} \, s_{\, Y_2} = 
\sum_{\, Y} c^{\, Y}_{\, Y_1 \, Y_2} \, s_{\, Y}
\end{equation}

defines the integers $c^{\, Y}_{\, Y_1 \, Y_2}$, and 
the skew Schur symmetric functions are then defined as
\footnote{\, 
Ch. I, p. 69, in \cite{macdonald.book}
},

\begin{equation}
s_{\, Y / Y_1} = 
\sum_{\, Y_2} c^{\, Y}_{\, Y_1 \, Y_2} s_{\, Y_2}
\end{equation}

By definition, 

\begin{equation}\label{schur.id.4}
s_{\emptyset / \emptyset} \ll \bx \rr = 1, 
\quad 
s_{\emptyset/\eta}\ll \bx \rr = 0, \textup{\, if $\eta \neq \emptyset$},
\quad 
Q^{|\lambda| - |\eta|} \, 
s_{\lambda/\eta} \ll \bx \rr = 
\, s_{\lambda/\eta} \ll Q \bx \rr
\end{equation}

\subsection{Cauchy identities} 
The (skew) Schur functions satisfy the Cauchy identities
\footnote{\, 
Ch. I, p. 93, in \cite{macdonald.book}
}, 
	
\begin{equation}
\sum_Y 
s_{Y    /   Y_2} \ll\,\bx\, \rr \, 
s_{Y    /   Y_1} \ll\,\by\, \rr
=
\prod_{i,\, j \, = \, 1}^\infty
\frac{
1}{
1\, -\, x_i\, y_j 
}
\sum_Y  
 \, 
s_{Y_1 / Y} \ll\,\bx\, \rr\, 
s_{Y_2 / Y} \ll\,\by\, \rr
\label{schur.id.1}
\end{equation}

\begin{equation}
\sum_Y 
s_{Y^{\, \intercal} / Y_2^{\, \intercal}} \ll\,\bx\, \rr 
s_{Y        / Y_1  } \ll\,\by\, \rr
=
\prod_{i, j = 1}^\infty \ll 1 + x_i\, y_j \rr
\sum_Y 
s_{Y_1^{\, \intercal} / Y^{\, \intercal}} \ll\,\bx\,\rr\,
s_{Y_2        / Y  } \ll\,\by\,\rr 
\label{schur.id.2}
\end{equation}

\subsection{The Heisenberg algebra}

In this work, a Heisenberg algebra $\cH$ is the infinite-dimensional algebra generated 
by 
$\ba_+ = \ll a_{+1}, a_{+2},\cdots \rr$ and 
$\ba_- = \ll a_{-1}, a_{-2},\cdots \rr$,  
that satisfy the commutation relations, 

\begin{equation}
[a_m, a_n] = m \, \delta_{n + m, 0}
\label{heisenberg.commutation.relations}
\end{equation}

act on the left-state as creation and annihilation 
operators, respectively, and on the  right-state as 
annihilation and creation operators, respectively.

\subsection{The left- and right-Heisenberg states}
The left-Heisenberg state  $ \langle \, \ba_{ Y} \, |$, and 
the right-Heisenberg state $ |       \, \ba_{ Y} \, \rangle$, 
$ Y = \ll y_1, y_2, \cdots \rr$, 
are generated from the left- and the right-vacuum states,

\begin{equation}
\langle\, \ba_Y\, | =  \langle\, 0\, | \, 
a_{\, y_1}\, 
a_{\, y_2}\, 
\cdots,
\quad 
|\, \ba_Y\, \rangle = 
\cdots
a_{ - y_2}\, 
a_{ - y_1}\, 
|\, 0\, \rangle
\end{equation} 

Using the Heisenberg commutation relations,
the inner product of 
$ \langle\, \ba_{\, Y_1}\, |$ 
and 
$ |\, \ba_{\, Y_2}\,\rangle$ 
is, 

\begin{equation}
\langle\, a_{\, Y_1}\, |\, a_{\, Y_2}\,\rangle = 
z_{ Y_1}\, \delta_{ Y_1 Y_2}, 
\quad
z_Y 
=
1^{n_1} \ll n_1 ! \rr
2^{n_2} \ll n_2 ! \rr 
\cdots.
\label{inner.product.right.left.states}
\end{equation}

\subsection{The power-sum/Heisenberg correspondence}
	
From the inner products 
\ref{young.diagram.power.sum.inner.product}
and 
\ref{inner.product.right.left.states}, 
we deduce that the power-sum basis
spanned by $p \ll \bx \rr$ is isomorphic 
to the left-state Heisenberg basis spanned 
by $\langle\,a_Y\, |$, as well as 
the right-state Heisenberg basis 
spanned by $|\, a_Y\, \rangle$, 
	
\begin{equation}
p_n \ll \bx \rr \rightleftharpoons - \, a_{ n},  
\quad 
p_n \ll \bx \rr \rightleftharpoons - \, a_{-n}, \quad 
n \geq 1.
\label{power.sum.heisenberg.correspondence.dual}
\end{equation}

\subsection{The left- and right-Schur states}
	
Expanding the Schur functions in terms of the power-sum functions, 
then formally replacing the latter with Heisenberg generators, we
obtain operator-valued Schur functions that act on left- and right-vacuum states to produce left- and right-Schur states,

\begin{equation}
\langle \, s_{\, Y} \, | =  \langle\, 0\, | \, s_{\, Y} \, \ll \ba_{\, +} \rr,
\quad 
| \, s_{\, Y} \, \rangle = \, s_{\, Y} \ll \ba_{\, -} \rr  |\, 0 \, \rangle.
\end{equation} 

The left- and the right-Schur states satisfy the orthogonality condition, 

\begin{equation}
\langle\,
s_{\, Y_1}  \ll\, \bx\, \rr
\, |\, 
s_{\, Y_2}  \ll\, \bx\, \rr 
\rangle = 
\delta_{Y_1 Y_2}.
\label{schur.inner.product}
\end{equation}

\subsection{The $\Gamma$-operators}
To the Heisenberg algebra 
$\cH$, 
we associate the operators $\Gamma_{\pm}$, 

\begin{equation}\label{q.t.vertex.operators}
\Gamma_{\pm} \ll x \rr = 
\exp 
\ll 
- \sum_{n=1}^\infty 
\frac{
x^{\mp n}
}{
n
}
\, 
a_{\pm\, n} 
\rr
\end{equation}

Using the Heisenberg commutation relations Equation 
\ref{heisenberg.commutation.relations}, we obtain

\begin{equation}
\label{q.t.vertex.operator.commutation.relation}
\Gamma_{+} \ll x^{-1} \rr
\Gamma_{-} \ll y      \rr = 
\ll 
\frac{
1
}{
1 - xy 
} 
\rr 
\Gamma_{-} \ll y      \rr 
\Gamma_{+} \ll x^{-1} \rr.
\end{equation}

\subsection{The action of the $\Gamma_{\pm}$ operators on Schur states}
In \cite{foda.wu.02, foda.zhu}, identities that describe the action of 
the $\Gamma_{\pm}$ operators on states labelled by Macdonald and
$q$-Whittaker symmetric functions, respectively, were derived. In the 
Schur limit, these identities reduce to,  

\begin{equation}
\langle\, s_{\, Y_1}\, |\, 
\prod_{i \, =\, 1}^\infty 
\Gamma_{-} \ll y_j \rr  =  
\sum_Y 
\langle\, s_Y  \, |\, 
s_{Y_1 / Y} \ll \by \rr, 
\quad 
\prod_{i \, =\, 1}^\infty 
\Gamma_+  \ll x^{\, -1}_i \rr\, |\, s_{\, Y_1}\, \rangle =  
\sum_Y  
s_{Y_1 / Y} \ll\, \bx \,\rr\, 
|\, s_Y \rangle .
\label{two.identities}
\end{equation}

\section{The $n$-coloured refined topological vertex}
\label{section.04}

\textit{We construct the $n$-coloured refined topological 
vertex using $n$ free bosons.}

The $n$-coloured refined topological vertex is 
a trivalent vertex with 
an incoming leg labelled by $n$ Young diagrams 
$\bY_1 = \ll Y_{\, 1, 0},\cdots,Y_{\, 1, \, n-1} \rr$,
an outgoing leg labelled by
$n$ Young diagrams 
$\bY_2 = \ll Y_{\, 2, 0},\cdots,Y_{\, 2, \, n-1} \rr$, 
and a preferred leg labelled by a single 
charged Young diagram $\ll Y, c_Y \rr$, 
$c_{\, Y} \in \ll 0, 1, \cdots, n - 1 \rr$. 
We construct the $n$-coloured vertex in six steps
\footnote{\,
The construction of the $n$-coloured vertex and the conjugate
vertex, in this section, was guided by the form of 
the instanton partition function on 
$\CC^{\, 2} / \ZZ_{\, n}$ of the Landau School  
\cite{
alfimov.belavin.tarnopolsky,
alfimov.tarnopolsky, 
belavin.belavin.bershtein, 
belavin.bershtein.feigin.litvinov.tarnopolsky,
belavin.bershtein.tarnopolsky, 
belavin.feigin, 
belavin.mukhametzhanov}
}.

\subsection{\1 Introduce $n$ species of free bosons}
Instead of a single Heisenberg algebra, as in the case of the refined 
topological vertex, we work in terms of $n$ commuting Heisenberg algebras 
$\cH_{\, m}$, 
$m \in \ll 0, 1, \cdots, n-1 \rr$. From now on, all operators that belong 
to $\cH_{\, m}$ will carry a subscript $m$. In particular, we 
have $\Gamma_{\, m \, \pm}$, $m \in \ll 0, 1, \cdots, n -1 \rr$. 

\subsection{\2 Introduce an excited-state, charged Maya diagram} 

We start from a ground-state, 0-charge Maya diagram and use a charged 
Young diagram $\ll Y, c_Y \rr$ as in Section \ref{Maya.diagram.construction} 
to produce an excited-state, charge-$c_{\, Y}$ Maya diagram. 
The position of each stone is shifted from $c_{\, p}$ to  
$\ll c_{\, p} + c_{\, Y} \rr$.

\subsection{\3 Introduce the $\Gamma$-operators}
We map the excited-state, charge-$c_{\, Y}$ Maya diagram to an infinite 
sequence of $\Gamma$-vertex operators $\prod_{Maya \ll Y \rr} \Gamma_{\pm}$,
using the bijections

\begin{equation}
\diagup   \rightleftharpoons \Circle \rightleftharpoons \Gamma_{\, c_H \, +}, 
\quad
\quad
\diagdown \rightleftharpoons \CIRCLE \rightleftharpoons \Gamma_{\, c_H \, -},   
\quad
\quad 
c_{\, H} = \ll c_{\, p} + c_Y \rr \, \mod{\, n},      
\end{equation}

for each stone at position $\ll c_{\, p} + c_Y \rr$. We call $c_H$ the Heisenberg 
charge.

\subsection{\4 Assign arguments to the vertex operators}
We take the arguments of the $\Gamma$-operators to be,

\begin{equation}
\Gamma_{\, c_H \, +} \ll x^{  - i}\, y^{  Y_{ i}            }   \rr, 
\quad 
\Gamma_{\, c_H \, -} \ll y^{j - 1}\, x^{- Y^{\intercal}_{j}}   \rr, 
\label{full.arguments}
\end{equation}

where 
$Y_{\, i}$ is the length of the $i$-row of the Young diagram $Y$ that labels the 
preferred leg of the vertex, and 
$Y^{\intercal}_{j}$ is the length of the $j$-column of the Young diagram 
$Y$. The sum of the exponents of the arguments of $\Gamma_{\pm}$ is related to $c_{p}$ as follows.
For $\Gamma_{\, c_H \, +} \ll x^{  - i}\, y^{  Y_{i}            }   \rr$ at position $\ll c_{\, p} + c_{\, Y} \rr$, 

\begin{equation}
c_{\, p} = Y_{i} - i, 
\end{equation}

and for $\Gamma_{\, c_H \, -} \ll y^{j - 1}\, x^{- Y^{\, \intercal}_j}   \rr$ at
position $\ll c_{p} + c_{\, Y} \rr$, 

\begin{equation}
    c_{p} = -Y^{\, \intercal}_{j} + j - 1
\end{equation}

\subsection{\5 From the infinite sequence of $\Gamma$-operators to an expectation value} 
We evaluate the sequence 
$\prod_{Maya \ll Y \rr} \Gamma_{\pm}$ 
between a left-state,

\begin{equation}
\bra{s_{\mathbf{Y_1}}} = 
\bra{s_{Y_{1, \, 0}}} \otimes \cdots \otimes \bra{s_{Y_{1, \, (n-1)}}} = 
\bra{0} 
s_{Y_{1, \, 0}} 
\ll \ba_{1+} \rr
\cdots 
s_{Y_{1, \, (n-1)}}
\ll 
\ba_{n+} 
\rr
\end{equation}

and a right-state,  

\begin{equation}
\ket{s_{\mathbf{Y_2}}} = 
\ket{s_{Y_{2, \, 0}}} \otimes \cdots \otimes \ket{s_{Y_{2, \, (n-1)}}} = 
s_{Y_{2, \, 0}} \ll \ba_{1-} \rr \cdots 
s_{Y_{2, \, (n-1)}} 
\ll \ba_{n-} \rr
\ket{0}
\end{equation}

to get the unnormalized $n$-coloured refined topological vertex   

\begin{equation}\label{unnormalized.vertex}
\cC_{\bY_1 \, \bY_2 \, Y}^{\, p, \,  c_{\, Y}, \, unnorm}
\ll x, y \rr = 
\bra{s_{\bY_1}}
\ll 
\prod_{
Maya \ll Y \rr
}
\Gamma_{\pm}
\rr 
\ket{s_{\bY_2}}
\end{equation}

For $n = 3$ and  $Y = \ll 4, 3, 3, 2 \rr$
(see Figure  \ref{young.maya.04}),
the sequences for the possible 
$c_{\, Y}$ are 
\footnote{\,
We list all three possibilities, then we choose one.
},  

\begin{multline}
\cC^{\, 3, \, c_{\, Y} = 0, \, unnorm}_{\pmb{Y_1} \pmb{Y_2} Y} \ll x, y \rr = 
\bra{s_{\pmb{Y_1}}} 
\cdots 
\Gamma_{1+} \ll x^{-5}        \rr 
\Gamma_{2-} \ll x^{-4}        \rr
\Gamma_{0-} \ll y      x^{-4} \rr 
\Gamma_{1+} \ll x^{-4} y^2    \rr
\Gamma_{2-} \ll y^2    x^{-3} \rr 
\\
\Gamma_{0+} \ll x^{-3} y^{ 3} \rr 
\Gamma_{1+}\ll  x^{-2} y^{ 3} \rr 
\Gamma_{2-}\ll  y^{3}  x^{-1} \rr 
\Gamma_{0+}\ll  x^{-1} y^{ 4} \rr 
\Gamma_{1-}\ll  y^4           \rr
\cdots
\ket{s_{\pmb{Y_2}}}
\end{multline}

\begin{multline}
\cC^{\, 3, \, c_{\, Y} = 1, \, unnorm}_{\pmb{Y_1} \pmb{Y_2} Y} \ll x,y \rr = 
\bra{s_{\pmb{Y_1}}} 
\cdots 
\Gamma_{2+}\ll x^{-5}         \rr 
\Gamma_{0-}\ll x^{-4}         \rr
\Gamma_{1-}\ll y       x^{-4} \rr 
\Gamma_{2+}\ll x^{-4}  y^2    \rr
\Gamma_{0-}\ll y^2     x^{-3} \rr 
\\
\Gamma_{1+} \ll x^{-3}y^{3} \rr 
\Gamma_{2+} \ll x^{-2}y^{3} \rr 
\Gamma_{0-} \ll y^{3}x^{-1} \rr 
\Gamma_{1+} \ll x^{-1}y^{4} \rr 
\Gamma_{2-} \ll y^4 \rr
\cdots
\ket{s_{\pmb{Y_2}}}
\end{multline}

\begin{multline}
\cC^{\, 3, \, c_{\, Y} = 2, \, unnorm}_{\pmb{Y_1} \pmb{Y_2} Y} \ll x,y \rr = 
\bra{s_{\pmb{Y_1}}} 
\cdots 
\Gamma_{0+}\ll x^{-5}         \rr 
\Gamma_{1-}\ll x^{-4}         \rr
\Gamma_{2-}\ll y       x^{-4} \rr 
\Gamma_{0+}\ll x^{-4}  y^2    \rr
\Gamma_{1-}\ll y^2     x^{-3} \rr 
    \\
\Gamma_{2+} \ll x^{-3} y^{3} \rr 
\Gamma_{0+}\ll x^{-2}  y^{3} \rr 
\Gamma_{1-}\ll y^{3}  x^{-1} \rr 
\Gamma_{2+}\ll x^{-1} y^{4}  \rr 
\Gamma_{0-}\ll y^4           \rr
\cdots
\ket{s_{\pmb{Y_2}}}
\end{multline}

To evaluate the expectation value in Equation \ref{unnormalized.vertex}, we 
normal-order the sequence  
$\ll \prod_{Maya \ll Y \rr} \Gamma_{\pm} \rr$ 
to put 
all $\Gamma_{\, c_H \, +}$ vertex operators on the right, and 
all $\Gamma_{\, c_H \, -}$ vertex operators on the left. 
Since $\Gamma$-operators with different Heisenberg charges commute, the $n$-coloured vertex is a product of $n$ components labelled by $c_H \in \ZZ_{\, n}$, 

\begin{equation}\label{unnormalized.vertex.factored}
\cC_{\bY_1 \, \bY_2 Y}^{\, n, \,  c_{\, Y}, \, unnorm}
\ll x, y \rr = 
\prod_{c_H = 0}^{n-1}
\bra{s_{Y_{1c_H}}}
\ll 
\prod_{
\substack{
Maya\ll Y \rr
\\
c_{\, p} + c_Y = c_H\, \mod{\, n}
}
}
\Gamma_{\pm c_H}
\rr 
\ket{s_{Y_{2c_H}}},    
\end{equation}

where 
$\ll
\prod_{
\substack{
Maya\ll Y \rr
\\
c_{\, p} + c_Y = c_H\, \mod{\, n}
}
}
\Gamma_{\pm c_H} \rr$ 
is restricted to the sub-Maya diagram of $Maya\ll Y \rr$ with stones at positions $\ll c_{\, p} + c_{\, Y} = c_H \, \mod{\, n} \rr$. Within each factor, we commute $\Gamma$-operators (all of which are built from the same Heisenberg algebra) using, 

\begin{multline}
\Gamma_{\, c_H \, +} \ll x^{  - i}\, y^{  Y_{\, i}} \rr  \,
\Gamma_{\, c_H \, -} \ll y^{j - 1}\, x^{- Y^{\intercal}_{\, j}} \rr =
\\
\ll 
\frac{
1
}{
1 - 
x^{\, - \, Y^{\intercal}_{\, 3, \, j} + i} \, y^{- Y_{\, 3, \, i} + j - 1}
} 
\rr
\Gamma_{\, c_H \, -} \ll  y^{\, j \, - \, 1} \, x^{\, - \, Y^{\intercal}_{\, 3, \, j}} \rr\,
\Gamma_{\, c_H \, +} \ll  x^{\,      - \, i} \, y^{\,      Y_{\, 3, \, i}} 
\rr
\label{q.t.vertex.operator.commutation.relation.with.arguments}
\end{multline}

Since $\Gamma_{\, c_H \, +}$ is attached to a segment 
$\diagup$ in the borderline of $Y$, and 
$\Gamma_{\, c_H \, -}$ is attached to an adjacent segment 
$\diagdown$ to the right of the former, 
the commutation relation, 
Equation \ref{q.t.vertex.operator.commutation.relation.with.arguments}, is the same as adding a strip of length $n$ to $Y$. Repeating this process, 

\begin{multline}
\cC_{
\mathbf{Y_1}
\mathbf{Y_2} Y}^{\, n, \, c_{\, Y}, \, unnorm}\ll x,y \rr = 
\prod_{c_H = 0}^{n-1}
\prod_{
\substack{
(i,j) \notin Y
\\ 
Y_{\, i} - i + c_{\, Y} = c_H\, \mod{\, n}
\\
-Y^{\, \intercal}_{\, j}+j-1+c_{\, Y} = c_H\, \mod{\, n}}}
\frac{
1 
}{
1 - x^{-Y^{\, \intercal}_{\, j} + i} y^{-Y_{\, i} + j - 1}} 
\\
\bra{s_{\mathbf{Y_1}}}
\prod_{c_H = 0}^{n-1} 
\ll 
\prod_{
\substack{
j=1
\\
-Y^{\, \intercal}_{\, j} + j - 1 + c_{\, Y} = c_H \, \mod{\, n}}}^\infty 
\Gamma_{c_H-} \ll y^{\, j - 1}x^{-Y^{\, \intercal}_{\, j}} \rr 
\prod_{
\substack{
i=1
\\
Y_{\, i} - i + c_{\, Y} = c_H\, \mod{\, n}}}^\infty
\Gamma_{c_H+} \ll x^{-i} y^{Y_{\, i}} \rr 
\rr
\ket{s_{\mathbf{Y_2}}}
\\
= 
\ll 
\prod_{
\substack{
\wsquare \notin Y
\\ 
A_{\wsquare,Y} + L_{\wsquare, Y}+1 = 0\, \mod{\, n}}
}
\frac{
1 
}{
1 - x^{-L_{\wsquare,Y}} y^{-A^{++}_{\wsquare,Y}}} \rr 
\prod_{c_H = 0}^{n-1}
\sum_{Y} s_{Y_{1c_H}/Y}
\ll [y^{\, \bj - 1} x^{\, - Y^{\, \intercal}}]_{c_H - c_{\, Y}} \rr 
s_{Y_{2 c_H}/Y} 
\ll [x^{\, \bi} y^{ - Y}]_{-c_H+c_{\, Y}} \rr
\end{multline}

where, to obtain the last equality, we used $\bi = \bj = \ll 1, 2, \cdots \rr$, 
Equations \ref{two.identities} and \ref{schur.inner.product}, and the definition of 
$[x^{\, \ba}y^{\, \bb}]_c$, for any two sequences of integers $\ba, \bb$, in Equation 
\ref{subsequence.symbol}.

\subsection{6. Normalization of the vertex}

Using the unnormalized vertex 
$\cC^{\, n, \, c_{\, Y}, \, unnorm}_{\mathbf{Y_1} \mathbf{Y_2} Y} \ll x, y \rr$, 
we obtain the normalization  
$\cC^{\, n, \, c_{\, Y}, \, unnorm}_{\emptyset\emptyset\emptyset} \ll x, y \rr$,
from which we then obtain the normalized vertex:

\begin{multline}
\cC_{\mathbf{Y_1}\mathbf{Y_2}Y}^{\, n, \, c_{\, Y}}\ll x,y \rr = 
\frac{
\cC^{\, n, \, c_{\, Y}, \, unnorm}_{\mathbf{Y_1} \mathbf{Y_2} Y} \ll x, y \rr
}{
\cC^{\, n, \, c_{\, Y}, \, unnorm}_{\emptyset\emptyset\emptyset} \ll x, y \rr}
\\
= 
\prod_{
\substack{
i, j = 1
\\
1 - i - j = 0 \, \mod{\, n}
}
}^\infty
\ll 1 - x^{i} y^{\, j - 1} \rr 
\ll 
\prod_{
\substack{
\wsquare \notin Y
\\ 
A_{\wsquare,Y} + L_{\wsquare, Y}+1 = 0\, \mod{\, n}}
}
\frac{
1   
}{
1 - x^{-L_{\wsquare,Y}} y^{-A_{\wsquare,Y}^{++}}} \rr 
\\
\prod_{c_H = 0}^{n-1} 
\sum_{Y} s_{Y_{1c_H}/Y}
\ll [y^{\, \bj-1} x^{-Y^{\, \intercal}}]_{c_H-c_{\, Y}} \rr 
    s_{Y_{2c_H}/Y} \ll [x^{\, \bi} y^{-Y}]_{-c_H+c_{\, Y}} \rr
\\
= 
\ll 
\prod_{
\substack{
\wsquare \in Y
\\ 
A_{\wsquare,Y} + L_{\wsquare, Y}+1 = 0\, \mod{\, n}}
}
\frac{
1 
}{
1 - x^{L_{\wsquare,Y}^{++}} y^{A_{\wsquare,Y}}} \rr 
\prod_{c_H = 0}^{n-1}
\sum_{Y} s_{Y_{1c_H}/Y}
\ll [y^{\, \bj-1} x^{-Y^{\, \intercal}}]_{c_H-c_{\, Y}} \rr 
s_{Y_{2c_H}/Y} \ll [x^{\, \bi} 
y^{-Y}]_{-c_H+c_{\, Y}} \rr.
\end{multline}

Then, on defining 

\begin{equation}
Z^{\, n}_{\, Y} \ll x, y \rr = 
\prod_{c_H = 0}^{n-1}
\prod_{
\substack{
(i,j) \in Y
\\ 
Y_{\,i} - i + c_{\, Y} = c_H\, \mod{\, n}
\\
-Y^{\, \intercal}_{\,j}+j-1+c_{\, Y} = c_H\, \mod{\, n}}}
\frac{
1 
}{
1 - x^{Y^{\, \intercal}_{\,j}-i+1}y^{Y_{\,i}-j}} 
= 
\prod_{
\substack{
\wsquare \in Y
\\
A_{\wsquare, Y}+L_{\wsquare, Y} + 1 = 0\, \mod{\, n}
}}
\frac{
1
}{
1 - x^{L^{++}_{\wsquare, Y}}y^{A_{\wsquare, Y}}
}, 
\end{equation}

we obtain 

\begin{equation}
\boxed{
\cC_{\mathbf{Y_1} \mathbf{Y_2}Y}^{\, n, \, c_{\, Y}} \ll x,y \rr = 
Z^{\, n}_{\, Y}\ll x,y \rr 
\prod_{c_H = 0}^{n-1}
\sum_{Y} s_{Y_{1c_H}/Y}
\ll [y^{\, \bj-1} x^{-Y^{\, \intercal}}]_{c_H-c_{\, Y}} \rr 
s_{Y_{2c_H}/Y} \ll [x^{\, \bi} y^{-Y}]_{-c_H+c_{\, Y}} \rr
}
\end{equation}

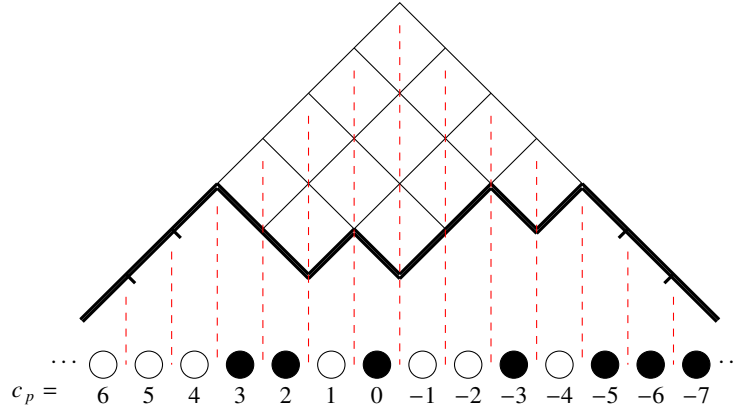
\begin{figure}
\begin{center}
\begin{scriptsize}
\begin{tikzpicture}[scale=.6]
\draw [thin] (-1.5,1.0)--(5.5,8.0)--(12.5,1.0);
\draw [thin] ( 2.5,3.0)--(6.5,7.0);
\draw [thin] ( 3.5,2.0)--(7.5,6.0);
\draw [thin] ( 5.5,2.0)--(8.5,5.0);
\draw [thin] ( 8.5,3.0)--(9.5,4.0);
\draw [thin] ( 1.5,4.0)--(3.5,2.0);
\draw [thin] ( 2.5,5.0)--(5.5,2.0);
\draw [thin] ( 3.5,6.0)--(6.5,3.0);
\draw [thin] ( 4.5,7.0)--(8.5,3.0);

\draw [very thick] (-1.50, 1.00)--(1.5, 4.0);
\draw [very thick] (-1.45, 0.95)--(1.5, 3.9);

\draw [very thick] ( 1.5,4.0)--(3.5,2.0);
\draw [very thick] ( 1.5,3.9)--(3.5,1.9);

\draw [very thick] ( 3.5,2.0)--(4.5,3.0);
\draw [very thick] ( 3.5,1.9)--(4.5,2.9);

\draw [very thick] ( 4.5,3.0)--(5.5,2.0);
\draw [very thick] ( 4.5,2.9)--(5.5,1.9);

\draw [very thick] ( 5.5,2.0)--(7.5,4.0);
\draw [very thick] ( 5.5,1.9)--(7.5,3.9);

\draw [very thick] ( 7.5,4.0)--(8.5,3.0);
\draw [very thick] ( 7.5,3.9)--(8.5,2.9);

\draw [very thick] ( 8.5,3.0)--(9.5,4.0);
\draw [very thick] ( 8.5,2.9)--(9.5,3.9);

\draw [very thick] ( 9.5, 4.00)--(12.50, 1.00);
\draw [very thick] ( 9.5, 3.90)--(12.45, 0.95);

\draw [very thick] ( 10.3,2.8)--(10.5,3.0);
\draw [very thick] ( 11.3,1.8)--(11.5,2.0);
\draw [very thick] ( -0.5,2.0)--(-0.3,1.8);
\draw [very thick] ( 0.5,3.0)--(0.7,2.8);

\foreach \i in {0,...,6} 
{
\draw [dashed, red] (\i - .5, \i + 1.5)--(\i - .5, 0);
}
\foreach \j in {7,...,12} 
{
\draw [dashed, red](\j -.5, 13.5 -\j)--(\j - .5, 0);
}
\node [left]  at (-1.3,0) {$\cdots$};
\node [right] at (12.3,0) {$\cdots$};
\foreach \x in {0,...,2}
{
\draw                 (\x- 1,0) circle (0.3);
\draw [fill=black!100] (\x+10,0) circle (0.3);
}

\draw [fill=black!100] (2,0) circle (0.3);
\draw [fill=black!100] (3,0) circle (0.3);
\draw (4,0) circle (0.3);
\draw [fill=black!100] (5,0) circle (0.3);
\draw (6,0) circle (0.3);
\draw  (7,0) circle (0.3);
\draw [fill=black!100] (8,0) circle (0.3);
\draw (9,0) circle (0.3);

\foreach \a in {6, 5, ..., 0} 
{
\node [below] at (- \a + 5.0, -.3) {$\a$};
}
\foreach \a in {1, 2, ..., 7} 
{
\node [below] at (  \a + 5.0, -.3) {$ - \a$};
}				

\node [below] at (-2.5, -.3) {$c_{\, p} = $};

\end{tikzpicture}
\end{scriptsize}
\end{center}
\caption{\it An $n$-coloured Maya diagram for conjugate vertex that corresponds to a partition $Y = \ll 4,3,3,2 \rr$. Note that the position coordinates are reversed with respect to those in Figure \ref{young.maya.02}.
}
\label{conjugate.p.coloured.maya.diagram}
\end{figure}

\subsection{The conjugate vertex}
To form strips, we need the conjugate vertex which is obtained by reversing the positions of the Maya diagram as shown in Figure \ref{conjugate.p.coloured.maya.diagram}, 
that is, $c_{p} \mapsto - \, c_{p} - 1$, 
and repeat the construction of the vertex 
outlined above, but now the sum of the exponents of the argument of $\Gamma_{\pm}$ is related to the initial position $c_{p}$ in a different way. 
For 
$\Gamma_{\, c_H \, +} 
\ll x^{\, - i}\, y^{\, Y_{\, i}} \rr$, 
at position 
$\ll c_{\, p} + c_{\, Y} \rr$, 

\begin{equation}
    c_{  p} = -Y_{i} + i - 1, 
\end{equation}

and for $\Gamma_{\, c_H \, -} \ll y^{j - 1}\, x^{- Y^{\, \intercal}_{j}}   \rr$, at position $\ll c_{\, p} + c_{\, Y} \rr$,

\begin{equation}
c_{\, p} = Y^{\, \intercal}_{j} - j, 
\end{equation}

so that,

\begin{multline}
\cC_{
\mathbf{Y_1}
\mathbf{Y_2}Y}^{\, n, \, c_{\, Y}, \, unnorm \, *}
\ll x, y \rr = 
\prod_{c_H = 0}^{n-1}
\bra{s_{\mathbf{Y_1}}} 
\ll 
\prod_{
\substack{
Maya\ll Y \rr
\\
c_{\, p}+c_Y = c_{\, H} \, \mod{\, n}
}
}
\Gamma_{c_H \pm}
\rr 
\ket{s_{\mathbf{Y_2}}}
\\
= 
\prod_{c_{\, H} = 0}^{\, n - 1}
\prod_{
\substack{
(i,j) \notin Y
\\ 
-Y_{\,i} + i - 1 + c_{\, Y} = 
c_{\, H} \, \mod{\, n}
\\
Y^{\, \intercal}_{\, j} - j + c_{\, Y} = c_{\, H} \, \mod{\, n}
}
}
\frac{
1 
}{
1 - x^{-Y^{\, \intercal}_{\,j}+i}y^{-Y_{\,i}+j-1}} 
\\
\bra{s_{\mathbf{Y_1}}}
\prod_{c_{\, H} = 0}^{n-1} 
\ll 
\prod_{
\substack{
j=1
\\
Y^{\, \intercal}_{\,j}-j+c_{\, Y} = c_H \, \mod{\, n}}}^\infty 
\Gamma_{\, c_H \, -} 
\ll 
y^{\, j - 1}
x^{\, - Y^{\, \intercal}_{\,j}} 
\rr 
\prod_{
\substack{
i=1
\\
-Y_{\, i} + i - 1 + c_{\, Y} = c_H
\, \mod{\, n}}}^\infty
\Gamma_{\, c_H \, +} 
\ll 
x^{-i} y^{Y_{\,i}} 
\rr 
\rr
\ket{s_{\mathbf{Y_2}}}
\\
= 
\ll 
\prod_{
\substack{
\wsquare \notin Y
\\ 
A_{\wsquare,Y} + L_{\wsquare, Y}+1 = 0\, \mod{\, n}}
}
\frac{
1 
}{
1 - x^{-L_{\wsquare,Y}} y^{-A^{++}_{\wsquare,Y}}} \rr 
\prod_{c_H = 0}^{n-1}\sum_{Y} s_{Y_{1c_H}/Y}
\ll [y^{\, \bj-1} x^{-Y^{\, \intercal}}]_{-c_H+c_{\, Y}-1} \rr 
s_{Y_{2c_H}/Y} \ll [x^{\, \bi} y^{-Y}]_{c_H-c_{\, Y}+1} \rr, \end{multline}

and after normalization,

\begin{equation}
\boxed{
\cC_{
\mathbf{Y_1} 
\mathbf{Y_2} Y}^{\, n, \, c_{\, Y}*} 
\ll x, y \rr = 
Z^{\, n}_{\, Y} 
\ll x,y \rr 
\prod_{c_H = 0}^{n-1}
\sum_{Y} 
s_{Y_{1c_H} / Y} 
\ll [y^{\, \bj-1} x^{-Y}]_{-c_H + c_{\, Y}-1} \rr 
s_{Y_{2c_H} / Y} 
\ll [x^{\, \bi}  \, y^{-Y^{\, \intercal}}]_{c_H - c_{\, Y} + 1} \rr
}
\end{equation}

\begin{scriptsize}
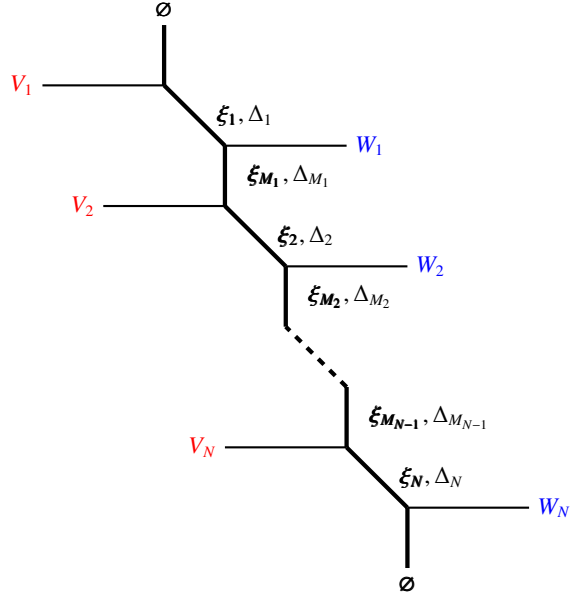
\begin{figure}
\begin{tikzpicture} [scale=.8]

\draw [ultra thick] (0,0)--(0,1);
\node [below] at (0,0) {$\pmb{\emptyset}$};
\draw [thick] (0,1)--(2,1);
\node [right, blue] at (2,1) {$W_N$}; 
\draw [ultra thick] (0,1)--(-1,2);
\node [right] at (-0.25,1.5) {$\pmb{\xi_N}, \Delta_N$};
\draw [thick] (-1,2)--(-3,2);
\node [left, red] at (-3,2) {$V_N$};
\draw [ultra thick] (-1,2)--(-1,3);
\node [right] at (-0.75,2.5) {$\pmb{\xi_{M_{N-1}}}, \Delta_{M_{N-1}}$};
\draw [ultra thick, dashed] (-1,3)--(-2,4);
\draw [ultra thick] (-2,4)--(-2,5);
\node [right] at (-1.75,4.5) {$\pmb{\xi_{M_2}}, \Delta_{M_2}$};
\draw [thick] (-2,5)--(0,5);
\node [right, blue] at (0,5) {$W_2$};
\draw [ultra thick] (-2,5)--(-3,6);
\node [right] at (-2.25,5.5) {$\pmb{\xi_2}, \Delta_2$};
\draw [thick] (-3,6)--(-5,6);
\node [left, red] at (-5,6) {$V_2$};
\draw [ultra thick] (-3,6)--(-3,7);
\node [right] at (-2.75,6.5) {$\pmb{\xi_{M_1}}, \Delta_{M_1}$};
\draw [thick] (-3,7)--(-1,7);
\node [right, blue] at (-1,7) {$W_1$};
\draw [ultra thick] (-3,7)--(-4,8);
\node [right] at (-3.25,7.5) {$\pmb{\xi_1}, \Delta_1$};
\draw [thick] (-4,8)--(-6,8);
\node [left, red] at (-6,8) {$V_1$};
\draw [ultra thick] (-4,8)--(-4,9);
\node [above] at (-4,9) {$\pmb{\emptyset}$};

\end{tikzpicture} 
\caption{
\textit{
The strip diagram made of $N$ pairs of $n$-coloured vertices 
and conjugate vertices. Each internal line is labeled by 
a partition and a K{\"a}hler parameter. Each external 
(horizontal) line is labeled by a partition. The external 
legs are preferred and labelled by Young's diagrams 
$\bV = 
\ll 
\ll 
V_1, \, c_{V_1} 
\rr, \cdots, 
\ll 
V_N,\, c_{V_N}
\rr 
\rr$ and 
$\bW = 
\ll 
\ll W_1,\, c_{W_1}
\rr, \cdots, 
\ll 
W_N, \, c_{W_N}
\rr 
\rr$.
}
} 
\label{strip.diagram} 
\end{figure}
\end{scriptsize}

\begin{figure} 
\begin{tikzpicture} [scale=.7]
\draw [thick] (-6,4)--(-4,4);
\draw [ultra thick] (-4,6)--(-4,4);
\draw [ultra thick] (-4,4)--(-3,3);
\node [left] at (-6,4) {$Y$} ;
\node [right] at (-3,3) {$\pmb{Y_2}$};
\node [above] at (-4,6) {$\pmb{Y_1}$};

\draw [ultra thick] (1,6)--(2,5);
\draw [ultra thick] (2,5)--(2,3);
\draw [thick] (2,5)--(4,5);
\node [left] at (1,6) {$\pmb{Y_2}$} ;
\node [below] at (2,3) {$\pmb{Y_1}$};
\node [right] at (4,5) {$Y$};
\end{tikzpicture} 
\caption{
\textit{
The vertex $\cC^{\, p, \, c_{\nu}}_{\pmb{Y_1Y_2}Y} \ll x, y \rr$ is on the left and
the conjugate vertex
$\cC^{\, p, \, c_{\nu}*}_{\pmb{Y_1}^{\intercal} \pmb{Y_2}^{\intercal} Y^{\intercal}} \ll y, x \rr $ 
is on the right.
} 
} 
\label{topological.vertices} 
\end{figure} 

\section{From $n$-coloured vertices to $n$-coloured $N$-strip partition functions}
\label{section.05}
\textit{We construct 5D strip partition functions from 
$N$ pairs of $n$-coloured refined topological vertices, 
where a pair of vertices consists of a vertex and a conjugate vertex.}

We consider an $N$-strip partition function on 
$\CC^{\, 2}/\ZZ_{\, n}$ shown in Figure \ref{strip.diagram}. 
The horizontal external legs are labelled by Young diagrams 
with $\mathbb{Z}_n$-charges. Denote a Young diagram 
$Y$ carrying $\mathbb{Z}_n$-charge $c_Y \in \mathbb{Z}_n$ by 
$\ll Y,c_Y \rr$. There are $N$ horizontal external legs to the left 
labelled by,  

\begin{equation}
    \ll \bV,\, \bc_{\bV}\rr = \ll \ll V_1,\, c_{V_1} \rr,\cdots, \ll V_N,\, c_{V_N} \rr \rr
\end{equation}

There are $N$ horizontal external legs to the right labelled by, 

\begin{equation}
\ll \bW,\, \bc_{\bW}\rr = \ll \ll W_1,\, c_{W_1} \rr,\cdots, \ll W_N,\, c_{W_N} \rr \rr
\end{equation}

Each internal edge is labelled by uncharged
$n$-Young diagrams   
$\pmb{\xi} = \ll \xi_0,\cdots, \xi_{n-1} \rr$.
For an internal edge with diagram $\pmb{\xi}_i$ we define, 

\begin{equation}
\text{Edge factor} = \ll - Q_{\, i} \rr^{|\, \pmb{\xi}_i \, |}, 
\quad 
Q_{\, i} = e^{- R \, \Delta_i}, 
\quad i = 1, M_1, 2, M_2, \cdots, M_{N-1}, N 
\end{equation}

The external vertical legs are labelled by $n$ null Young 
diagrams $\pmb{\emptyset}$.

\subsection{The unnormalized strip partition function}
The unnormalized $N$-strip $\CC^{\, 2}/\ZZ_{\, n}$ partition 
function can be written directly from Figure 
\ref{strip.diagram}, 

\begin{multline}
\cS^{\, n, \, unnorm
}_{
\ll \bV, \, \bc_{\bV} \rr
\ll \bW, \, \bc_{\bW} \rr
\bDelta
}
\ll x, y, R \rr
=
\\
\sum_{
\pmb{\xi_{\, 1}}, \, \cdots, \, \pmb{\xi_{\, N}}
}
\sum_{
\pmb{\xi_{\, M_1}}, \, \cdots, \, \pmb{\xi_{\, M_{\, N - 1}
}
}
}
\prod_{I = 1}^N
\ll 
\ll -Q_{\, i}        \rr^{\, | \, \pmb{\xi_I}     \, |} 
\ll -Q_{\, M_{\, I}} \rr^{\, | \, \pmb{\xi_{M_I}} \, |} 
\cC^{\, n, \,  V_I
}_{
\pmb{\xi_{\, M_{I-1}}}^{\, \intercal} 
\pmb{\xi_{\, I      }} 
V^\intercal_I
} 
\ll x, y \rr 
\cC^{\, n, \, W_I \, *
}_{
\pmb{\xi_{\, M_I}} 
\pmb{\xi_{\, I}}^{\, \intercal} 
W_I
}
\ll y, x \rr 
\rr
\end{multline}

where the summations are over all possible internal $n$-Young diagrams.
Expanding the definition of the vertices and  the conjugate vertices gives

\begin{multline}
\label{uneval.s.N}
\cS^{\, n, \, unnorm}_{\ll \, \bV, \, \bc_{\bV} \rr
\ll \bW, \, \bc_{\bW} \rr \bDelta
} 
\ll x, y, R \rr
\\ 
= 
\prod_{I=1}^N
\ll 
Z^{\, n}_{V_I} \ll x, y \rr 
Z^{\, n}_{W^{\, \intercal}_I} 
\ll y, x \rr 
\rr
\sum_{
\pmb{\xi_{1  }}, \, \cdots, \, \pmb{\xi_{N}    }
}
\sum_{
\pmb{\xi_{M_1}}, \, \cdots, \, \pmb{\xi_{M_{N-1}
}
 }
}
\prod_{I=1}^N 
\ll 
\ll - Q_{\,   i} \rr^{\, | \, \pmb{\xi_I}     \, |} 
\ll - Q_{\, M_I} \rr^{\, | \, \pmb{\xi_{M_I}} \, |}
\rr
\\
\prod_{c_H= 0 }^{n-1} 
\sum_{\eta'_I}
s_{\, \xi^{\, \intercal}_{\, M_{I-1} c_H} /\eta'_I}
\ll 
[x^{-V_I} y^{\, \bj-1}]_{c_H - c_{V_I}} 
\rr
s_{\, \xi_{I c_H} / \eta'_I}
\ll 
[x^{\, \bi} y^{-V^{\, \intercal}_I}]_{- c_H + c_{V_I}} 
\rr
\\
\prod_{c_H= 0 }^{n-1}
\sum_{\eta''_I}
s_{\xi_{\, M_I c_H}/\eta''_I}
\ll 
[y^{- W^{\, \intercal}_I} x^{\, \bi - 1}]_{- c_H + c_{W_I} - 1}
\rr
s_{\xi^{\, \intercal}_{I c_H} / \eta''_I}
\ll 
[y^{\, \bj} x^{- W_I}]_{c_H - c_{W_I} + 1} 
\rr
\\
= 
\prod_{I=1}^N
\ll Z^{\, n}_{V_I}\ll x,y \rr 
Z^{\, n}_{W^{\, \intercal}_I}
\ll y, x \rr 
\rr
\prod_{c_H= 0 }^{n-1}
\sum_{\ll \xi_{1  }, \cdots, \xi_{N      } \rr}
\sum_{\ll \xi_{M_1}, \cdots, \xi_{M_{N-1}} \rr}
\prod_{I=1}^N 
\ll 
\ll -Q_{\, i} \rr^{|\xi_I|} 
\ll -Q_{M_I} \rr^{|\xi_{M_I}|} 
\rr
\\
\sum_{\eta'_I}
s_{\xi^{\, \intercal}_{M_{I-1}}/\eta'_I}
\ll [x^{-V_I}y^{\, \bj-1}]_{c_H-c_{V_I}} \rr
s_{\xi_{I}/\eta'_I}
\ll 
[x^{\, \bi}
y^{-V^{\, \intercal}_I}]_{-c_H+c_{V_I}} \rr
\\
\sum_{\eta''_I}
s_{\xi_{M_I}/\eta''_I}
\ll 
[ y^{-W^{\, \intercal}_I} x^{\, \bi-1}]_{-c_H+c_{W_I}-1} 
\rr
s_{\xi^{\, \intercal}_{I}/\eta''_I}
\ll 
[ y^{\, \bj} x^{-W_I}]_{c_H-c_{W_I}+1} 
\rr
\end{multline}

where 
$\xi_{M_0} = \xi_{M_N} = \emptyset$. 
Using the result of Appendix \ref{appendix.B} then gives

\begin{multline}
\label{5d.N.strip.partition.function}
\cS^{\, n, \, unnorm}_{\ll \bV,\, \bc_{\bV} \rr
\ll \bW,\, \bc_{\bW}\rr \bDelta} 
\ll x, y, R \rr = 
\prod_{I=1}^N 
\ll 
Z^{\, n}_{V_I}
\ll x,y \rr 
Z^{\, n}_{W^{\, \intercal}_I}
\ll y,x \rr 
\rr
\\
\prod_{c_H= 0 }^{n-1} 
\prod_{J=1}^N \prod_{I=1}^J 
\prod_{
\substack{
i=1
\\
-V^{\, \intercal}_{I, i} + i = - c_H + c_{V_I}\, \mod{\, n}
}
}^\infty
\prod_{
\substack{
j=1
\\
-W_{J,j}+j-1= c_H - c_{W_J} \, \mod{\, n} 
}
}^\infty
\ll 1 - \prod_{K=I}^{J-1}Q_{M_K}\prod_{K=I}^J Q_K x^{-W_{J,j}+i}y^{-V^{\, \intercal}_{I,i}+j} \rr
\\
\prod_{c_H= 0 }^{n-1}\prod_{J=1}^N\prod_{I=1}^{J-1}
\prod_{
\substack{
i=1
\\
-W^{\, \intercal}_{I,i}+i-1 = -c_H+c_{W_I}-1 \, \mod{\, n}
}
}^\infty
\prod_{
\substack{
j=1
\\
-V_{J,j}+j-1 = c_H-c_{V_J}\, \mod{\, n}
}
}^\infty
\ll 1 - \prod_{K=I}^{J-1} Q_{M_K} \prod_{K=I+1}^{J-1} Q_K x^{-V_{J,j}+i-1}y^{-W^{\, \intercal}_{I,i}+j-1} \rr
\\
\prod_{c_N= 0 }^{n-1}\prod_{J=1}^N\prod_{I=1}^{J-1}
\prod_{
\substack{
i=1
\\
-V^{\, \intercal}_{I,i}+i = -c_H + c_{V_I}\, \mod{\, n}
}
}^\infty
\prod_{
\substack{
j=1
\\
-V_{J,j}+j-1 = c_H - c_{V_J} \, \mod{\, n}
}
}^\infty
\ll 1 - \prod_{K=I}^{J-1} Q_{M_K} 
\prod_{K=I}^{J-1} Q_K x^{-V_{J,j}+i}y^{-V^{\, \intercal}_{I,i}+j-1} \rr^{-1}
\\
\prod_{c_H= 0 }^{n-1}
\prod_{J=1}^N
\prod_{I=1}^{J-1}
\prod_{
\substack{
i=1
\\
-W^{\, \intercal}_{I,i}+i-1 = -c_H + c_{W_I}-1\, \mod{\, n}
}
}^\infty
\prod_{
\substack{j=1\\
    -W_{J,j}+j-1 = c_H - c_{W_J} \, \mod{\, n}
    }
    }^\infty
    \ll 1 - \prod_{K=I}^{J-1} Q_{M_K} \prod_{K=I+1}^J Q_K x^{-W_{J,j}+i-1}y^{-W^{\, \intercal}_{I,i}+j} \rr^{-1}
    \\
    = \prod_{I=1}^N \ll Z^{\, n}_{V_I}\ll x,y \rr Z^{\, n}_{W^{\, \intercal}_I}\ll y,x \rr \rr
    \\
    \prod_{J=1}^N\prod_{I=1}^J
    \prod_{\substack{i,j=1\\
    (-V^{\, \intercal}_{I,i}+j) + (-W_{J,j}+i) - 1 + c_{W_J}-c_{V_I} = 0\, \mod{\, n}
    }}^\infty
    \ll 1 - \prod_{K=I}^{J-1}Q_{M_K}\prod_{K=I}^J Q_K x^{-W_{J,j}+i}y^{-V^{\, \intercal}_{I,i}+j} \rr
    \\
    \prod_{J=1}^N\prod_{I=1}^{J-1}
    \prod_{\substack{i,j=1\\
    (-W^{\, \intercal}_{I,i}+j) + (-V_{J,j}+i) - 1 + c_{V_J}-c_{W_I} = 0\, \mod{\, n}
    }}^\infty
    \ll 1 - \prod_{K=I}^{J-1} Q_{M_K} \prod_{K=I+1}^{J-1} Q_K x^{-V_{J,j}+i-1}y^{-W^{\, \intercal}_{I,i}+j-1} \rr
    \\
    \prod_{J=1}^N\prod_{I=1}^{J-1}
    \prod_{\substack{i,j=1\\
    (-V^{\, \intercal}_{I,i}+j) + (-V_{J,j}+i) - 1 + c_{V_J}-c_{V_I} = 0\, \mod{\, n}
    }}^\infty
    \ll 1 - \prod_{K=I}^{J-1} Q_{M_K} \prod_{K=I}^{J-1} Q_K x^{-V_{J,j}+i}y^{-V^{\, \intercal}_{I,i}+j-1} \rr^{-1}
    \\
    \prod_{J=1}^N\prod_{I=1}^{J-1}
    \prod_{\substack{i,j=1\\
    (-W^{\, \intercal}_{I,i}+j) + (-W_{J,j}+i) - 1 + c_{W_J}-c_{W_I} = 0\, \mod{\, n}
    }}^\infty
    \ll 1 - \prod_{K=I}^{J-1} Q_{M_K} \prod_{K=I+1}^J Q_K x^{-W_{J,j}+i-1}y^{-W^{\, \intercal}_{I,i}+j} \rr^{-1} 
\end{multline}

\subsection{The normalization of the strip}

For $n=1$, the normalized $N$-strip partition 
function on $\mathbb{R}^4$ is obtained by normalizing 
$\cS_{\, \bV  \, \bW  \, \bDelta}\ll x,y,R \rr$ by 
$\cS_{\, \pmb{\emptyset}  \, \pmb{\emptyset}  \, \bDelta}\ll x, y, R \rr$, 
so that $\cS^{\, norm}_{\, \pmb{\emptyset}  \, \pmb{\emptyset}  \, \bDelta} \ll x, y, R \rr = 1$. Similarly, for $n > 1$, we write the normalized strip partition 
function on $\mathbb{R}^4/\mathbb{Z}_n$ as,

\begin{equation}
\label{normalized.5d.strip.partition.function}
\cS_{
\,  \ll \bV,\, \bc_{\bV} \rr \, \ll \bW, \, \bc_{\bW} \rr \, \bDelta
}^{
\, n, \, norm} \ll x, y, R \rr = 
\frac{
\cS_{\, \ll \bV,\, \bc_{\bV}\rr \, \ll \bW,\, \bc_{\bW}\rr \, \bDelta}^{\, n} \ll x, y, R \rr
}{
\cS_{
\, 
\ll \bemptyset,\, \bc_{\bV} \rr \, 
\ll \bemptyset,\, \bc_{\bW} \rr \, 
\bDelta}^{\, n} \ll x, y, R \rr
}
\end{equation}

From Equation \ref{5d.N.strip.partition.function} 
and identity \ref{R.0.normalized.product.formula},  

\begin{multline}
\label{after.identity}
\cS^{\, n, \,norm}_{
\ll \bV, \, \bc_{\bV} \rr 
\ll \bW, \, \bc_{\bW} \rr 
\bDelta}
\ll x, y, R \rr = 
\prod_{I = 1}^N
\ll 
Z^{\, n}_{V_I}            \ll x, y \rr 
Z^{\, n}_{W^{\, \intercal}_I}  \ll y, x \rr 
\rr
\\
\prod_{I = 1}^N
\prod_{J = 1}^I
\prod_{
\substack{
\bsquare \in W_I
\\
A_{\bsquare,W_I} + 
L_{\bsquare,V_J} + 1 - c_{W_I} +c_{V_J} = 0  \, \mod{\, n}
}
} 
\ll 
1 - \prod_{K = J}^{I - 1} Q_{M_K} 
\prod_{K=J}^I 
Q_K 
x^{- A_{\bsquare,W_I}} \, y^{- L_{\bsquare,V_J}} 
\rr
\\
\prod_{
\substack{
\wsquare \in V_J
\\
A_{\wsquare, V_J} + 
L_{\wsquare, W_I} + 1 + c_{W_I} - c_{V_J} = 0 \, \mod{\, n}
}
}
\ll 1 - \prod_{K = J}^{I - 1} 
Q_{M_K}
\prod_{K=J}^I 
Q_K 
x^{A^{++}_{\wsquare, V_J}} \, y^{L^{++}_{\wsquare, W_I}} 
\rr
\\
\prod_{I=1}^N 
\prod_{J=1}^{I-1} 
\prod_{
\substack{
\wsquare \in  V_I
\\
A_{\wsquare, V_I} + 
L_{\wsquare, W_J} + 1 - c_{V_I} + c_{W_J} = 0  \, \mod{\, n}
}
}
\ll 1 - 
\prod_{K=J}^{I-1} Q_{M_K} 
\prod_{K=J+1}^{I-1} 
Q_K 
x^{-A_{\wsquare, V_I}^{++}} \, y^{-L_{\wsquare, W_J}^{++}}  
\rr
\\
\prod_{
\substack{
\bsquare \in W_J
\\
A_{\bsquare, W_J} + 
L_{\bsquare, V_I}  + 1 + c_{V_I} - c_{W_J} = 0 \, \mod{\, n}
}} 
\ll 1 - 
\prod_{K=J}^{I-1} Q_{M_K} 
\prod_{K=J+1}^{I-1} 
Q_K 
x^{A_{\bsquare,  W_J}} \, y^{L_{\bsquare,   V_I}}  
\rr
\\
\prod_{I=1}^N 
\prod_{J = 1}^{I-1} 
\prod_{
\substack{
\wsquare \in V_I \\
A_{\wsquare,V_I} + 
L_{\wsquare,V_J} + 1 - c_{V_I} + c_{V_J} = 0 \, \mod{\, n}
}} 
\ll 1 - 
\prod_{K=J}^{I-1} Q_{M_K} 
\prod_{K=J}^{I-1} Q_K 
x^{-A_{\wsquare,   V_I}^{++}} \, y^{-L_{\wsquare,   V_J}}  
\rr^{-1}
\\
\prod_{
\substack{
\wsquare \in V_J
\\
A_{\wsquare,V_J} + 
L_{\wsquare,V_I} + 1 + c_{V_I} - c_{V_J} = 0  \, \mod{\, n}
}} 
\ll 1 - 
\prod_{K=J}^{I-1} 
Q_{M_K} 
\prod_{K=J}^{I-1} 
Q_K x^{A_{\wsquare,  V_J}} \, y^{L_{\wsquare,   V_I}^{++}}  
\rr^{-1}
\\
\prod_{I=1}^N 
\prod_{J=1}^{I-1} 
\prod_{
\substack{
\bsquare \in  W_I
\\
A_{\bsquare, W_I} + 
L_{\bsquare, W_J} + 1 - c_{W_I} + c_{W_J} = 0 \, \mod{\, n}
}} 
\ll 
1 - 
\prod_{K=J}^{I-1} 
Q_{M_K} 
\prod_{K=J+1}^I 
Q_K x^{-A_{\bsquare,   W_I}} \, y^{-L_{\bsquare, W_J}^{++}}  
\rr^{-1}
\\
\prod_{
\substack{
\bsquare \in  W_J
\\
A_{\bsquare, W_J} + 
L_{\bsquare, W_I} + 1 + c_{W_I} - c_{W_J} = 0\, \mod{\, n}
}} 
\ll 
1 - 
\prod_{K=J}^{I-1} Q_{M_K} 
\prod_{K=J+1}^I 
Q_K x^{A^{++}_{\bsquare,   W_J}} \, y^{L_{\bsquare, W_I}} 
\rr^{-1}
\end{multline}

\section{From $n$-coloured 5D strips to  $n$-coloured 4D instanton 
partition functions}
\label{section.06}
\textit{We take the 4D limit of the 5D strip partition function 
to obtain the corresponding 4D instanton partition function.
}

\subsection{Parameters} 
The parameters $x, y$ of the $n$-coloured topological
vertex and the parameters $\eo, \et$ of the instanton partition
functions are related by 

\begin{equation}
x = e^{\, + R \, \et},
\quad 
y = e^{\, - R \, \eo}.
\end{equation}

Further, recall that 
$Q_{\, i} = e^{- R \, \Delta_{\, i}}, i = 1, M_1, 2, M_2, \cdots, M_{N-1}, N$.

\subsection{The 4D limit}
For $n = 1$, taking the 4D limit is straightforward because both the numerator 
and the denominator of 
$\cS^{
\, n, \, norm
}_{
\ll \bV, \, \bc_{\bV} \rr
\ll \bW, \, \bc_{\bW} \rr \bDelta}
\ll x, y, R \rr$ 
approach zero as 
$R^{\, N \sum_{I=1}^N \ll \, | \, V_I \, | + \, | \, W_I \, | \, \rr}$, in the limit $R \rightarrow 0$.   
For $n > 1$, this is no longer the case and 
$\cS^{\, n, \, norm
}_{
\ll \bV, \, \bc_{\bV} \rr \ll \bW, \, \bc_{\bW} \rr \bDelta} 
\ll x, y, R \rr$ 
approaches zero as $R^{\, K}$, in the limit $R \rightarrow 0$, 
for some integer $K$ which depends on 
$\ll \bV,\, \bc_{\bV} \rr$ and $\ll \bW,\, \bc_{\bW} \rr$.
So, to take the 4D limit, we must cancel the excess factors of $R$ by

\begin{equation}
\cS^{\, n, \, norm}_{\ll \bV,\, \bc_{\bV} \rr \ll \bW, \, \bc_{\bW} \rr \bDelta}\ll \eo, \et \rr =
\lim_{R \, \rightarrow \, 0}
\ll 
R^{-K} \, 
\cS^{\, n, \, norm
}_{
\ll \bV, \, \bc_{\bV} \rr 
\ll \bW, \, \bc_{\bW} \rr \bDelta}
\ll x, y, R \rr
\rr, 
\end{equation}

to obtain
\footnote{\,
We do not need the closed form expression of the factor
$K$, and it suffices for our purposes to compute it on 
a case by case basis.}, 

\begin{multline}\label{sn.norm}
\cS^{\, n, \,norm
}_{
\ll \bV, \, \bc_{\bV} \rr 
\ll \bW, \, \bc_{\bW} \rr \bDelta} 
\ll \eo, \, \et       \rr
= 
\\
\prod_{I = 1}^N
\prod_{J = 1}^I 
\prod_{
\substack{
\bsquare \in W_I
\\
A_{\bsquare, W_I} + L_{\bsquare, V_J} + 1 - c_{W_I} + c_{V_J} = 0 \, \mod{\, n}
}
}
\ll 
\sum_{K=J}^{I-1}
\Delta_{M_K} + 
\sum_{K=J}^I
\Delta_K + A_{\bsquare, W_I} \et - L_{\bsquare, V_J} \eo \rr
\\
\prod_{
\substack{
\wsquare, V_J
\\
A_{\wsquare, V_J} + L_{\wsquare, W_I} + 1 + c_{W_I} - c_{V_J} = 0 \, \mod{\, n}
}
}
\ll 
\sum_{K=J}^{I-1}
\Delta_{M_K} + \sum_{K=J}^I \Delta_K - A^{++}_{\wsquare, V_J} \et + L^{++}_{\wsquare, W_I} \eo 
\rr
\\
\prod_{I=1}^N
\prod_{J=1}^{I-1}
\prod_{
\substack{
\wsquare \in V_I
\\
A_{\wsquare, V_I} + L_{\wsquare, W_J} + 1 - c_{V_I} + c_{W_J} = 0 \, \mod{\, n}
}
}
\ll 
\sum_{K=J}^{I-1}
\Delta_{M_K} + 
\sum_{K=J+1}^{I-1}
\Delta_K + A^{++}_{\wsquare, V_I} \et - L^{++}_{\wsquare, W_J} \eo 
\rr
\\
\prod_{
\substack{
\bsquare \in W_J
\\
A_{\bsquare, W_J}+L_{\bsquare, V_I}+1+ c_{V_I}-c_{W_J} = 0\, \mod{\, n}
}
}
\ll \sum_{K=J}^{I-1}\Delta_{M_K} + \sum_{K=J+1}^{I-1}\Delta_K - A_{\bsquare, W_J}\et + L_{\bsquare, V_I}\eo \rr
    \\
    \prod_{I=1}^N
    \ll 
    \prod_{\substack{\wsquare \in V_I\\
    A_{\wsquare, V_I}+L_{\wsquare, V_I}+1 = 0\, \mod{\, n}}}\ll -A^{++}_{\wsquare, V_I}\et + L_{\wsquare, V_I}\eo \rr^{-1}
    \prod_{\substack{\bsquare \in W_I\\
    A_{\bsquare, W_I}+L_{\bsquare, W_I}+1 = 0\, \mod{\, n}}}\ll -A_{\bsquare, W_I}\et + L^{++}_{\bsquare, W_I}\eo \rr^{-1}
    \rr
    \\
    \prod_{I=1}^N\prod_{J=1}^{I-1}\prod_{\substack{\wsquare \in V_I\\
    A_{\wsquare, V_I}+L_{\wsquare, V_J} + 1 -c_{V_I}+c_{V_J} = 0\, \mod{\, n}}}\ll \sum_{K=J}^{I-1}\Delta_{M_K} + \sum_{K=J}^{I-1}\Delta_K + A^{++}_{\wsquare, V_I}\et - L_{\wsquare, V_J}\eo \rr^{-1}
    \\
    \prod_{\substack{\wsquare \in V_J\\
    A_{\wsquare, V_J}+L_{\wsquare, V_I}+1 + c_{V_I} - c_{V_J} = 0\, \mod{\, n}}}\ll \sum_{K=J}^{I-1}\Delta_{M_K} + \sum_{K=J}^{I-1}\Delta_K - A_{\wsquare, V_J}\et + L^{++}_{\wsquare, V_I} \rr^{-1}
    \\
    \prod_{I=1}^N 
    \prod_{J=1}^{I-1} 
    \prod_{\substack{\bsquare \in  W_I\\
    A_{\bsquare, W_I}+L_{\bsquare, W_J}+1-c_{W_I}+c_{W_J} = 0\, \mod{\, n}}} 
    \ll 
    \sum_{K=J}^{I-1} \Delta_{M_K} + 
    \sum_{K=J+1}^I \Delta_K + A_{\bsquare,   W_I} \et - L_{\bsquare,    W_J}^{++} \eo 
    \rr^{-1}
    \\
    \prod_{\substack{\bsquare \in  W_J\\
    A_{\bsquare, W_J}+L_{\bsquare, W_I} + 1 + c_{W_I} - c_{W_J} = 0\, \mod{\, n}}} 
    \ll 
    \sum_{K=J}^{I-1} \Delta_{M_K} + 
    \sum_{K=J+1}^I \Delta_K - A_{\bsquare,   W_J}^{++} \et + L_{\bsquare,    W_I} \eo 
    \rr^{-1}
\end{multline}

\subsection{Comparison with the $\mathbb{R}^4/\mathbb{Z}_n$ 4D instanton partition function}

The $\mathbb{R}^4/\mathbb{Z}_n$ 4D instanton 
partition function is,  

\begin{equation}
\label{zp.z}
\cZ^{\, n}      \ll \ba, \bV,\bc_{\bV} \, | \, \mu \, | \,  \bb, \bW, \bc_{\bW} \rr 
= 
\frac{
\cZ^{\, n,  \, num} \ll \ba, \bV, \bc_{\bV} \, | \, \mu \, | \, \bb, \bW, \bc_{\bW} \rr
}{
\cZ_{\, n,  \, den} \ll \ba, \bV, \bc_{\bV} \, |             \, \bb, \bW, \bc_{\bW} \rr
}
\end{equation}

where

\begin{multline}
\label{zp.z.num}
\cZ^{\, n, \, num}
\ll \ba,\bV,\bc_{\bV} \,  | \, \mu \, | \,  \bb, \bW, \bc_{\bW} \rr = 
\prod_{I,J = 1}^N
\prod_{
\substack{
\wsquare \in V_I 
\\
A_{\wsquare,V_I} + L_{\wsquare,W_J} + 1 - c_{V_I} + c_{W_J} = 0 \, \mod{\, n}
}
}
\ll a_I - b_J - \mu + A^{++}_{\wsquare, V_I} \et - L_{\wsquare, W_J} \eo \rr
\\
\prod_{
\substack{
\bsquare \in W_J
\\
A_{\bsquare,W_J} + L_{\bsquare,V_I} + 1 + c_{V_I} - c_{W_J} = 0 \, \mod{\, n}
}
}
\ll a_I - b_J - \mu - A_{\bsquare, W_J} \et + L^{++}_{\bsquare, V_I} \eo \rr
\end{multline}

and,

\begin{multline}\label{zp.z.den}
\cZ_{den}^{\, n} \ll \ba, \bV, \bc_{\bV}\, | \, \bb, \bW, \bc_{\bW} \rr = 
\\ 
\ll 
\cZ^{\, n, \, num}\ll \ba, \bV, \bc_{\bV} \, | \, 0 \, | \,  \ba,\bV,\bc_{\bV} \rr
\cZ^{\, n, \, num}\ll \bb, \bW, \bc_{\bW} \, | \, 0 \, | \,  \bb,\bW,\bc_{\bW} \rr
\rr^{\frac12}
\\
=
\prod_{I,J=1}^N
\prod_{
\substack{
\wsquare \in V_I
\\
A_{\wsquare, V_I} + L_{\wsquare, V_J} + 1 - c_{V_I} + c_{V_J} = 0\, \mod{\, n}}
}
\ll 
a_I - a_J + A^{++}_{\wsquare,V_I} \et - L_{\wsquare,V_J} \eo 
\rr^{\frac12}
\\ 
\prod_{
\substack{
\wsquare \in V_J
\\
A_{\wsquare, V_J} + L_{\wsquare, V_I} + 1 + c_{V_I} - c_{V_J} = 0 \, \mod{\, n}}
}
\ll a_I - a_J - A_{\wsquare, V_J} \et + L^{++}_{\wsquare,V_I}\eo \rr^{\frac12}
\\
\prod_{I,J = 1}^N
\prod_{
\substack{
\bsquare \in W_I
\\
A_{\bsquare, W_I} + L_{\bsquare, W_J} + 1 - c_{W_I} + c_{W_J} = 0\, \mod{\, n}}
}
\ll 
b_I - b_J + A^{++}_{\bsquare, W_I} \et - L_{\bsquare,W_J}\eo 
\rr^{\frac12}
\\ 
\prod_{
\substack{
\bsquare \in W_J
\\
A_{\bsquare, W_J} + L_{\bsquare, W_I} + 1 + c_{W_I} - c_{W_J} = 0\, \mod{\, n}}
}
\ll 
b_I - b_J -A_{\bsquare, W_J} \et + L^{++}_{\bsquare, W_I} \eo 
\rr^{\frac12}, 
\end{multline}

where,
$\ba = \ll a_1,...,a_N \rr$ and $\bb = \ll b_1,...,b_N \rr$ are Coulomb parameters, $\mu$ is a mass parameter and $\eo, \et$ are deformation parameters.

\subsubsection{Parameter identification}

To compare the 4D strip partition function to the 4D instanton partition functions, we substitute

\begin{equation}
\label{delta.K} 
\boxed{
\Delta_i = -a_i + b_i + \mu - \eo, 
\quad
\Delta_{M_i}  = a_{i+1} - b_i - \mu + \eo,
\quad
i = 1, \cdots, N}
\end{equation} 

where 
$a_i = a_{i+N}$, and  
$b_I = b_{i+N}$, 
into Equation \ref{sn.norm}, to obtain, 

\begin{multline}\label{s.n}
\cS^{\, n, \,norm}_{\ll \bV,\, \bc_{\bV}\rr \ll\bW,\, \bc_{\bW}\rr \bDelta}\ll \eo,\et \rr
= 
\\
\prod_{I=1}^N
\prod_{J=1}^I
\prod_{
\substack{
\wsquare \in 
V_J
\\
A_{\wsquare, V_J} + L_{\wsquare, W_I} + 1 + c_{W_I} - c_{V_J} = 0  \,
\mod{\, n}}}
-
\ll 
a_J - b_I - \mu + A^{++}_{\wsquare, V_J} \et - 
                  L^{  }_{\wsquare, W_I} \eo 
\rr
\\
\prod_{
\substack{
\bsquare \in W_I
\\
A_{\bsquare, W_I} + L_{\bsquare, V_J} + 1 - c_{W_I} + c_{V_J} = 0  \, 
\mod{\, n}}}
-\ll a_J - b_I - \mu - A_{\bsquare, W_I}\et + L^{++}_{\bsquare, V_J}\eo \rr
    \\
    \prod_{I=1}^N\prod_{J=1}^{I-1}\prod_{\substack{\wsquare \in V_I\\
    A_{\wsquare, V_I}+L_{\wsquare, W_J}+1 -c_{V_I}+c_{W_J} = 0\, \mod{\, n}}}\ll a_I - b_J - \mu + A^{++}_{\wsquare, V_I}\et - L_{\wsquare, W_J}\eo \rr
    \\
    \prod_{\substack{\bsquare \in W_J\\
    A_{\bsquare, W_J}+L_{\bsquare, V_I}+1 + c_{V_I}-c_{W_J} = 0\, \mod{\, n}}}\ll a_I - b_J - \mu - A_{\bsquare, W_J}\et + L^{++}_{\bsquare, V_I}\eo \rr
    \\
    \prod_{I=1}^N
    \ll 
    \prod_{\substack{\wsquare \in V_I\\
    A_{\wsquare, V_I}+L_{\wsquare, V_I}+1 = 0\, \mod{\, n}}}-\ll A^{++}_{\wsquare, V_I}\et - L_{\wsquare, V_I}\eo \rr^{-1}
    \prod_{\substack{\bsquare \in W_I\\
    A_{\bsquare, W_I}+L_{\bsquare, W_I}+1 = 0\, \mod{\, n}}}\ll -A_{\bsquare, W_I}\et + L^{++}_{\bsquare, W_I}\eo \rr^{-1}
    \rr
    \\
    \prod_{I=1}^N\prod_{J=1}^{I-1}\prod_{\substack{\wsquare \in V_I\\
    A_{\wsquare, V_I}+L_{\wsquare, V_J} + 1 -c_{V_I}+c_{V_J} = 0\, \mod{\, n}}}\ll a_I-a_J + A^{++}_{\wsquare, V_I}\et - L_{\wsquare, V_J}\eo \rr^{-1}
    \\
    \prod_{\substack{\wsquare \in V_J\\
    A_{\wsquare, V_J}+L_{\wsquare, V_I}+1 + c_{V_I} - c_{V_J} = 0\, \mod{\, n}}}\ll a_I-a_J - A_{\wsquare, V_J}\et + L^{++}_{\wsquare, V_I} \eo \rr^{-1}
    \\
    \prod_{I=1}^N 
    \prod_{J=1}^{I-1} 
    \prod_{\substack{\bsquare \in  W_I\\
    A_{\bsquare, W_I}+L_{\bsquare, W_J}+1-c_{W_I}+c_{W_J} = 0\, \mod{\, n}}} 
    -\ll 
    b_J-b_I - A_{\bsquare,   W_I} \et + L_{\bsquare,    W_J}^{++} \eo 
    \rr^{-1}
    \\
    \prod_{\substack{\bsquare \in  W_J\\
    A_{\bsquare, W_J}+L_{\bsquare, W_I} + 1 + c_{W_I} - c_{W_J} = 0\, \mod{\, n}}} 
    -\ll 
    b_J-b_I + A_{\bsquare,   W_J}^{++} \et - L_{\bsquare,    W_I} \eo 
    \rr^{-1}
    \\
    = \frac{\cS^{\, n, \,num}_{\ll \bV,\, \bc_{\bV} \rr \ll \bW,\, \bc_{\bW} \rr\bDelta}\ll \eo,\et \rr}{\cS^{\, n}_{den\ll \bV,\, \bc_{\bV} \rr \ll \bW,\, \bc_{\bW} \rr\bDelta}\ll \eo,\et \rr}
\end{multline}

\subsubsection{Sign factors and abbreviations}
We define, 

\begin{equation}
F \ll \bV, \bc_{\bV} \, |\, \bW, \bc_{\bW} \rr = 
\frac{
F^{num} \ll \bV, \bc_{\bV}\, |\, \bW, \bc_{\bW} \rr
}{
F_{den} \ll \bV, \bc_{\bV}\, |\, \bW, \bc_{\bW} \rr
}, 
\end{equation}

\begin{multline}
F^{\, num}
\ll \bV, \bc_{\bV} \, |\, \bW, \bc_{\bW} \rr = 
\prod_{I=1}^N
\prod_{J=1}^I
\prod_{
\substack{
\wsquare \in V_{J}
\\
A_{\wsquare, V_J} + 
L_{\wsquare, W_I} + 1 + c_{W_I} - c_{V_J} = 0\, \mod{\, n}
}
}
\ll -1 \rr
\prod_{
\substack{
\bsquare \in W_I
\\
A_{\bsquare, W_I} + 
L_{\bsquare, V_J} + 1 - c_{W_I} + c_{V_J} = 0\, \mod{\, n}
}
}
\ll -1 \rr
\end{multline}

\begin{multline}
F_{den}
\ll \bV, \bc_{\bV} \, | \, \bW, \bc_{\bW} \rr = 
\prod_{I=1}^N
\prod_{
\substack{
\wsquare \in V_I
\\
A_{\wsquare, V_I} + 
L_{\wsquare, V_J} + 1 - c_{V_I} - c_{V_J} = 0 \, \mod{\, n}
}
}
\ll -1 \rr
\\
\prod_{I=1}^N
\prod_{J=1}^{I-1}
\prod_{
\substack{
\bsquare \in W_I
\\
A_{\bsquare, W_I} + 
L_{\bsquare, W_J} + 1 - c_{W_I} + c_{W_J} = 0 \, \mod{\, n}
}
}
\ll -1 \rr
\prod_{
\substack{
\bsquare \in W_J
\\
A_{\bsquare, W_J} + 
L_{\bsquare, W_I} + 1 + c_{W_I} - c_{W_J} = 0 \, \mod{\, n}
}
}
\ll -1 \rr    
\end{multline}

and introduce the abbreviations, 

\begin{equation}
\cH_{Y_{IJ}}  \ll \bx, \bc_{\bY} \rr   
=   
\cA_{Y_{IJ}}  \ll \bx, \bc_{\bY} \rr \, \cL_{Y_{IJ}}  
\ll \bx, \bc_{\bY} \rr,
\end{equation} 

where,

\begin{equation} 
\cA_{V_{IJ}} \ll \bx, \bc_{\bV} \rr =   
\prod_{
\substack{
\wsquare \in V_I
\\
A_{\wsquare, V_I} + 
L_{\wsquare, V_J} + 1 - c_{V_I} + c_{V_J} = 0 \, \mod{\, n}
}
} 
\ll 
x_I - x_J 
+ A^{++}_{\wsquare,  V_I} \et
- L^{  }_{\wsquare,  V_J} \eo 
\rr,
\quad
\cA_{\emptyset, \emptyset}  \ll \bx,\bc_{\bV} \rr = 1,
\end{equation}

\begin{equation}
\cL_{W_{IJ}}  \ll \bx, \bc_{\bW} \rr =   
\prod_{
\substack{
\bsquare \in  W_J
\\
A_{\bsquare, W_J} +
L_{\bsquare, W_I} + 1 + c_{W_I} - c_{W_J} = 0 \, \mod{\, n}
}
} 
\ll x_I - x_J 
- A^{  }_{\bsquare,    W_J} \et
+ L^{++}_{\bsquare,    W_I} \eo 
\rr,
\quad
\cL_{\emptyset, \emptyset}  \ll \bx, \bc_{\bW} \rr = 1,
\end{equation}

\subsubsection{Comparing the numerators}
The numerators
$\cS^{
\, n, \,num
}_{
\ll \bV,\, \bc_{\bV} \rr \ll \bW,\, \bc_{\bW} \rr \bDelta
}
\ll \eo,\et \rr
$ 
and 
$\cZ^{\, n, \, num}
\ll \ba,\bV,\bc_{\bV} \,  | \, \mu \, | \,  \bb, \bW, \bc_{\bW} \rr$ 
are identical up to possible overall signs, 

\begin{equation}
\label{abbrev.s.num}
\cS^{\, n, \,num}_{\ll \, \bV, \, \bc_{\bV} \rr 
                   \ll \, \bW, \, \bc_{\bW} \rr \bDelta} 
                   \ll \, \eo, \, \et \rr = 
F^{num} \ll \bV, \bc_{\bV} \, | \, \bW, \bc_{\bW} \rr 
\cZ^{\, n, \, num} 
\ll \ba, \bV, \bc_{\bV} \, | \, \mu \, | \, \bb, \bW, \bc_{\bW} \rr
\end{equation}

\subsubsection{Comparing the denominators}
The denominators are, 

\begin{multline}
\label{abbrev.s.den}
\cS^{\, n
}_{
\, den\, \ll \bV,\, \bc_{\bV} \rr \ll \bW,\, \bc_{\bW} \rr \bDelta
}
\ll \eo,\et \rr =
\\ 
F_{den}\ll \bV,\bc_{\bV} \, | \,  \bW, \bc_{\bW} \rr
\prod_{I=1}^N
\cA_{V_{II}} \ll \ba, \bc_{\bV} \rr \cL_{W_{II}} \ll \bb, \bc_{\bW} \rr
\prod_{J=1}^N
\prod_{I=1}^{J-1} 
\cH_{V_{JI}} \ll \ba, \bc_{\bV} \rr
\cH_{W_{IJ}} \ll \bb, \bc_{\bW} \rr
\end{multline} 

and

\begin{equation}
\label{abbrev.z.den}
\cZ^{\, n}_{\, den} 
\ll \ba, \bV, \bc_{\bV} \, | \, \bb, \bW, \bc_{\bW} \rr  = 
\prod_{I=1}^N
\prod_{J=1}^N
\ll 
\cH_{V_{IJ}} \ll \ba, \bc_{\bV} \rr
\cH_{W_{IJ}} \ll \bb, \bc_{\bW} \rr
\rr^{1/2} 
\end{equation} 

In other words,  
$\cS^{\, n, \, norm
}_{
\ll \bV,\, \bc_{\bV} \rr 
\ll \bW,\, \bc_{\bW} \rr \bDelta
}
\ll \eo,\et \rr$ and 
$\cZ^{\, n, \, norm
}
\ll 
\ba, \bV,\bc_{\bV} 
\, | \, 
\mu 
\, | \,  
\bb, \bW, \bc_{\bW} \rr$ 
have different denominators. 
However, we show in the sequel that in computing  
linear as well as cyclic 4D instanton partition 
functions, we get exactly the same result 
by using $\cS^{\, n, \, norm}$ (with an appropriate 
choice of framing factors) and by using $\cZ^{\, n, \, norm}$.

\section{From $n$-coloured 5D strips to $n$-coloured 5D webs}
\label{section.07}
\textit{We glue 5D $n$-coloured strip diagrams to form 5D $n$-coloured web diagrams.}

\begin{scriptsize}
\begin{figure}
\begin{tikzpicture} [scale=.8]

\draw [ultra thick] (0,0)--(0,1);
\node [below] at (0,0) {$\pmb{\emptyset}$};
\draw [thick] (0,1)--(2,1);
\node [above] at (1,1) {$V_{1N},\Delta_{C_{1,N}}$}; 
\draw [ultra thick] (0,1)--(-1,2);
\draw [thick] (-1,2)--(-2,2);
\node [left] at (-2,2) {$\emptyset$};
\draw [ultra thick] (-1,2)--(-1,3);
\draw [ultra thick, dashed] (-1,3)--(-2,4);
\draw [ultra thick] (-2,4)--(-2,5);
\draw [thick] (-2,5)--(0,5);
\node [above] at (-1,5) {$V_{12},\Delta_{C_{1,2}}$};
\draw [ultra thick] (-2,5)--(-3,6);
\draw [thick] (-3,6)--(-4,6);
\node [left] at (-4,6) {$\emptyset$};
\draw [ultra thick] (-3,6)--(-3,7);
\draw [thick] (-3,7)--(-1,7);
\node [above] at (-2,7) {$V_{11},\Delta_{C_{1,1}}$};
\draw [ultra thick] (-3,7)--(-4,8);
\draw [thick] (-4,8)--(-5,8);
\node [left] at (-5,8) {$\emptyset$};
\draw [ultra thick] (-4,8)--(-4,9);
\node [above] at (-4,9) {$\pmb{\emptyset}$};

\draw [ultra thick] (2,1)--(2,2);
\draw [ultra thick] (2,1)--(3,0);
\draw [ultra thick] (3,0)--(3,-1);
\node [below] at (3,-1) {$\pmb{\emptyset}$};
\draw [thick] (3,0)--(5,0);
\node [above] at (4,0) {$V_{2N}, \Delta_{C_{2,N}}$};
\draw [ultra thick,dashed] (2,2)--(1,3);
\draw [ultra thick] (1,3)--(1,4);
\draw [thick] (1,4)--(3,4);
\node [above] at (2,4) {$V_{22}, \Delta_{C_{2,2}}$};
\draw [ultra thick] (1,4)--(0,5);
\draw [ultra thick] (0,5)--(0,6);
\draw [thick] (0,6)--(2,6);
\node [above] at (1,6) {$V_{21}, \Delta_{C_{2,1}}$};
\draw [ultra thick] (0,6)--(-1,7);
\draw [ultra thick] (-1,7)--(-1,8);
\node [above] at (-1,8) {$\pmb{\emptyset}$};

\draw [thick,dashed] (5,0)--(6,0);
\draw [thick,dashed] (3,4)--(4,4);
\draw [thick,dashed] (2,6)--(3,6);

\draw [thick] (6,0)--(8,0);
\node [above] at (7,0) {$V_{m1}, \Delta_{C_{m,N}}$};
\draw [thick] (4,4)--(6,4);
\node [above] at (5,4) {$V_{m2}, \Delta_{C_{m,2}}$};
\draw [thick] (3,6)--(5,6);
\node [above] at (4,6) {$V_{mN}, \Delta_{C_{m,1}}$};

\draw [ultra thick] (5,6)--(5,7);
\node [above] at (5,7) {$\pmb{\emptyset}$};
\draw [ultra thick] (5,6)--(6,5);
\draw [ultra thick] (6,5)--(6,4);
\draw [thick] (6,5)--(7,5);
\node [right] at (7,5) {$\emptyset$};
\draw [ultra thick] (6,4)--(7,3);
\draw [ultra thick] (7,3)--(7,2);
\draw [thick] (7,3)--(8,3);
\node [right] at (8,3) {$\emptyset$};
\draw [ultra thick,dashed] (7,2)--(8,1);
\draw [ultra thick] (8,1)--(8,0);
\draw [ultra thick] (8,0)--(9,-1);
\draw [ultra thick] (9,-1)--(9,-2);
\node [below] at (9,-2) {$\pmb{\emptyset}$};
\draw [thick] (9,-1)--(10,-1);
\node [right] at (10,-1) {$\emptyset$};

\end{tikzpicture} 
\caption{
\textit{
The linear web diagram constructed by gluing $(m+1)$ $N$-strip
diagrams. Each $N$-strip is linked to the next by a series 
of internal horizontal edges assigned with K\"ahler parameters
$\Delta_{C_{iI}}$ and sum over all possible associated Young  
diagrams $\bV_i = \ll V_{i1}, \cdots, V_{iN} \rr$ in a given 
series.   
}
} 
\label{linear.web.diagram} 
\end{figure}
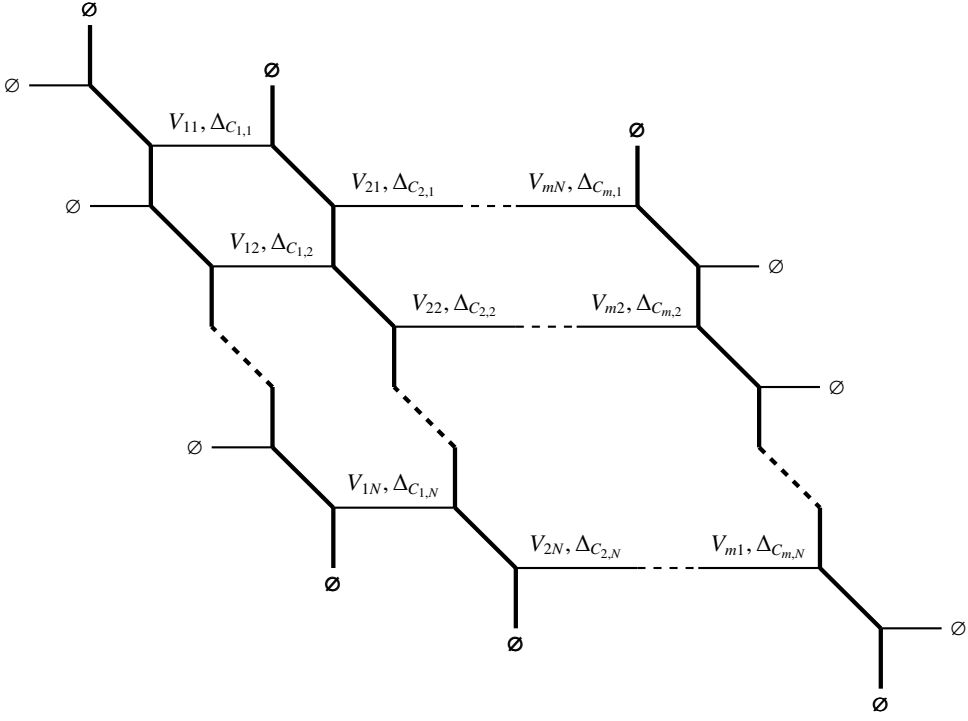
\end{scriptsize}

\begin{scriptsize}
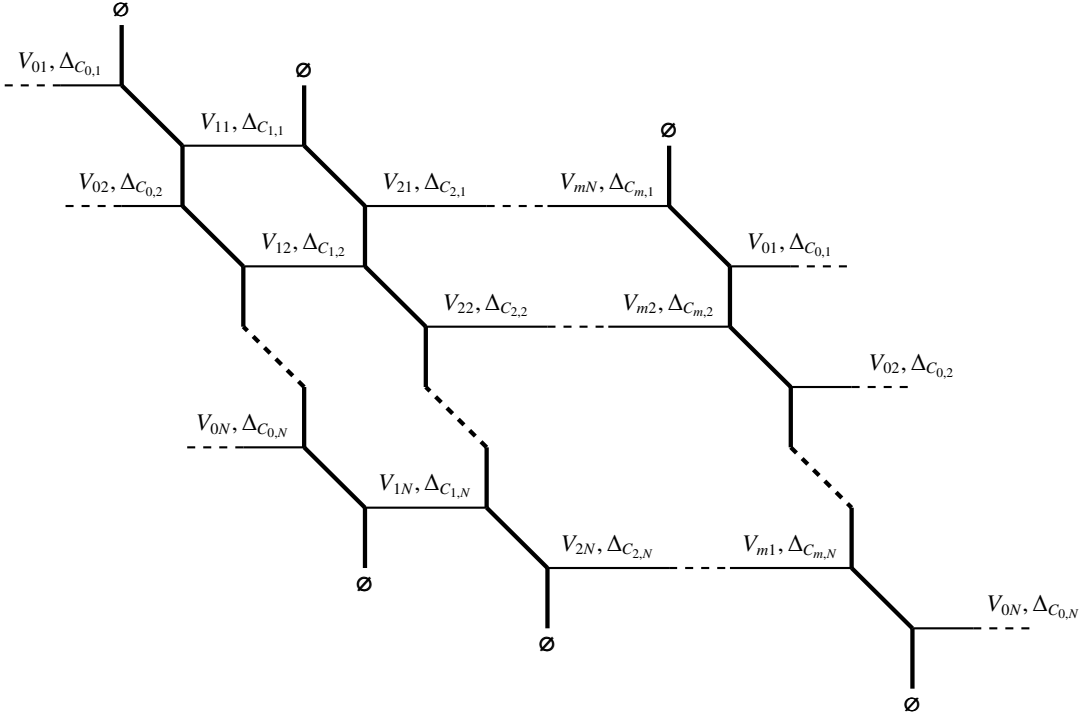
\begin{figure} 
\begin{tikzpicture} [scale=.8]

\draw [ultra thick] (0,0)--(0,1);
\node [below] at (0,0) {$\pmb{\emptyset}$};
\draw [thick] (0,1)--(2,1);
\node [above] at (1,1) {$V_{1N},\Delta_{C_{1,N}}$}; 
\draw [ultra thick] (0,1)--(-1,2);
\draw [thick] (-1,2)--(-2,2);
\draw [ultra thick] (-1,2)--(-1,3);
\draw [ultra thick, dashed] (-1,3)--(-2,4);
\draw [ultra thick] (-2,4)--(-2,5);
\draw [thick] (-2,5)--(0,5);
\node [above] at (-1,5) {$V_{12},\Delta_{C_{1,2}}$};
\draw [ultra thick] (-2,5)--(-3,6);
\draw [thick] (-3,6)--(-4,6);
\draw [ultra thick] (-3,6)--(-3,7);
\draw [thick] (-3,7)--(-1,7);
\node [above] at (-2,7) {$V_{11},\Delta_{C_{1,1}}$};
\draw [ultra thick] (-3,7)--(-4,8);
\draw [thick] (-4,8)--(-5,8);
\draw [ultra thick] (-4,8)--(-4,9);
\node [above] at (-4,9) {$\pmb{\emptyset}$};

\draw [thick,dashed] (-5,8)--(-6,8);
\node [above] at (-5,8) {$V_{01}, \Delta_{C_{0,1}}$};
\draw [thick,dashed] (-4,6)--(-5,6);
\node [above] at (-4,6) {$V_{02}, \Delta_{C_{0,2}}$};
\draw [thick,dashed] (-2,2)--(-3,2);
\node [above] at (-2,2) {$V_{0N}, \Delta_{C_{0,N}}$};

\draw [ultra thick] (2,1)--(2,2);
\draw [ultra thick] (2,1)--(3,0);
\draw [ultra thick] (3,0)--(3,-1);
\node [below] at (3,-1) {$\pmb{\emptyset}$};
\draw [thick] (3,0)--(5,0);
\node [above] at (4,0) {$V_{2N}, \Delta_{C_{2,N}}$};
\draw [ultra thick,dashed] (2,2)--(1,3);
\draw [ultra thick] (1,3)--(1,4);
\draw [thick] (1,4)--(3,4);
\node [above] at (2,4) {$V_{22}, \Delta_{C_{2,2}}$};
\draw [ultra thick] (1,4)--(0,5);
\draw [ultra thick] (0,5)--(0,6);
\draw [thick] (0,6)--(2,6);
\node [above] at (1,6) {$V_{21}, \Delta_{C_{2,1}}$};
\draw [ultra thick] (0,6)--(-1,7);
\draw [ultra thick] (-1,7)--(-1,8);
\node [above] at (-1,8) {$\pmb{\emptyset}$};

\draw [thick,dashed] (5,0)--(6,0);
\draw [thick,dashed] (3,4)--(4,4);
\draw [thick,dashed] (2,6)--(3,6);

\draw [thick] (6,0)--(8,0);
\node [above] at (7,0) {$V_{m1}, \Delta_{C_{m,N}}$};
\draw [thick] (4,4)--(6,4);
\node [above] at (5,4) {$V_{m2}, \Delta_{C_{m,2}}$};
\draw [thick] (3,6)--(5,6);
\node [above] at (4,6) {$V_{mN}, \Delta_{C_{m,1}}$};

\draw [ultra thick] (5,6)--(5,7);
\node [above] at (5,7) {$\pmb{\emptyset}$};
\draw [ultra thick] (5,6)--(6,5);
\draw [ultra thick] (6,5)--(6,4);
\draw [thick] (6,5)--(7,5);
\draw [ultra thick] (6,4)--(7,3);
\draw [ultra thick] (7,3)--(7,2);
\draw [thick] (7,3)--(8,3);
\draw [ultra thick,dashed] (7,2)--(8,1);
\draw [ultra thick] (8,1)--(8,0);
\draw [ultra thick] (8,0)--(9,-1);
\draw [ultra thick] (9,-1)--(9,-2);
\node [below] at (9,-2) {$\pmb{\emptyset}$};
\draw [thick] (9,-1)--(10,-1);

\draw [thick,dashed] (7,5)--(8,5);
\node [above] at (7,5) {$V_{01}, \Delta_{C_{0,1}}$};
\draw [thick,dashed] (8,3)--(9,3);
\node [above] at (9,3) {$V_{02}, \Delta_{C_{0,2}}$};
\draw [thick,dashed] (10,-1)--(11,-1);
\node [above] at (11,-1) {$V_{0N}, \Delta_{C_{0,N}}$};

\end{tikzpicture} 
\caption{
\textit{
The cyclic web diagram constructed by gluing the first 
and the last $N$-strips in the linear web diagram.  
}
} 
\label{cyclic.web.diagram} 
\end{figure} 
\end{scriptsize}

The linear and cyclic web diagram
topological partition functions are obtained from Figures 
\ref{linear.web.diagram} and \ref{cyclic.web.diagram}, once 
the horizontal edge factors are specified. The answer takes 
the general form,

\begin{equation}
\label{generic.web.partition}
\cS^{\, n}\ll x,y,R \rr = 
\sum_{\ll \bV_1, \, 
\bc_{\bV_1} \rr, \cdots, \ll \bV_p, \, 
\bc_{\bV_p} \rr}
\cS^{\, n, \, block}_{\ll \bV_1,\, \bc_{\bV_1} \rr, \ll \bV_2,\, 
\bc_{\bV_2} \rr, \cdots, \ll \bV_m,\, 
\bc_{\bV_m} \rr
}
\ll x, y, R \rr
\end{equation}

where 
$\cS^{\, n, \, block}_{
\ll \bV_1,\, \bc_{\bV_1} \rr, 
\ll \bV_2,\, \bc_{\bV_2} \rr, \cdots,
\ll \bV_p,\, \bc_{\bV_m} \rr} \ll x, y, R \rr$ 
is a product of factors of type 
$\cS^{\, p, \,norm}_{
\ll \bV_i, \bc_{\bV_i} \rr \ll \bV_{i+1},\, 
\bc_{\bV_{i+1}}\rr \bDelta_{\, i, i+1}}$ 
for every strip, and 
an edge factor for every internal 
horizontal edge.

\subsection{The summation in Equation 
\ref{generic.web.partition}}

For $n=1$, the summation is over all $N$-Young diagrams because the moduli space 
of instantons on $\CC^{\, 2}$ is connected, and the localization theorem requires that we sum over 
the contributions of all fixed points, each of which  
is labelled by $N$-Young diagrams. 
For $n = 2, 3, \cdots$, the moduli space of instantons on     
$\CC^{\, 2}/\ZZ_{\, n}$ is a union of disjoint smaller 
spaces \cite{fucito.morales.poghossian}, 
and the summation in Equation \ref{generic.web.partition} 
is not over all possible charged $N$-Young diagrams, but 
restricted to certain series of charged Young diagrams. 

\subsection{
$\ll \bN,\bk \rr$-type
Young diagrams
}
Following \cite{alfimov.tarnopolsky,ito.maruyoshi.okuda}, given
$\ll \bY,\, \bc_{\bY}\rr = 
\ll \ll Y_1,\, c_{Y_1} \rr, \cdots, \ll Y_N, \, c_{Y_N} \rr \rr$, 
we define $\bN = \ll N_{\, 0}, \cdots, N_{\, n-1} \rr$, where
$N_c$ is the number of Young diagrams of charge $c$. Further, 
we define $\bk = \ll k_{\, 0}, \cdots, k_{\, n-1} \rr$, where  
$k_c$ is the number of cells with coordinates 
$\ll \iota, \jmath \rr \in Y_i$, $i = 1, \cdots, N$ (that is, in all Young diagrams), such that, 

\begin{equation}
c_{\, Y_i} + (\iota - 1) - (\jmath -  1) \, =  \, c,
\quad 
c \in \ll 0, \cdots, n-1 \rr
\end{equation}

and say that $\ll \bY,\bc_{\bY} \rr$ is of $\ll \bN,\bk \rr$-type. 
Clearly, 
$\sum_{c = 0}^{n-1} N_c = N$, and 
$\sum_{c = 0}^{n-1} k_c = \sum_{I=1}^N \, | \, Y_I\, | \, $.  
  
\subsection{$\ll \bN, \bu \rr$-series of solutions}
Let $\bu = \ll u_1,\cdots, u_{n-1} \rr$ be given by
\footnote{\, 
$\bu$ is related to the first Chern class 
of the instanton gauge bundle 
$c_1 ( E ) = 
\sum_{i=1}^{n-1} 
u_i \, 
c_1 \ll \mathcal{T}_i \rr$,  
\cite{fucito.morales.poghossian, alfimov.tarnopolsky}.
}, 

\begin{equation}
\label{first.chern.class}
u_{\, i} = 
N_{\, i}   
+   
\ll
k_{\, i-1}  
- 2 k_{\, i} 
+   k_{\, i+1}
\rr, 
\end{equation}

then, for fixed $\bu$, solve 
Equation \ref{first.chern.class} to 
find all possible values of $\ll \bN, \, \bk \rr$. 
For fixed $\bN$, there may be no solutions, 
or 
there may be infinitely-many solutions 
$\bk$ that satisfy  
Equation \ref{first.chern.class}. 
In the latter (relevant) case, for a corresponding fixed $\bN$,
$\bk$ is a sequence in $\mathbb{Z}^{\, n}_{\, \geq \, 0}$ indexed 
by $k_{\, 0} = r \in \mathbb{Z}_{\, \geq \, 0}$  which we denote 
by $\bk \, (r)$. We define,  

\begin{align}
\text{$\ll \bN, \bu \rr$-series} 
&
= 
\ll 
\ll \bY, \bc_{\bY}\rr \, | \, 
\textrm{ 
$\ll \bY, \bc_{\bY}\rr$ is of $\ll \bN, \bk \rr$-type 
for $\bk$ such that $\bN$ and $\bk$ 
satisfy \eqref{first.chern.class}
} 
\rr 
\\
&
= 
\ll 
\ll \bY, \bc_{\bY}\rr \, | \, 
\textrm{ 
$\ll \bY, \bc_{\bY}\rr$ is of $\ll \bN, \bk \, (r) \rr$-type 
for some $r \in \mathbb{Z}_{\, \geq \, 0}$ 
} 
\rr, 
\end{align}

and restrict the summation in 
$\ll \bY, \bc_{\bY}\rr$ to one or more of 
these series. 

\subsection{Example}
For 
$N = 2$, $n = 4$, and 
$\bu = \ll 0, 0, 0 \rr$,
there are three  
$\ll \bN, \, \bk\rr$-series
\footnote{\, 
We use this example, which can be found in
\cite{alfimov.tarnopolsky}, 
to reproduce the same result.
}, 

\begin{align}
\text{Series 1:} & \quad \bN = \ll 2, 0, 0, 0 \rr, 
\quad \bk = \ll r, r, r, r \rr
\label{u.0.series.1}
\\
\text{Series 2:} & \quad \bN = \ll 0, 1, 0, 1 \rr, 
\quad \bk = \ll r, r+1, r+1, r+1 \rr
\label{u.0.series.2}
\\
\text{Series 3:} & \quad \bN = \ll 0, 0, 2, 0 \rr, 
\quad \bk = \ll r, r+1, r+2, r+1 \rr
\label{u.0.series.3}
\end{align}

for $r \in \mathbb{Z}_{\, \geq \, 0}$. 
This gives rise to three series of charged 
Young diagrams 
$\ll \bY,\, \bc_{\bY} \rr$. The charges $\bc_{\bY}$ are fixed within the series by $\bN$,

\begin{align*}
\text{Series 1 : } &   \quad \bc_{\bY} = \ll 0, 0 \rr
\\
\text{Series 2 : } &   \quad \bc_{\bY} = \ll 1, 3 \rr 
\quad 
\text{or equivalently} \quad \bc_{\bY} = \ll 3, 1 \rr
\\
\text{Series 3 : } &   \quad \bc_{\bY} = \ll 2, 2 \rr
\end{align*}

Switching $\bc_{\bY} = \ll 1, 3 \rr$ and 
$\bc_{\bY} = \ll 3,1 \rr$ is the same as switching the labels 
of $Y_1$ and $Y_2$ in the summation, hence they are equivalent. 
The summation in Equation \ref{generic.web.partition} is 
restricted to one or more of these series. 

\subsection{The edge factors}
For the $I$-th line between
$\cS^{\, n, \, norm}_{\ll \bV_{i-1}, \, \bc_{\bV_{i-1}} \rr \ll 
\bV_i,     \, \bc_{\bV_i}     \rr \bDelta_{\, i-1,\, i}}$ 
and 
$\cS^{\, n, \, norm}_{\ll \bV_i,     \, \bc_{\bV_i}     \rr \ll \bV_{i+1}, \,    \bc_{\bV_{i+1}} \rr \bDelta_{\, i,\, i+1}}$, 
we define, 

\begin{equation}
\text{Edge factor} = 
\ll 
- Q_{C_{iI}} \rr^{\, | \, V_{iI}\, | \, } \, 
f^{\, n}_{i-1, i, i+1, I}, 
\end{equation}

with $Q_{C_{i I}} = z_i \, e^{- R \, \Delta_{\, C_{i I}}}$, 
where $z_i$ is an instanton 
expansion parameter, and the framing 
factor $f^{\, n}_{i-1,i,i+1,I}$ is,  

\begin{equation}
f^{\, n}_{i-1,i,i+1,I} = 
\ll -1 \rr^{\, | \, V_{iI}\, | \, }
f^{\, n, \,left}_{i - 1, i, I} f^{\, p, \,right}_{i,i+1,I}
\end{equation}

\begin{equation}
f^{\, n, \,right}_{i,i+1,I} = 
\prod_{J=I}^N
\prod_{
\substack{
\wsquare \in V_{iI}
\\
A_{\wsquare, V_{iI}} + L_{\wsquare, V_{i+1,J}} + 1
\\ 
+ c_{V_{i+1,J}} - c_{V_{iI}} = \, 0 \, \mod{\, n}
}
}
\ll -1 \rr
\prod_{J=1}^I
\prod_{
\substack{
\wsquare \in V_{iI}
\\
A_{\wsquare, V_{iI}} + L_{\wsquare, V_{iJ}} + 1 
\\ 
- c_{V_{iI}} - c_{V_{iJ}} = \, 0 \, \mod{\, n}
}
}
\ll -1 \rr
\end{equation}

\begin{equation}
f^{\, n, \,left}_{i-1,i,I} = 
\prod_{J=1}^I
\prod_{
\substack{
\wsquare \in V_{iI}
\\
A_{\wsquare, V_{iI}} + L_{\wsquare, V_{i-1,J}} + 1 
\\
- c_{V_{iI}} + c_{V_{i-1,J}} = \, 0 \, \mod{\, n}
}
}
\ll -1 \rr
\prod_{J=1}^{I-1}
\prod_{
\substack{
\wsquare \in V_{iI}
\\
A_{\wsquare, V_{iI}} + L_{\wsquare, V_{iJ}} + 1 
\\
- c_{V_{iI}} + c_{V_{iJ}} = \, 0 \, \mod{\, n}
}
}
\ll -1 \rr
\prod_{J=I+1}^N
\prod_{
\substack{
\wsquare \in V_{iI}
\\
A_{\wsquare, V_{iI}} + L_{\wsquare, V_{iJ}} + 1 
\\
+ c_{V_{iJ}} - c_{V_{iI}} = \, 0 \, \mod{\, n}
}
}
\ll -1 \rr
\end{equation}

The framing factor is designed to take care of the sign difference between the conformal blocks calculated from $\cZ^{\, n}$ and $\cS^{\, n, \, norm}$. 
When $n = 1$, the framing factor is, 

\begin{multline}
f^{p=1}_{i-1, i, i+1, I} = 
\ll -1 \rr^{\, | \, V_{iI}\, | \, } 
\prod_{J=I}^N
\prod_{
\wsquare \in V_{iI}}
\ll -1 \rr
\prod_{J=1}^I
\prod_{
\wsquare \in V_{iI}}
\ll -1 \rr
\prod_{
\wsquare \in V_{iI}}
\ll -1 \rr
\prod_{J=1}^{I-1}
\prod_{\wsquare \in V_{iI}}
\ll -1 \rr \prod_{J=I+1}^N
\prod_{
\wsquare \in V_{iI}}
\ll -1 \rr
\\
= 
\ll -1 \rr^{\, | \,  V_{iI} \, | \, } 
\ll -1 \rr^{(N-I+1) \, | \, V_{iI}\, | \, }
\ll -1 \rr^{I\, | \, V_{iI}\, | \, } 
\ll -1 \rr^{\, | \, V_{iI}\, | \, } 
\ll -1 \rr^{(I-1)\, | \, V_{iI}\, | \, } 
\ll -1 \rr^{(N-I)\, | \, V_{iI}\, | \, } = 1
\end{multline}

$\cS^{\, 5D}_{\, N}$ is given in terms of $N$-strip partition functions 
for linear and cyclic cases as follows, 

\subsubsection{Linear conformal blocks}
(See Figure \ref{linear.web.diagram})
\begin{multline}
\cS^{
\, n, \,linear\, block
}_{
\ll
\bV_1,\, \bc_{\bV_1}
\rr, 
\ll 
\bV_2,\, \bc_{\bV_2} 
\rr, \cdots, 
\ll
\bV_m, \, \bc_{\bV_m}\rr} 
\ll x, y, R 
\rr
\\
=
\cS_{
\ll
\pmb{\emptyset}, \, \bc_{\pmb{\emptyset}_0} 
\rr 
\ll
\bV_1, \, \bc_{\bV_1} 
\rr
\bDelta_{\, 0 1}}^{\, n, \, norm}
\ll x, y, R 
\rr 
\ll 
\prod_{I=1}^N
\ll -Q_{C_{1I}} \rr^{\, | \, V_{1I}\, | \, }f^{\, n}_{012,I}
\rr
\cS_{
\ll 
\bV_1,\, \bc_{\bV_1}
\rr
\ll 
\bV_2,\, \bc_{\bV_2}
\rr 
\bDelta_{\, 1 2}}^{\, n, \, norm}
\ll x, y, R \rr
\cdots
\\
\cdots 
\cS_{
\ll 
\bV_{m-1},\, \bc_{\bV_{m-1}}
\rr 
\ll 
\bV_{m},\, \bc_{\bV_{m}}
\rr
\bDelta_{\, m-1, m}
}^{
\, n, \, norm
}
\ll x, y, R \rr 
\ll
\prod_{I = 1}^N
\ll 
- Q_{C_{m,I}} \rr^{\, | \, V_{m,I}\, | \, } f^{\, n}_{m-1, m, m+1, I}
\rr
\cS_{
\ll \bV_{p},\, \bc_{\bV_{p}} \rr 
\ll 
\pmb{\emptyset}, \, \bc_{\pmb{\emptyset}_{m+1}}
\rr 
\bDelta_{\, m, m+1}}^{\, n, \, norm}
\ll 
x, y, R 
\rr
\end{multline}

\subsubsection{Cyclic conformal blocks}
(See Figure \ref{cyclic.web.diagram})

\begin{multline}
    \cS^{\, n, \,cyclic\,  block}_{\ll \bV_0,\, \bc_{\bV_0}\rr,\ll \bV_1,\, \bc_{\bV_1}\rr,\ll \bV_2,\, \bc_{\bV_2}\rr,\cdots,\ll \bV_m,\, \bc_{\bV_m}\rr}\ll x,y,R \rr\\
    = 
    \cS_{\ll\bV_0,\, \bc_{\bV_0}\rr \ll \bV_1,\, \bc_{\bV_1}\rr\bDelta_{\, 0 1}}^{\, n, \, norm}\ll x,y,R \rr 
    \ll\prod_{I=1}^N\ll -Q_{C_{1I}} \rr^{\, | \, V_{1I}\, | \, }f^{\, n}_{012,I}\rr
    \cS_{\ll\bV_1,\, \bc_{\, \bV_1} \rr \ll \bV_2,\, \bc_{\bV_{2}}\rr\bDelta_{\, 1 2}}^{\, n, \, norm}\ll x,y,R \rr\cdots
    \\
    \cdots \cS_{\ll\bV_{m-1},\, \bc_{\bV_{m-1}} \rr \ll\bV_{m},\, \bc_{\bV_{m}}\rr\bDelta_{\, m-1, m}}^{\, n, \, norm}\ll x,y,R \rr 
    \ll \prod_{I=1}^N\ll -Q_{C_{m,I}} \rr^{\, | \, V_{m,I}\, | \, }f^{\, n}_{m-1,m,0,I}\rr
    \cS_{\ll\bV_{m},\, \bc_{\bV_{m}} \rr \ll \bV_0,\, \bc_{\bV_{0}}\rr \bDelta_{\, m 0}}^{\, n, \, norm}\ll x,y,R \rr
    \\
    \ll\prod_{I=1}^N\ll -Q_{C_{0,I}} \rr^{\, | \, V_{0,I}\, | \, }f^{\, n}_{m01,I}\rr
\end{multline}

\section{The 4D limit of the 5D $n$-coloured web diagrams}
\label{section.08}
\textit{We take the 4D limit of the 5D $n$-coloured web diagrams, and show that 
$\cS^{\, n,\, norm}$ and
$\cZ^{\, n,\, 4D}_{\, N}$
lead to the same 
4D instanton partition functions.
}

Given a pair of charged $N$-Young diagrams 
$\ll \bV, \bc_{\bV} \rr$ and $\ll \bW, \bc_{\bW}\rr$ of 
$\ll \bN_{\bV}, \bk_{\bV} \rr$-type and $\ll \bN_{\bW}, \bk_{\bW} \rr$-type respectively, 
the number of factors that appear in 
$\cZ^{\, n, \, num}\ll \ba,\bV,\bc_{\bV}\, |\, \mu\, |\, \bb, \bW, \bc_{\bW} \rr$ (as given in Equation \ref{zp.z.num}) is,

\begin{equation}
K_n\ll \bN_{\bV},\, \bk_{\bV}\, |\, \bN_{\bW}, \bk_{\bW} \rr = 
\sum_{c = 0}^{n-1}
\ll 
   k_{V,c+1} \, k_{W,c  }
- 2k_{V,c  } \, k_{W,c  }
+  k_{V,c  } \, k_{W,c+1} 
+ N_{V,c} \, k_{W,c} + N_{W,c} \,     k_{V,c}
\rr
\end{equation}

For the web diagram partition function in Equation 
\ref{generic.web.partition}, where 
$\ll \bV_i,\, \bc_{\bV_i}\rr$ is a $\ll \bN_i,\, \bu_i \rr$-series,

\begin{multline}
\cS^{\, n}\ll x,y,R \rr = 
\sum_{\ll \bV_1,\, \bc_{\bV_1} \rr \in \ll \bN_1,\, \bu_1 \rr-series}
\cdots
\sum_{\ll \bV_m,\, \bc_{\bV_m} \rr \in \ll \bN_m,\, \bu_m \rr-series}
\cS^{\, n, \, block}_{\ll \bV_1,\, \bc_{\bV_1}\rr,\ll\bV_2,\, \bc_{\bV_2} \rr, \cdots,
\ll \bV_m,\, \bc_{\bV_m} \rr}
\ll x, y, R \rr
\\
= 
\sum_{m_1,\cdots, m_n = 0}^\infty
\sum_{\text{ $\ll \bV_1,\bc_{\bV_1}\rr$ is $\ll\bN_1, \bk_1(m_1)\rr$-type } }
\cdots
\sum_{\text{ $\ll \bV_n,\bc_{\bV_n}\rr$ is $\ll\bN_m, \bk_m(r_m)\rr$-type }}
\cS^{\, n, \, block}_{\ll\bV_1,\, \bc_{\bV_1}\rr,\ll\bV_2,\, \bc_{\bV_2}\rr,
\cdots,
\ll \bV_m,\, \bc_{\bV_m}\rr}\ll x,y,R \rr
\end{multline}

we multiply each summand of $\cS^{\, n}\ll x,y,R \rr$ with 
$\ll r_1, \cdots, r_m \rr$ by a factor of, 

\begin{equation}
R^{\sum_{i= 0 }^{p}
\ll 
K_n 
\ll \bN_i, \, \bk_i(r_i) \, | \, \bN_i,     \, \bk_i (r_i)         \rr 
- K_n 
\ll \bN_i, \, \bk_i(r_i) \, | \, \bN_{i+1}, \, \bk_{i+1} (r_{i+1}) \rr
\rr
}
\end{equation}

with either linear or cyclic boundary conditions for $\ll \bN_i, \bk_i \rr$ 
for a linear or a cyclic web diagram, respectively. 
For linear conformal blocks, 

\begin{equation}
\ll\bN_0,\, \bk_0\rr = 
\ll 
\bN_{\pmb{\emptyset}_0}, \pmb{0} 
\rr,
\quad 
\ll \bN_{m+1},\, \bk_0 \rr = 
\ll \bN_{\pmb{\emptyset}_{m+1}}, \pmb{0} \rr, 
\end{equation}

where $N_{\emptyset_i,c}$ is the number of $\mathbb{Z}_n$-charges in $\bc_{\pmb{\emptyset}_i}$ that are equal to $c$, $i = 0, m+1$.
For cyclic conformal blocks,

\begin{equation}
\ll \bN_{0}, \, \bk_{0} \rr = 
\ll \pmb{0}, \, \pmb{0} \rr,
\quad
\ll 
\bN_{m+1}, \, 
\bk_{m+1} 
\rr = 
\ll 
\bN_1, \bk_1   
\rr
\end{equation}

In the limit $R \rightarrow 0$, 

\begin{multline}
\cS^{\, n}
\ll \eo,\et \rr = 
\lim_{R\rightarrow 0}
\sum_{m_1,\cdots,m_p = 0}^\infty
R^{\sum_{i= 0 }^{n}
\ll K_n
\ll \bN_i,\bk_i(r_i)\, |\, \bN_i,\bk_i(r_i)             \rr - K_n
\ll \bN_i,\bk_i(r_i)\, |\, \bN_{i+1},\bk_{i+1}(r_{i+1}) \rr
\rr
}
\\
\sum_{\text{ $\ll \bV_1,\bc_{\bV_1}\rr$ is $\ll\bN_1, \bk_1(m_1)\rr$-type } }
\cdots
\sum_{\text{ $\ll \bV_m,\bc_{\bV_m}\rr$ is $\ll\bN_n, \bk_p(r_m)\rr$-type }}
\cS^{\, n, \, block}_{\ll\bV_1,\, \bc_{\bV_1}\rr,\ll\bV_2,\, 
\bc_{\bV_2}\rr,\cdots,\ll\bV_m,\, \bc_{\bV_m}\rr}\ll x,y,R \rr
\\
= 
\sum_{\text{ $\ll \bV_1,\, \bc_{\bV_1} \rr \in \ll \bN_1,\, \bu_1 \rr$-series }}
\cdots
\sum_{\text{ $\ll \bV_m,\, \bc_{\bV_m} \rr \in \ll \bN_m,\, \bu_m \rr$-series }}
\cS^{\, n, \, block}_{\ll \bV_1,\, \bc_{\bV_1} \rr,
\ll \bV_2,\, \bc_{\bV_2} \rr,
\cdots,
\ll \bV_m,\, \bc_{\bV_m} \rr
}
\ll \eo,\et \rr.
\end{multline}

Since the 4D limit of the edge factors are, 

\begin{equation}
\lim_{R\rightarrow 0} 
\ll - Q_{C_{iI}} \rr^{\, | \, V_{iI}\, | \, } f^{\, n}_{i-1,i,i+1,I} = 
\ll -1 \rr^{\, | \, V_{iI}\, | \, } f^{\, n}_{i-1,i,i+1,I}, 
\end{equation}

$\cS^{\, n, \, block
}_{
\ll\bV_1,\, \bc_{\bV_1}\rr,\ll\bV_2,\, \bc_{\bV_2}\rr,\cdots,\ll
\bV_m,\, \bc_{\bV_m}\rr}\ll \eo,\et \rr$ for linear and cyclic 
building blocks is

\begin{multline}
\label{4d.bb.linear}
\cS^{\, n, \,linear\, block}_{\ll\bV_1,\, \bc_{\bV_1}\rr, 
\ll \bV_2,\, \bc_{\bV_2}\rr, \cdots, \ll \bV_m,\, \bc_{\bV_m}\rr}
\ll \eo, \et \rr
\\
=
\cS_{
\ll \pmb{\emptyset}, \, \bc_{\pmb{\emptyset}_0} \rr 
\ll \bV_1,\, \bc_{\bV_{1}} \rr \bDelta_{\, 0 1}}^{\, n, \, norm}
\ll \eo, \et \rr 
\ll -1 \rr^{\sum_{I=1}^N \, | \, V_{1I}\, | \, }
\prod_{I=1}^N f^{\, n}_{0 1 2, I}
\cS_{
\ll \bV_1, \, \bc_{\bV_1} \rr 
\ll \bV_2, \, \bc_{\bV_2} \rr \bDelta_{\, 1 2}}^{\, n, \, norm} 
\ll \eo, \et \rr
\cdots
\\
\cdots 
\cS_{
\ll \bV_{m-1},\, \bc_{\bV_{m-1}} \rr 
\ll \bV_{m},  \, \bc_{\bV_{m}}   \rr \bDelta_{\, m-1, m}}^{\, n, \, norm}
\ll \eo, \et \rr 
\ll -1 \rr^{\sum_{I = 1}^N \, | \, V_{mI}\, | \, }
\prod_{I=1}^N f^{\, n}_{m-1, m, m+1, I}
\cS_{
\ll \bV_{m},\, \bc_{\bV_{m}} \rr
\ll \pmb{\emptyset},\, \bc_{\pmb{\emptyset}_{m+1}} \rr 
\bDelta_{\, m, m+1}}^{\, n, \, norm}
\ll \eo, \et \rr
\end{multline}

\begin{multline}
\label{4d.bb.cyclic}
\cS^{
\, n, \,cyclic\,  block
}_{
\ll \bV_0, \, \bc_{\bV_0} \rr, 
\ll \bV_1, \, \bc_{\bV_1} \rr,
\ll \bV_2, \, \bc_{\bV_2} \rr,
\cdots,
\ll \bV_m, \, \bc_{\bV_m} \rr}
\ll \eo, \et \rr 
\\
= 
\cS_{
\ll \bV_0, \, \bc_{\bV_0} \rr
\ll \bV_1, \, \bc_{\bV_1} \rr
\bDelta_{\, 0 1}
}^{
\, n, \, norm
}
\ll \eo, \et \rr
\ll -1 \rr^{\sum_{I=1}^N \, | \, V_{1I} \, | \, }
\prod_{I=1}^N 
f^{\, n}_{012,I}
\cS_{
\ll
\bV_1,\, \bc_{\bV_1}
\rr 
\ll 
\bV_2,\, \bc_{\bV_2}
\rr
\bDelta_{\, 1 2}}^{\, n, \, norm}
\ll \eo, \et \rr
\cdots
\\
\cdots 
\cS_{
\ll \bV_{m-1},\, \bc_{\bV_{m-1}} \rr 
\ll \bV_{m},  \, \bc_{\bV_m}     \rr
\bDelta_{\, m-1, m}}^{\, n, \, norm}
\ll \eo, \et \rr 
\ll -1 \rr^{\sum_{I=1}^N \, | \, V_{mI}\, | \, } 
\prod_{I=1}^N 
f^{\, n}_{m-1,m,0,I}
\\ 
\cS_{
\ll \bV_m, \, \bc_{\bV_m} \rr 
\ll \bV_0, \, \bc_{\bV_0} \rr
\bDelta_{\, n 0}}^{\, n, \, norm}
\ll \eo, \et \rr
\ll -1 \rr^{\sum_{I=1}^N \, | \, V_{0I}\, | \, }
\prod_{I=1}^N f^n_{m01,I}
\end{multline}

Observing that, 

\begin{equation}
F \ll \bV_i, \bc_{\bV_i} \, | \,  \bV_{i+1}, \bc_{\bV_{i+1}} \rr = 
\prod_{I = 1}^N 
f^{\, n, \, right}_{i,i+1,I} f^{\, n, \,left}_{i,i+1,I}, 
\end{equation}

the relation between 
$F\ll \bV, \bc_{\bV} \, | \, \bW, \bc_{\bW} \rr$ and 
the framing factors $f^{\, p}_{\, i-1, i, i+1, I}$ for 
the linear conformal blocks is,

\begin{multline}
\ll 
\prod_{I = 1}^N \ll -1 \rr^{\, | \, V_{1I}\, | \, } 
f^{\, n}_{\, 012, I} \rr
\ll 
\prod_{I=1}^N 
\ll -1 \rr^{\, | \, V_{2I}\, | \, } f^{\, n}_{\, 123, I} 
\rr
\cdots
\\ 
\ll 
\prod_{I=1}^N 
\ll - 1 \rr^{\, | \, V_{m-1,I}\, | \, } 
f^{\, n}_{\, m-2, m-1, m, I} \rr
\ll 
\prod_{I=1}^N 
\ll -1 \rr^{\, | \, V_{mI} \, | \, } f^{\, n}_{m-1,m,m+1, I} \rr
\\
= 
\prod_{I=1}^N
\ll 
f^{\, p, \,  left}_{01, I} 
f^{\, p, \, right}_{12, I} 
f^{\, p, \,  left}_{12, I} 
f^{\, p, \, right}_{23, I} 
\cdots 
f^{\, p,  \, left}_{m-2, m-1, I} 
f^{\, p, \, right}_{m-1, m,   I} 
f^{\, p, \,  left}_{m-1, m,   I} 
f^{\, p, \, right}_{m,   m+1, I} 
\rr
\\
= 
\prod_{I=1}^N
\ll 
f^{\, n, \,right}_{01, I} 
f^{\, n, \,left}_{01,I} 
f^{\, n, \, right}_{12,I} 
f^{\, n, \,left}_{12,I} 
f^{\, n, \,right}_{23,I} 
\cdots 
f^{\, n, \,left}_{m-2,m-1,I} 
f^{\, n, \,right}_{m-1,m,I} 
f^{\, n, \, left}_{m-1,m,I} 
f^{\, n, \, right}_{m,m+1,I} 
f^{\, n, \,left}_{m,m+1,I} 
\rr
\\
= 
\prod_{i = 0}^m
F \ll \bV_i, \bc_{\bV_i} \, | \,  \bV_{i+1}, \bc_{\bV_{i+1}} \rr
\end{multline}

Since 
$\bV_{0} = \bV_{m+1} = \emptyset$, 
we have 
$f^{\, n, \,right}_{01,I} = f^{\, n, \,left}_{m,m+1,I} = 1$, 
and it is justified to multiply the right hand side in  
the second last equality by
$f^{\, n, \,right}_{01,I}f^{\, n, \,left}_{m,m+1,I}$.
For the cyclic conformal blocks, the relation is,

\begin{multline}
\ll \prod_{I=1}^N 
\ll -1 \rr^{\, | \, V_{1I}\, | \, } f^{\, n}_{012, I} \rr 
\ll 
\prod_{I=1}^N 
\ll -1 \rr^{\, | \, V_{1I}\, | \, } f^{\, n}_{123, I} \rr 
\cdots
\\ 
\ll 
\prod_{I=1}^N 
\ll -1 \rr^{\, | \, V_{mI}\, | \, } f^{\, n}_{m-1,m,0, I} 
\rr
\ll 
\prod_{I=1}^N 
\ll -1 \rr^{\, | \, V_{0I}\, | \, } f^{\, n}_{m01, I} \rr
\\
= \prod_{I=1}^N
\ll 
f^{\, n,  left}_{01,I} 
f^{\, n, right}_{12,I} 
f^{\, n,  left}_{12,I} 
f^{\, n, right}_{23,I} 
\cdots
f^{\, n,  left}_{m-2,m-1,I} 
f^{\, n, right}_{m-1,m,I} 
f^{\, n,  left}_{m-1,m,I} 
f^{\, n, right}_{m,0,I} 
f^{\, n,  left}_{m,0,I} 
f^{\, n, right}_{0 1,I} 
\rr
\\
= 
\prod_{i = 0}^m
F \ll \bV_i, \bc_{\bV_i} \, | \,  \bV_{i+1}, \bc_{\bV_{i+1}} \rr
\end{multline}

and we can write the linear and cyclic conformal blocks in Equation 
\ref{4d.bb.linear} and Equation \ref{4d.bb.cyclic} as, 

\begin{multline}
\label{4d.bb.linear.v2}
\cS^{\, n, \,linear\, block}_{\ll\bV_1,\, \bc_{\bV_1}\rr,\ll\bV_2,\, \bc_{\bV_2}\rr,\cdots,\ll\bV_m,\, \bc_{\bV_m}\rr}\ll \eo, \et \rr
\\
=
F \ll \pmb{\emptyset}, \bc_{\pmb{\emptyset}_0} \, | \,  \bV_1, \bc_{\bV_1} \rr
\cS_{\pmb{\emptyset} \bV_1 \bDelta_{\, 0 1}}^{\, n, \, norm}
\ll \eo, \et \rr 
F \ll \bV_1, \bc_{\bV_1} | \bV_{2}, \bc_{\bV_{2}} \rr
\cS_{\ll\bV_1,\, \bc_{\bV_1}\rr
\ll\bV_2, \, \bc_{\bV_2} \rr
\bDelta_{\, 12}}^{\, n, \, norm}
\ll \eo, \et \rr
\cdots
\\
\cdots 
F 
\ll \bV_{m-1}, \bc_{\bV_{m-1}} \, | \,  \bV_{m}, \bc_{\bV_{m}} \rr
\cS_{\ll \bV_{m-1},\, \bc_{\bV_{m-1}} \rr 
\ll \bV_{m}, \, \bc_{\bV_{m}} \rr 
\bDelta_{\, m-1,m}}^{\, n, \, norm}
\ll \eo, \et \rr
F 
\ll 
\bV_m, \bc_{\bV_m} \, | \,  \pmb{\emptyset}, \bc_{\pmb{\emptyset}_{m+1}} 
\rr
\\ 
\cS_{
\ll \bV_{m},\, \bc_{\bV_{m}} \rr
\ll \pmb{\emptyset},\, \bc_{\pmb{\emptyset}_{n+1}} \rr \bDelta_{\, m,m+1}}^{\, n, \, norm}
\ll \eo, \et \rr
\end{multline}

\begin{multline}\label{4d.bb.cyclic.v2}
\cS^{
\, n, \,cyclic\,  block
}_{
\ll \bV_0,\, \bc_{\bV_0} \rr,
\ll \bV_1,\, \bc_{\bV_1} \rr,
\ll \bV_2,\, \bc_{\bV_2} \rr,
\cdots,
\ll \bV_m,\, \bc_{\bV_m} \rr}
\ll \eo, \et \rr 
\\
= 
F 
\ll \bV_0, \bc_{\bV_0} \, | \,  \bV_1, \bc_{\bV_{1}} \rr
\cS_{
\ll \bV_0, \, \bc_{\bV_0} \rr
\ll \bV_1, \, \bc_{\bV_1} \rr
\bDelta_{\, 0 1}}^{\, n, \, norm}
\ll \eo, \et \rr
F
\ll \bV_1, \bc_{\bV_1} \, | \,  \bV_2, \bc_{\bV_2} \rr
\cS_{
\ll \bV_1, \, \bc_{\bV_1} \rr
\ll \bV_2, \, \bc_{\bV_2} \rr
\bDelta_{\, 12}}^{\, n, \, norm}
\ll \eo, \et \rr
\cdots
\\
\cdots 
F
\ll 
\bV_{m-1}, \bc_{\bV_{m-1}} \, | \,  \bV_{m}, \bc_{\bV_{m}} 
\rr
\cS_{
\ll
\bV_{m-1}, \, \bc_{\bV_{m-1}} 
\rr
\ll
\bV_{m},   \, \bc_{\bV_{m}}
\rr
\bDelta_{\, m-1,m}}^{\, n, \, norm}
\ll \eo, \et \rr 
F
\ll \bV_m, \bc_{\bV_m} \, | \,  \bV_0, \bc_{\bV_0} \rr
\\
\cS_{
\ll
\bV_m, \, \bc_{\bV_m} \rr
\ll 
\bV_0, \, \bc_{\bV_0} \rr
\bDelta_{\, m 0}}^{\, n, \, norm}
\ll \eo, \et \rr
\end{multline}

\subsection{
$\cS^{\, n,\, norm}$ and
$\cZ^{\, n,\, 4D}_{\, N}$
lead to the same 
4D instanton partition functions} 
Let, 

\begin{multline} 
\cF
\ll \ba, \bV \, | \,  \bb, \bW \rr = 
F
\ll \bV, \bc_{\bV} \, | \,  \bW, \bc_{\bW} \rr
\frac{
\cS^{\, n,\, norm}_{\, \ll \bV,\, \bc_{\bV} \rr 
\ll \bW,\, \bc_{\bW} \rr\bDelta} \ll \eo, \et \rr
}{
\cZ^{\, n,\, 4D}_{\, N}   
\ll  \ba, \bV, \bc_{\bV} \, | \, \mu \, | \, \bb, \bW, \bc_{\bW} \rr
}
\\
= \frac{
\prod_{I=1}^N
\prod_{J=1}^N
\ll 
\cH_{V_{IJ}} \ll \ba, \bc_{\bV} \rr
\cH_{W_{IJ}} \ll \bb, \bc_{\bW} \rr
\rr^{1/2} 
}{
\prod_{I=1}^N
\cA_{V_{II}} \ll \ba, \bc_{\bV} \rr \cL_{W_{II}} \ll \bb, \bc_{\bW} \rr
\prod_{J=1}^N
\prod_{I=1}^{J-1} 
\cH_{V_{JI}} \ll \ba, \bc_{\bV} \rr
\cH_{W_{IJ}} \ll \bb, \bc_{\bW} \rr
} 
\\
=
\cF_{left}  \ll \ba,\bV \rr
\cF_{right} \ll \bb,\bW \rr
\end{multline} 

where we have used Equations \ref{abbrev.s.num}, \ref{abbrev.s.den} and \ref{abbrev.z.den} for the second equality. We also define, 

\begin{equation}
    \cF_{left} \ll \ba, \bV \rr = 
\frac{
\prod_{I=1}^N 
\prod_{J=1}^N
\cH_{V_{IJ}} \ll \ba, \bc_{\bV} \rr^{\frac12}
}{
\prod_{I=1}^N 
\cA_{V_{II}} \ll \ba, \bc_{\bV} \rr
\prod_{J=1}^N
\prod_{I = 1}^{J-1} 
\cH_{V_{JI}} \ll \ba, \bc_{\bV} \rr}
\end{equation}

\begin{equation}
    \cF_{right} \ll\bb, \bW\rr = 
\frac{
\prod_{I=1}^N
\prod_{J=1}^N 
\cH_{W_{IJ}} \ll \bb, \bc_{\bW} \rr^{\frac12}
}{
\prod_{I=1}^N \cL_{W_{II}} \ll \bb, \bc_{\bW} \rr
\prod_{J=1}^N
\prod_{I = 1}^{J-1} \cH_{W_{IJ}} \ll \bb, \bc_{\bW} \rr
} 
\end{equation}

The linear and cyclic conformal blocks take the form, 

\begin{multline}
\label{linear.quiver}
\cZ^{\, n,\, linear\, block} 
= \cZ^{\, n,\, 4D}_{\, N} 
\ll \ba_0, \emptyset, \bc_{\pmb{\emptyset}} \, | \, \mu_{0 1} \, | \, \ba_1, \bV_1, \bc_{\bV_1} \rr
\cZ^{\, n,\, 4D}_{\, N} 
\ll \ba_1, \bV_1, \bc_{\bV_1} \, | \, \mu_{12} \, | \, \ba_2, \bV_2,\bc_{\bV_2} \rr  
\cdots
\\ 
\cZ^{\, n,\, 4D}_{\, N} 
\ll \ba_m, \bV_m, \bc_{\bV_m} \, | \, \mu_{m,m+1} \, | \, \ba_{m+1}, 
\emptyset, \bc_{\pmb{\emptyset}} \rr
\end{multline}

\begin{multline}
\label{cyclic.quiver}
\cZ^{\, n,\, cyclic\, block} 
= \cZ^{\, n,\, 4D}_{\, N} 
\ll \ba_0, \bV_0, \bc_{\bV_0} \, | \, \mu_{0 1} \, | \, \ba_1, \bV_1, \bc_{\bV_1} \rr
\cZ^{\, n,\, 4D}_{\, N} 
\ll \ba_1, \bV_1, \bc_{\bV_1} \, | \, \mu_{12} \, | \, \ba_2, \bV_2, \bc_{\bV_2} \rr  
\cdots 
\\ 
\cZ^{\, n,\, 4D}_{\, N} 
\ll \ba_m,\bV_m, \bc_{\bV_m} \,| \, \mu_{m0} \, |\, \ba_0, \bV_0, \bc_{\bV_0} \rr
\end{multline}

We claim that 
\begin{equation}
    \boxed{\cZ^{\, n,\, linear\, block} = \cS^{\, n,\, linear\, block}\qquad \text{and} \qquad \cZ^{\, n,\, cyclic\, block} = \cS^{\, n,\, cyclic\, block}}
\end{equation}
Comparing Equation 
\ref{4d.bb.linear.v2} to \ref{linear.quiver} and 
\ref{4d.bb.cyclic.v2} to \ref{cyclic.quiver}, 
the claim is equivalent to, 

\begin{equation} 
\cF_{left} \ll\ba_0, \mathbf{\emptyset}\rr 
\cF_{right} \ll\ba_1, \bV_1\rr
\cdots 
\cF_{left} \ll\ba_m, \bV_m\rr
\cF_{right} \ll\ba_{m+1}, \mathbf{\emptyset}\rr = 1, 
\end{equation}

\begin{equation}
\cF_{left}  \ll \ba_0,\bV_0\rr
\cF_{right} \ll \ba_1,\bV_1\rr
\cdots
\cF_{left}  \ll \ba_m,\bV_m\rr 
\cF_{right} \ll \ba_0, \bV_0\rr = 1
\end{equation} 

Since 
$\cF\ll \ba, \pmb{\emptyset} \, | \,  \bb, \pmb{\emptyset} \rr = 1$, 
it follows that 
$\cF_{left} \ll \ba, \emptyset \rr = 
\cF_{right} \ll \ba, \emptyset \rr = 1$, 

\begin{multline} 
\cF_{left} \ll \ba, \bV \rr \cF_{right} \ll \ba, \bV \rr = 
\\
\frac{
\prod_{I=1}^N
\prod_{J=1}^N \cH_{V_{IJ}} \ll \ba, \bc_{\bV} \rr
}
{
\prod_{I=1}^N \ll \cA_{V_{II}}\ll \ba, \bc_{\bV} \rr \cL_{V_{II}}\ll \ba, \bc_{\bV} \rr \rr
\prod_{J=1}^N
\prod_{I=1}^{J-1}\cH_{V_{IJ}}\ll \ba, \bc_{\bV} \rr
\prod_{I=1}^N
\prod_{J=1}^{I-1}\cH_{V_{IJ}}\ll \ba, \bc_{\bV} \rr
}
=\\
\frac{
\ll -1 \rr^{\sum_{I=1}^N\, | \, V_I\, | \, }
\prod_{I=1}^N
\prod_{J=1}^N \cH_{V_{IJ}}\ll \ba, \bc_{\bV} \rr
}
{
\prod_{I=1}^N\cH_{V_{II}}(0)
\prod_{I<J}\cH_{V_{IJ}}\ll \ba, \bc_{\bV} \rr
\prod_{I>J}\cH_{V_{IJ}}\ll \ba, \bc_{\bV} \rr
} = 1
\end{multline} 

proving the claim.

\section{A 4-point conformal block}
\label{section.09}
\textit{We recover the 4-point conformal block on $\mathbb{R}^4/\mathbb{Z}_n$ 
in Equation (2.10) of \cite{alfimov.tarnopolsky}
by gluing two strip partition functions. To help with the comparison, in this section, we use the notation of \cite{alfimov.tarnopolsky}.
} 

\subsection{Change of parameters and notation}
In previous sections, the case of computing 
4-point conformal blocks is the same as linear blocks with $\bV_0 = \pmb{\emptyset}, \bV_1 = \bY, \bV_2 = \pmb{\emptyset}$. We used the K\"{a}hler 
parameters
$\Delta_{\, 0 1,   K}, \cdots$,
the Coulomb parameters 
$\ba_0, \ba_1, \ba_2$, 
the mass parameters $\mu_{0 1}$ and $\mu_{12}$,
the deformation parameters $\epsilon_1$ and $\epsilon_2$,
and $c_I$ for the Young diagram charges.

In \cite{alfimov.tarnopolsky}, Alfimov and Tarnopolsky
use the Coulomb parameters 
$\pmb{P}_i = 
\ll P_{i,1}, \cdots, P_{i,N} \rr, 
\pmb{P} = 
\ll P_{1},   \cdots, P_{N} \rr$, 
$i = 1, 2$, 
the mass parameters $\alpha_1, \alpha_2$, 
the deformation parameters 
$b^{-1}, b$, and $\bq = \ll q_1,\cdots q_N \rr$ for the Young diagram charges. 
These two sets of parameters are related as, 

\begin{equation}
\ba_0 = \bP_1, \quad 
\ba_1 = \bP, \quad 
\ba_2 = \bP_2, \quad 
\mu_{0 1} = \alpha_1, \quad 
\mu_{12} = \alpha_2, \quad 
\eo = b^{-1}, \quad 
\et = b, \quad 
c_{Y_I} = -q_I, 
\end{equation}

and the K\"{a}hler parameters
$\ll \Delta_{\, 0 1,   K}, \cdots \rr$, 
in terms of the parameters of 
\cite{alfimov.tarnopolsky} are,   

\begin{multline}
\Delta_{\, 0 1,   K} = -P_{\, 1, K} + P_{\, K} + \alpha_{\, 1} - b^{\, -1}, 
\quad 
\Delta_{\, 0 1, M_K} = P_{\, 1, K + 1} - P_{\, K} + \alpha_{\, 1} - b^{\, -1}, 
\\ 
\Delta_{\, 12, K} = - P_{\, K} + P_{\, 2, K} + \alpha_{\, 2} - b^{\, -1}, 
\quad 
\Delta_{\, 12, M_K} = P_{\, K + 1} - P_{\, 2, K} + \alpha_{\, 2} - b^{\, -1}
\end{multline}

We will also use, 

\begin{equation}
\ll \bY,\, \bc_{\bY}\rr = \ll \bY,\, - \, \bq \rr    
\end{equation}

\subsection{The 4-point conformal block}
In the notation of \cite{alfimov.tarnopolsky}, introduced 
above, the 4-point $n$-coloured conformal block constructed from two $n$-coloured strip partition functions is, 

\begin{equation}
\cS^{\, n, \, 4-point} \ll b \rr = 
\sum_{\ll \bY,\, -\bq \rr} 
\cS^{\, n, \, block}_{\ll \bY, \, \bq \rr}
\ll b \rr z^{\, |\, \bY \, | \, / \, n},
\end{equation}

where $z$ is an instanton expansion parameter (a position parameter in 2D conformal field theory).
The summation is over each of the 
$\ll \bN, \bu = 
\ll 0, 0, 0 \rr \rr$-series in 
Equations  \ref{u.0.series.1}--\ref{u.0.series.3}. 
For example, if we choose Series 2 in Equation \ref{u.0.series.2} then 
$\bN = \ll 0, 1, 0, 1 \rr$ and 
$\bk (r) = \ll r, r + 1, r + 1, r + 1 \rr$ 
and the summation is over all 
$\ll \bY, \bc_{\bY} \rr = 
\ll \bY, - \bq \rr$ of 
$\ll \bN, \bk (r) \rr$-type. We obtain the expression for the 4-point conformal block $\cS^{\, n, \, block}_{\ll \bY, \, \bq \rr}\ll b \rr$ from the expression of linear conformal block in Equation \ref{4d.bb.linear.v2} by substituting $\bV_0 = \pmb{\emptyset}, \bV_1 = \bY, \bV_2 = \pmb{\emptyset}$. 
In the notation of \cite{alfimov.tarnopolsky}, 

\begin{multline}
\cS^{\, n, \, block}_{\ll \bY, \, \bq \rr}\ll b \rr 
=
F \ll \pmb{\emptyset}, \pmb{0} \, | \,  \bY, \pmb{c}_{\pmb{Y}} \rr
\cS^{
\, n, \,norm
}_{
\ll \pmb{\emptyset}, \,  \pmb{0} \rr
\ll \bY,             \, \pmb{c}_{\pmb{Y}}   \rr 
\bDelta_{\, 0 1}} 
\ll \eo, \, \et \rr
F \ll \pmb{Y}, \pmb{c}_{\pmb{Y}} \, | \,  \pmb{\emptyset}, \pmb{0} \rr
\cS^{
\, n, \,norm
}_{
\ll \bY,            \, \pmb{c}_{\pmb{Y}}    \rr 
\ll \pmb{\emptyset},\,  \pmb{0} \rr 
\bDelta_{\, 1 2}
} 
\ll \eo, \et \rr
\\
= 
F \ll \pmb{\emptyset}, \pmb{0} \, | \,  \bY, -\pmb{q} \rr
\cS^{
\, n, \,norm
}_{
\ll \pmb{\emptyset}, \,  \pmb{0} \rr
\ll \bY,             \, -\bq     \rr 
\bDelta_{\, 0 1}} 
\ll b^{-1}, \, b \rr
F \ll \pmb{Y}, - \pmb{q} \, | \,  \pmb{\emptyset}, \pmb{0} \rr
\cS^{
\, n, \,norm
}_{
\ll \bY,            \, -\bq     \rr 
\ll \pmb{\emptyset},\,  \pmb{0} \rr 
\bDelta_{\, 1 2}
} 
\ll b^{-1}, b \rr
\\
= 
\prod_{I=1}^N
\prod_{J=1}^I
\prod_{
\substack{
\wsquare \in Y_I
\\
A_{\wsquare, Y_I} + 
L_{\wsquare, \emptyset} + 1 + q_I \, = \,  0 \, \mod{\, n}
}
}
\ll 
  P_{1, J} - P_{\, I} - \alpha_1
- A^{  }_{\wsquare,       Y_I} b 
+ L^{++}_{\wsquare, \emptyset} b^{-1} 
\rr
\\
\prod_{I=1}^N
\prod_{J=1}^{I-1}
\prod_{
\substack{
\wsquare \in Y_J
\\
A_{\wsquare,       Y_J} + 
L_{\wsquare, \emptyset} + 1 + q_J \, = \, 0 \, \mod{\, n}
}
}
\ll 
P_{1,I} - P_J - \alpha_1
- A^{  }_{\wsquare,       Y_J} b 
+ L^{++}_{\wsquare, \emptyset} b^{-1} 
\rr
\\
\prod_{I\, = 1}^N
\prod_{
\substack{
\wsquare \in Y_I
\\
A_{\wsquare, Y_I} + 
L_{\wsquare, Y_I} + 1 \, =  \, 0 \, \mod{\, n}
}
}
\ll 
- A^{  }_{\wsquare,       Y_I} b 
+ L^{++}_{\wsquare, \emptyset} b^{-1} 
\rr^{-1}
\\
\prod_{I=1}^N 
\prod_{J=1}^{I-1} 
\prod_{
\substack{
\wsquare \in  Y_I
\\
A_{\wsquare, Y_I} + 
L_{\wsquare, Y_J} + 1 + q_I - q_J \, = \, 0 \, \mod{\, n}
}
} 
\ll  
P_J - P_{\, I} 
- A_{\wsquare,   Y_I}      b 
+ L_{\wsquare,   Y_J}^{++} b^{-1} 
\rr^{-1}
\\
\prod_{
\substack{
\wsquare \in  Y_J
\\
  A_{\wsquare, Y_J} 
+ L_{\wsquare, Y_I} + 1 - q_I + q_J\, = \, 0 \, \mod{\, n}
}
} 
\ll 
P_J - P_I  
+ A_{\wsquare,   Y_J}^{++} b 
- L_{\wsquare,   Y_I}      b^{-1} 
\rr^{-1}
\\
\prod_{I=1}^N
\prod_{J=1}^I
\prod_{
\substack{
\wsquare \in 
Y_J
\\
A_{\wsquare,       Y_J} + 
L_{\wsquare, \emptyset} + 1 + q_J \, = \, 0  \, \mod{\, n}
}
}
\ll 
P_J - P_{2,I} - \alpha_2
+ A^{++}_{\wsquare,       Y_J} b 
- L^{  }_{\wsquare, \emptyset} b^{-1} 
\rr
\\
\prod_{I=1}^N
\prod_{J=1}^{I-1}
\prod_{
\substack{
\wsquare \in Y_I
\\
A_{\wsquare,       Y_I} + 
L_{\wsquare, \emptyset} + 1 + q_I \, = \, 0 \, \mod{\, n}
}
}
\ll 
P_I - P_{2,J} - \alpha_2
+ A^{++}_{\wsquare,       Y_I} b 
- L^{  }_{\wsquare, \emptyset} b^{-1} 
\rr
\\
\prod_{I=1}^N
\prod_{
\substack{
\wsquare \in Y_I
\\
A_{\wsquare, Y_I} + L_{\wsquare, Y_I} + 1 \, = \, 0 \, \mod{\, n}
}
}
\ll 
  A^{++}_{\wsquare, Y_I} b 
- L^{  }_{\wsquare, Y_I} b^{-1} 
\rr^{-1}
\\
\prod_{I=1}^N
\prod_{J=1}^{I-1}
\prod_{
 \substack{
\wsquare \in Y_I
\\
  A_{\wsquare, Y_I} 
+ L_{\wsquare, Y_J} + 1 + q_I - q_J \, = \, 0 \, \mod{\, n}
}
}
\ll 
P_I - P_J 
+ A^{++}_{\wsquare, Y_I} b 
- L^{  }_{\wsquare, Y_J} b^{-1} 
\rr^{-1}
\\
\prod_{
\substack{
\wsquare \in Y_J
\\
  A_{\wsquare, Y_J} 
+ L_{\wsquare, Y_I} + 1 - q_I + q_J \, = \, 0 \, \mod{\, n}
}
}
\ll 
P_I - P_J 
- A^{  }_{\wsquare, Y_J} b 
+ L^{++}_{\wsquare, Y_I} b^{-1} 
\rr^{-1}
\\
= 
\prod_{I,J = 1}^N
\prod_{
\substack{
\wsquare \in Y_I
\\
  A_{\wsquare,       Y_I} 
+ L_{\wsquare, \emptyset} + 1 \, = \,  - q_I \, \mod{\, n}
}
}
\ll 
P_{1,J} - P_I - \alpha_1 
- A^{  }_{\wsquare,       Y_I} b 
+ L^{++}_{\wsquare, \emptyset} b^{-1} 
\rr
\\
\prod_{I, J \, = \,  1}^N
\prod_{
\substack{
\wsquare \in Y_J 
\\
  A_{\wsquare,       Y_J} 
+ L_{\wsquare, \emptyset} + 1 \, = \,  - q_J \, \mod{\, n}
}
}
\ll 
P_J - P_{2,I} - \alpha_2 
+ A^{++}_{\wsquare,       Y_J} b 
- L^{  }_{\wsquare, \emptyset} b^{-1} 
\rr
\\
\prod_{I,J}^N
\prod_{
\substack{
\wsquare \in Y_I
\\
A_{\wsquare,Y_I} + L_{\wsquare, Y_J} + 1 \, = \,  - q_I + q_J \, \mod{\, n}
}
}
\ll
P_I - P_J + A^{++}_{\wsquare, Y_I} b - L_{\wsquare, Y_J} b^{-1} \rr^{-1}
\\ 
\prod_{
\substack{
\wsquare \in Y_J \\
A_{\wsquare,Y_J} + L_{\wsquare, Y_I} + 1 \, = \, q_I - q_J \, \mod{\, n}
}
}
\ll P_I - P_J - A_{\wsquare, Y_J} b + L^{++}_{\wsquare, Y_I} b^{-1} \rr^{-1}
\\
\, = \,  
\frac{
\cZ^{\, num} 
\ll \bs_1, \pmb{\emptyset}, \pmb{0} 
\, | \, 
\alpha_1 \, | \, \bP, \bY,-\bq \rr
\cZ^{\, num} 
\ll \bP, \bY, -\bq \, | \, \alpha_2 \, | \, \bs_2, \pmb{\emptyset}, \pmb{0} \rr
}{
\cZ^{\, num} 
\ll \bP, \bY,-\bq  \, | \, 0 \, | \, \bP, \bY,-\bq \rr
}
\end{multline}

which is Equation (2.10) in \cite{alfimov.tarnopolsky}, 
for $N = 2, n = 4$.

\section{Comments and remarks}
\label{section.10}

\subsection{A formulation in terms of $n$-coloured Young diagrams}
We have refrained from discussing the $n$-coloured vertex in terms 
of $n$-coloured checkerboard Young diagrams, and in particular, we 
did not consider working in terms of $n$-cores and $n$-quotients
\cite{james.kerber, loehr, olsson}.
This is because, in this work, 
working in terms of 
$n$-cores and $n$-quotients was not needed and would have taken us far afield from our goal
which is to reproduce the 2D parafermion matrix elements in 
\cite{alfimov.tarnopolsky}. 

\subsection{The $n$-coloured vertex is not a product of $n$ arbitrary 
refined topological vertices}
The $n$-coloured vertex factorizes into a product of $n$ refined vertices.
However, it is not a product of $n$ arbitrary refined topological vertices 
unless the Young diagrams that label the preferred legs of the component 
vertices mesh together to form the Young diagram that labels the preferred 
leg of the $n$-coloured vertex.

\subsection{Counting parameters}
While the 5D strip partition function has $\ll 2 N - 1 \rr$ K\"ahler parameters, 
and the 4D instanton partition function also has $\ll 2 N - 1\rr$ parameters, 
there are $N^2$ relations among them which arise from the factors in 
$\cS^{\, n, \,num}_{\ll \bV,\, \bc_{\bV} \rr \ll \bW,\, \bc_{\bW} \rr\bDelta}\ll \eo,\et \rr$ (Equation \ref{s.n}) and 
$\cZ^{\, n,  \, num} \ll \ba, \bV, \bc_{\bV} \, | \, \mu \, | \, \bb, \bW, \bc_{\bW} \rr$ (Equation \ref{zp.z.num}).
The significance of the fact that $\ll 2 N - 1 \rr$ parameters satisfy 
$N^2$ relations is not clear at this stage. 

\subsection{Rational models}
In this work, we restricted our attention to generic, that is non-rational 
parameters and the corresponding conformal field theories. The relations 
between the parameters of the 5D strip partition function and the 2D matrix 
elements obtained in this work are expected to extend to the rational model,
after
\1 re-writing all parameters in terms of the screening charges $\alpha_+$ 
and $\alpha_-$ of the minimal 2D conformal field theory, and 
\2 restricting the Young diagrams that label the preferred legs to those
that satisfy Burge-type conditions (thereby eliminating the null states), 
as described in \cite{alkalaev.belavin, bershtein.foda, belavin.foda.santachiara}. 

\subsection{The orbifold vertex \cite{bryan.cadman.young}}
In \cite{bryan.cadman.young}, Bryan, Cadman and Young define Donaldson-Thomas 
invariants of Calabi-Yau orbifolds and introduce \textit{an orbifold topological 
vertex} to compute them. The orbifold vertex is $n$-coloured in the sense that 
it is the generating function of plane partitions that are made of interlacing 
checkerboard $n$-coloured Young diagrams.
However, the orbifold vertex differs (at least on a superficial level) from 
the $n$-coloured vertex introduced in this work in several technical respects.
\1 The orbifold vertex is constructed in terms of one type of 
$\Gamma_{\, \pm}$-operators, which is equivalent to using one Heisenberg algebra,
rather than $n$ Heisenberg algebras as in the $n$-coloured vertex, and consequently, 
the orbifold vertex is in the form of a single sum of a bilinear of Schur functions 
(as in the refined topological vertex), while the $n$-coloured vertex is in the form 
of a product of $n$ sums of bilinears of Schur functions, and 
\2 the cells of the plane partitions generated by the orbifold vertex are assigned 
$n$ colours, and cells of different colours are assigned different weights,
while in the case of the $n$-coloured vertex, all cells are assigned the same 
(trivial) weight
\footnote{\,
There can be an orbifold vertex that is formulated in terms of 
$n$ free bosons that assign different colours to different cells, but this is 
not clear at this stage.}, 
\3 the orbifold vertex is not refined, while the $n$-coloured vertex is
\footnote{\, 
Because of \textbf{1} and \textbf{2}, the orbifold vertex is 
parameterized in terms of $n$ parameters $\ll q_0, q_1, \cdots, q_{\, n-1} \rr$, 
while the $n$-coloured vertex is parameterized in terms of the equivariant 
deformation parameters $\ll \et, \eo \rr$, and the radius $R$. However, this 
is probably not a deep difference, and it is entirely possible that there is 
a refined version of the orbifold vertex.
}.
More generally, 
\4 The orbifold vertex computes topological string partition functions on 
orbifold Calabi-Yau (internal) spaces such as $\CC^{\, 3} / \ZZ_{\, n}$, 
while the $n$-coloured vertex computes topological string partition functions 
on the (Euclidean version of space-time) orbifolds $\CC^{\, 2} / \ZZ_{\, n}$.
It is possible that there is a more general operator formalism that produces
a more general vertex such that the orbifold vertex and the $n$-coloured vertex
are special cases of the same object, but this is not clear at this stage, and 
is unlikely given that these two vertices compute different objects.

\subsection{The intertwining operator of the Fock representations 
of the quantum toroidal algebras of type $A_n$
\cite{awata.kanno.mironov.morozov.suetake.zenkevich}}

Following the completion of this work, we learned that in
\cite{awata.kanno.mironov.morozov.suetake.zenkevich}, 
Awata, Kanno, Mironov, Morozov, Suetake and Zenkevich 
introduced an intertwining operator of the Fock representations 
of the quantum toroidal algebras of type $A_n$, and used these 
to obtain the instanton partition functions on 
$\CC^{\, 2} / \ZZ_{\, n}$
\footnote{\,
We thank M Bershtein and J-E Bourgine for bringing 
\cite{awata.kanno.mironov.morozov.suetake.zenkevich} 
to our attention.
}.
Using AGT, these instanton partition functions are identified 
with the matrix elements discussed in this work, and as such, 
the intertwiner of
\cite{awata.kanno.mironov.morozov.suetake.zenkevich}
is identified with the $n$-coloured vertex introduced 
in this work. 
The differences between the two works are that 
\1 The intertwiner of \cite{awata.kanno.mironov.morozov.suetake.zenkevich}
is in Awata-Feigin-Shiraishi operator form \cite{awata.feigin.shiraishi}, 
which bypasses working in terms of symmetric functions, while 
the $n$-coloured vertex in this work is in the (more conventional) 
Iqbal-Kozcaz-Vafa symmetric-function form \cite{iqbal.kozcaz.vafa},
\2 The focus of \cite{awata.kanno.mironov.morozov.suetake.zenkevich}
is on the representation theory of quantum toroidal algebras, while 
that of the present work is on making contact with the 2D matrix elements 
that the $\CC^{\, 2} / \ZZ_{\, n}$ topological vertex formalism computes,
with emphasis on the details, including the framing factors, the matching
of the various normalizations, \textit{etc.}

\subsection{More general orbifolds}
Following the completion of this work,
Bourgine and Jeong \cite{bourgine.jeong}
introduced an $n$-coloured topological vertex 
of the Awata-Feigin-Shiraishi type 
\cite{awata.feigin.shiraishi}, 
that allowed them to discuss new quantum toroidal algebras that 
are related to 5D supersymmetric gauge theories on 
$\ll \CC^{\, 2} / \, Z_n \rr \times S^1$ orbifolds, which were first discussed 
in \cite{bonelli.maruyoshi.tanzini.01, bonelli.maruyoshi.tanzini.02}. 
These orbifolds are more general than those discussed in this work,
and so far the corresponding 2D conformal field theories are not
known.

\subsection{Extracting the affine $A_n$, integral level-$N$, 
WZW model algebra}
In \cite{foda.macleod}, Macleod and the second author show
that there is a choice of the Nekrasov deformation parameters 
such that the coset component in the full $\cA \ll N, n \rr$ algebra trivializes and one obtains a formalism that involves only 
the affine $A_n$ WZW models, at integral level $N$, times 
a Heisenberg factor.

\section*{Acknowledgements}
We wish to thank V Belavin, M Bershtein, J-E Bourgine, J Bryan, 
H Kanno, N Macleod, M Manabe, A Morozov, R Santachiara, F Yagi 
and R-D Zhu for discussions on the subject of this work and related 
topics, and Trevor Welsh for a careful reading of the manuscript. 
WC is supported by an Australian Postgraduate Fellowship, and OF is supported by the Australian Research Council. 

\appendix 

\section{A proof of the $n$-coloured normalized product identity}
\label{appendix.A}
\textit{We prove the identity used to go from 
Equation \ref{normalized.5d.strip.partition.function} to 
Equation \ref{after.identity}}

In the following, $c$ stands for $c_V - c_W$, 
the difference of the charges of the Young diagrams $V$ 
and $W$.

\subsubsection{An identity}
For $c \in \ll 0, 1, \cdots, n-1 \rr$, 
we have the following identity:

\begin{multline}
\label{R.0.normalized.product.formula}
\frac{
\displaystyle
\prod^{\infty}_{
\substack{
i, j = 1
\\
\ll 
- V_i + j 
\rr + 
\ll 
- W^{\, \intercal}_j + i - 1
\rr 
\, = \, c \, \mod{\, n}
}
} 
\ll 1 - Q \, x^{- V_i + j} \, y^{-W^{\, \intercal}_j + i -1} \rr
}{
\displaystyle
\prod_{
\substack{
i, j = 1
\\
i + j - 1 = c \, \mod{\, n}
}
}^\infty
\ll 1 - Q \, x^j \, y^{i-1} \rr 
} 
\\ 
= \prod_{
\substack{
\bsquare \in V
\\
A_{\bsquare, V} + L_{\bsquare, W} + 1 \, = \, - c \, \mod{\, n}
}
}
\ll 
1 - Q \, x^{- A_{\bsquare, V}} \, y^{- L^{++}_{\bsquare, W}} 
\rr  
\prod_{
\substack{
\wsquare \in W
\\
A_{\wsquare, W} + L_{\wsquare, V} + 1 \, =  \, c \, \mod{\, n}}}
\ll 1 - Q \, x^{A^{++}_{\wsquare, W}} \, y^{L_{\wsquare, V}} \rr 
\\
= \prod_{
\substack{
\wsquare \in W
\\
A_{\wsquare, V} + L_{\wsquare, W} + 1 \, = \, - c \, \mod{\, n}
}
}
\ll 
1 - Q \, x^{- A_{\wsquare, V}} \, y^{- L^{++}_{\wsquare, W}} 
\rr  
\prod_{
\substack{
\bsquare \in V
\\
A_{\bsquare, W} + L_{\bsquare, V} + 1 \, =  \, c \, \mod{\, n}}}
\ll 1 - Q \, x^{A^{++}_{\bsquare, W}} \, y^{L_{\bsquare, V}} \rr 
\end{multline}

\subsection{A more general identity}

Here we prove an identity more general than 
Equation \ref{R.0.normalized.product.formula}, and from which the latter follows.

\begin{scriptsize}
\begin{figure}
\begin{tikzpicture}[scale=.6,every node/.style={scale= 0 .75}]

\foreach \i in {0,1,...,8}
{
\draw[thick] (1.5*\i,-2)--(1.5*\i,8);
}
\foreach \i in {-2,-1,...,8}
{
\draw[thick] (0,\i)--(1.5*8,\i);
}

\draw[->] (-1,8)--(-1,7);
\draw[->] (0,9)--(1,9);

\node [below,scale=1.2] at (-1,7) {$i$};
\node [right,scale=1.2] at (1,9) {$j$};

\foreach \i in {1, 2, ..., 8} 
{
\foreach \j in {1, 2, ..., 10} 
{
\node at (  -0.8 + 1.5*\i, 8.5 - \j ) {$\ll \i, \j \rr$};
}
}

\end{tikzpicture}
\caption{
{\it
$\psi_{\emptyset\emptyset}$
}
}
\label{A.Young.diagram.01}
\end{figure}
\end{scriptsize}

\begin{scriptsize}
\begin{figure}
\begin{tikzpicture}[scale=.6,every node/.style={scale= 0 .75}]

\begin{scope}[xshift=-10cm,yshift=8cm]
\draw[thick] (0,0) rectangle (1.5,-1); 
\draw[thick] (1.5,0) rectangle (3,-1); 
\draw[thick] (3,0) rectangle (4.5,-1); 
\draw[thick] (4.5,0) rectangle (6,-1); 

\draw[thick] (0,-1) rectangle (1.5,-2);
\draw[thick] (1.5,-1) rectangle (3,-2);
\draw[thick] (3,-1) rectangle (4.5,-2);

\draw[thick] (0,-2) rectangle (1.5,-3);
\draw[thick] (1.5,-2) rectangle (3,-3);
\draw[thick] (3,-2) rectangle (4.5,-3);

\draw[thick] (0,-3) rectangle (1.5,-4);
\draw[thick] (0,-4) rectangle (1.5,-5);
\draw[thick] (0,-5) rectangle (1.5,-6);

\color{red}

\node at (-0.8 + 1.5*1, -.5) {$\ll -3,-3 \rr$};
\node at (-0.8 + 1.5*2, -.5) {$\ll -2,-3 \rr$};
\node at (-0.8 + 1.5*3, -.5) {$\ll -1,-2 \rr$};
\node at (-0.8 + 1.5*4, -.5) {$\ll 0,-1 \rr$};

\node at (-0.8 + 1.5*1, -.5 - 1) {$\ll -2,-2 \rr$};
\node at (-0.8 + 1.5*2, -.5 - 1) {$\ll -1,-2 \rr$};
\node at (-0.8 + 1.5*3, -.5 - 1) {$\ll 0,-1 \rr$};

\node at (-0.8 + 1.5*1, -.5 - 2) {$\ll -2,-1 \rr$};
\node at (-0.8 + 1.5*2, -.5 - 2) {$\ll -1,-1 \rr$};
\node at (-0.8 + 1.5*3, -.5 - 2) {$\ll 0,0 \rr$};

\node at (-0.8 + 1.5*1, -.5 - 3) {$\ll 0,0 \rr$};
\node at (-0.8 + 1.5*1, -.5 - 4) {$\ll 0,1 \rr$};
\node at (-0.8 + 1.5*1, -.5 - 5) {$\ll 0,2 \rr$};

\node [scale=1.5] at (-1, -.5 - 2.5) {$V = $};

\color{black}

\end{scope}

\begin{scope}[xshift=-10cm,yshift= 0 cm]
\draw[thick] (0,0) rectangle (1.5,-1); 
\draw[thick] (1.5,0) rectangle (3,-1); 
\draw[thick] (3,0) rectangle (4.5,-1); 
\draw[thick] (4.5,0) rectangle (6,-1); 
\draw[thick] (6,0) rectangle (7.5,-1);

\draw[thick] (0,-1) rectangle (1.5,-2);
\draw[thick] (1.5,-1) rectangle (3,-2);
\draw[thick] (3,-1) rectangle (4.5,-2);
\draw[thick] (4.5,-1) rectangle (6,-2);
\draw[thick] (6,-1) rectangle (7.5,-2);

\draw[thick] (0,-2) rectangle (1.5,-3);
\draw[thick] (1.5,-2) rectangle (3,-3);
\draw[thick] (3,-2) rectangle (4.5,-3);

\draw[thick] (0,-3) rectangle (1.5,-4);
\draw[thick] (1.5,-3) rectangle (3,-4);

\color{blue}

\node at (-0.8 + 1.5*1, -.5) {$\ll 5,6 \rr$};
\node at (-0.8 + 1.5*2, -.5) {$\ll 4,3 \rr$};
\node at (-0.8 + 1.5*3, -.5) {$\ll 3,3 \rr$};
\node at (-0.8 + 1.5*4, -.5) {$\ll 2,1 \rr$};
\node at (-0.8 + 1.5*5, -.5) {$\ll 1,0 \rr$};

\node at (-0.8 + 1.5*1, -.5 - 1) {$\ll 5,5 \rr$};
\node at (-0.8 + 1.5*2, -.5 - 1) {$\ll 4,2 \rr$};
\node at (-0.8 + 1.5*3, -.5 - 1) {$\ll 3,2 \rr$};
\node at (-0.8 + 1.5*4, -.5 - 1) {$\ll 2,0 \rr$};
\node at (-0.8 + 1.5*5, -.5 - 1) {$\ll 1,-1 \rr$};

\node at (-0.8 + 1.5*1, -.5 - 2) {$\ll 3,4 \rr$};
\node at (-0.8 + 1.5*2, -.5 - 2) {$\ll 2,1 \rr$};
\node at (-0.8 + 1.5*3, -.5 - 2) {$\ll 1,1 \rr$};

\node at (-0.8 + 1.5*1, -.5 - 3) {$\ll 2,3 \rr$};
\node at (-0.8 + 1.5*2, -.5 - 3) {$\ll 1,0 \rr$};

\node [scale = 1.5] at (-1, -.5 - 1.5) {$W = $};

\color{blue}

\end{scope}

\begin{scope}
\foreach \i in {0,1,...,8}
{
\draw[thick] (1.5*\i,-2)--(1.5*\i,8);
}
\foreach \i in {-2,-1,...,8}
{
\draw[thick] (0,\i)--(1.5*8,\i);
}

\draw[->] (-1,8)--(-1,7);
\draw[->] (0,9)--(1,9);

\node [below,scale=1.2] at (-1,7) {$i$};
\node [right,scale=1.2] at (1,9) {$j$};

\color{red}
\node at (-0.8 + 1.5*1, 8.5 - 1) {$\ll -3,-3 \rr$};
\node at (-0.8 + 1.5*2, 8.5 - 1) {$\ll -2,-3 \rr$};
\node at (-0.8 + 1.5*3, 8.5 - 1) {$\ll -1,-2 \rr$};
\node at (-0.8 + 1.5*4, 8.5 - 1) {$\ll 0,-1 \rr$};
\color{black}
\color{blue}
\node at (-0.8 + 1.5*5, 8.5 - 1) {$\ll 1,-1 \rr$};
\color{black}
\node at (-0.8 + 1.5*6, 8.5 - 1) {$\ll 2,1 \rr$};
\node at (-0.8 + 1.5*7, 8.5 - 1) {$\ll 3,1 \rr$};
\node at (-0.8 + 1.5*8, 8.5 - 1) {$\ll 4,1 \rr$};

\color{red}
\node at (-0.8 + 1.5*1, 8.5 - 2) {$\ll -2,-2 \rr$};
\node at (-0.8 + 1.5*2, 8.5 - 2) {$\ll -1,-2 \rr$};
\node at (-0.8 + 1.5*3, 8.5 - 2) {$\ll 0,-1 \rr$};
\color{black}
\color{blue}
\node at (-0.8 + 1.5*4, 8.5 - 2) {$\ll 1,0 \rr$};
\node at (-0.8 + 1.5*5, 8.5 - 2) {$\ll 2,0 \rr$};
\color{black}
\node at (-0.8 + 1.5*6, 8.5 - 2) {$\ll 3,2 \rr$};
\node at (-0.8 + 1.5*7, 8.5 - 2) {$\ll 4,2 \rr$};
\node at (-0.8 + 1.5*8, 8.5 - 2) {$\ll 5,2 \rr$};

\color{red}
\node at (-0.8 + 1.5*1, 8.5 - 3) {$\ll -2,-1 \rr$};
\node at (-0.8 + 1.5*2, 8.5 - 3) {$\ll -1,-1 \rr$};
\node at (-0.8 + 1.5*3, 8.5 - 3) {$\ll 0,0 \rr$};
\color{black}
\node at (-0.8 + 1.5*4, 8.5 - 3) {$\ll 1,1 \rr$};
\color{blue}
\node at (-0.8 + 1.5*5, 8.5 - 3) {$\ll 2,1 \rr$};
\color{black}
\node at (-0.8 + 1.5*6, 8.5 - 3) {$\ll 3,3 \rr$};
\node at (-0.8 + 1.5*7, 8.5 - 3) {$\ll 4,3 \rr$};
\node at (-0.8 + 1.5*8, 8.5 - 3) {$\ll 5,3 \rr$};

\color{red}
\node at (-0.8 + 1.5*1, 8.5 - 4) {$\ll 0,0 \rr$};
\color{black}
\color{blue}
\node at (-0.8 + 1.5*2, 8.5 - 4) {$\ll 1,0 \rr$};
\node at (-0.8 + 1.5*3, 8.5 - 4) {$\ll 2,1 \rr$};
\color{black}
\color{blue}
\node at (-0.8 + 1.5*4, 8.5 - 4) {$\ll 3,2 \rr$};
\node at (-0.8 + 1.5*5, 8.5 - 4) {$\ll 4,2 \rr$};
\color{black}
\node at (-0.8 + 1.5*6, 8.5 - 4) {$\ll 5,4 \rr$};
\node at (-0.8 + 1.5*7, 8.5 - 4) {$\ll 6,4 \rr$};
\node at (-0.8 + 1.5*8, 8.5 - 4) {$\ll 7,4 \rr$};

\color{red}
\node at (-0.8 + 1.5*1, 8.5 - 5) {$\ll 0,1 \rr$};
`\color{blue}
\node at (-0.8 + 1.5*2, 8.5 - 5) {$\ll 1,1 \rr$};
\color{black}
\node at (-0.8 + 1.5*3, 8.5 - 5) {$\ll 2,2 \rr$};
\color{blue}
\node at (-0.8 + 1.5*4, 8.5 - 5) {$\ll 3,3 \rr$};
\node at (-0.8 + 1.5*5, 8.5 - 5) {$\ll 4,3 \rr$};
\color{black}
\node at (-0.8 + 1.5*6, 8.5 - 5) {$\ll 5,5 \rr$};
\node at (-0.8 + 1.5*7, 8.5 - 5) {$\ll 6,5 \rr$};
\node at (-0.8 + 1.5*8, 8.5 - 5) {$\ll 7,5 \rr$};

\color{red}
\node at (-0.8 + 1.5*1, 8.5 - 6) {$\ll 0,2 \rr$};
\color{black}
\node at (-0.8 + 1.5*2, 8.5 - 6) {$\ll 1,2 \rr$};
\node at (-0.8 + 1.5*3, 8.5 - 6) {$\ll 2,3 \rr$};
\node at (-0.8 + 1.5*4, 8.5 - 6) {$\ll 3,4 \rr$};
\node at (-0.8 + 1.5*5, 8.5 - 6) {$\ll 4,4 \rr$};
\node at (-0.8 + 1.5*6, 8.5 - 6) {$\ll 5,6 \rr$};
\node at (-0.8 + 1.5*7, 8.5 - 6) {$\ll 6,6 \rr$};
\node at (-0.8 + 1.5*8, 8.5 - 6) {$\ll 7,6 \rr$};

\node at (-0.8 + 1.5*1, 8.5 - 7) {$\ll 1,3 \rr$};
\color{blue}
\node at (-0.8 + 1.5*2, 8.5 - 7) {$\ll 2,3 \rr$};
\node at (-0.8 + 1.5*3, 8.5 - 7) {$\ll 3,4 \rr$};
\color{black}
\node at (-0.8 + 1.5*4, 8.5 - 7) {$\ll 4,5 \rr$};
\color{blue}
\node at (-0.8 + 1.5*5, 8.5 - 7) {$\ll 5,5 \rr$};
\color{black}
\node at (-0.8 + 1.5*6, 8.5 - 7) {$\ll 6,7 \rr$};
\node at (-0.8 + 1.5*7, 8.5 - 7) {$\ll 7,7 \rr$};
\node at (-0.8 + 1.5*8, 8.5 - 7) {$\ll 8,7 \rr$};

\node at (-0.8 + 1.5*1, 8.5 - 8) {$\ll 1,4 \rr$};
\node at (-0.8 + 1.5*2, 8.5 - 8) {$\ll 2,4 \rr$};
\node at (-0.8 + 1.5*3, 8.5 - 8) {$\ll 3,5 \rr$};
\node at (-0.8 + 1.5*4, 8.5 - 8) {$\ll 4,6 \rr$};
\color{blue}
\node at (-0.8 + 1.5*5, 8.5 - 8) {$\ll 5,6 \rr$};
\color{black}
\node at (-0.8 + 1.5*6, 8.5 - 8) {$\ll 6,8 \rr$};
\node at (-0.8 + 1.5*7, 8.5 - 8) {$\ll 7,8 \rr$};
\node at (-0.8 + 1.5*8, 8.5 - 8) {$\ll 8,8 \rr$};

\node at (-0.8 + 1.5*1, 8.5 - 9) {$\ll 1,5 \rr$};
\node at (-0.8 + 1.5*2, 8.5 - 9) {$\ll 2,5 \rr$};
\node at (-0.8 + 1.5*3, 8.5 - 9) {$\ll 3,6 \rr$};
\node at (-0.8 + 1.5*4, 8.5 - 9) {$\ll 4,7 \rr$};
\node at (-0.8 + 1.5*5, 8.5 - 9) {$\ll 5,7 \rr$};
\node at (-0.8 + 1.5*6, 8.5 - 9) {$\ll 6,9 \rr$};
\node at (-0.8 + 1.5*7, 8.5 - 9) {$\ll 7,9 \rr$};
\node at (-0.8 + 1.5*8, 8.5 - 9) {$\ll 8,9 \rr$};

\node at (-0.8 + 1.5*1, 8.5 - 10) {$\ll 1,6 \rr$};
\node at (-0.8 + 1.5*2, 8.5 - 10) {$\ll 2,6 \rr$};
\node at (-0.8 + 1.5*3, 8.5 - 10) {$\ll 3,7 \rr$};
\node at (-0.8 + 1.5*4, 8.5 - 10) {$\ll 4,8 \rr$};
\node at (-0.8 + 1.5*5, 8.5 - 10) {$\ll 5,8 \rr$};
\node at (-0.8 + 1.5*6, 8.5 - 10) {$\ll 6,10 \rr$};
\node at (-0.8 + 1.5*7, 8.5 - 10) {$\ll 7,10 \rr$};
\node at (-0.8 + 1.5*8, 8.5 - 10) {$\ll 8,10 \rr$};

\end{scope}

\end{tikzpicture}
\caption{
{\it
The Young diagram pair $W,V$ and their embedding into $\psi_{WV}$. The rest of $\psi_{WV}$ (colored black) exactly coincide with $\psi_{\emptyset\emptyset}$.
}
}
\label{A.Young.diagram.02}
\end{figure}
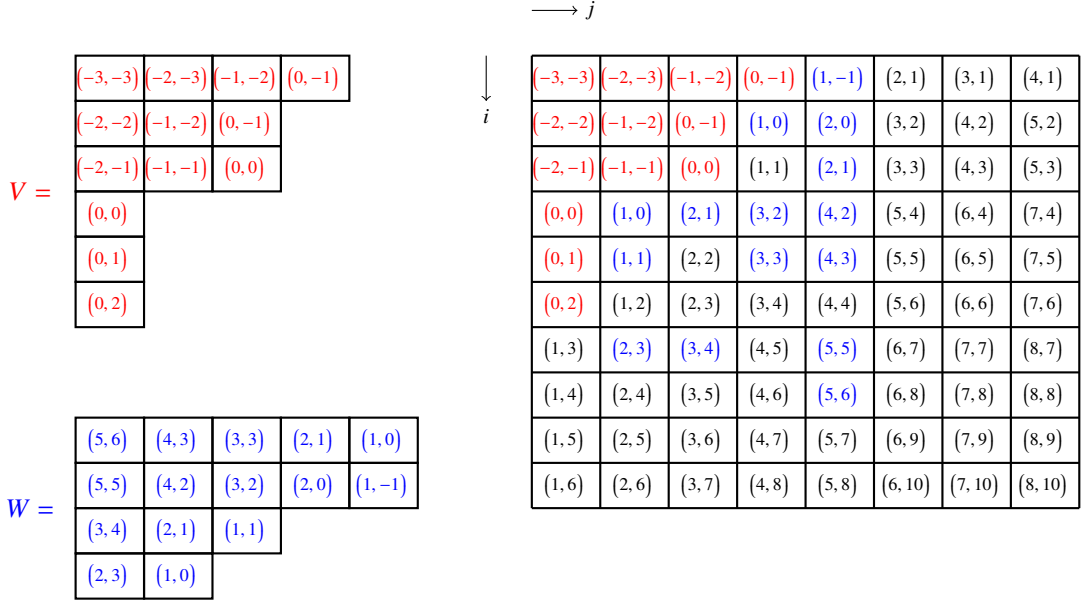
\end{scriptsize}

\subsubsection{Proposition}
Let $F : \mathbb{Z}\times \mathbb{Z} \rightarrow \mathbb{R}$ be a function such that $\prod_{i,j = 1}^\infty \ll 1 + F\ll j,i \rr \rr$ converges absolutely. Then, for any pair of Young diagrams $V$ and $W$,

\begin{multline}
\label{general.normalized.product.id}
\frac{
\displaystyle 
\prod_{i,j = 1}^\infty 
\ll 1 + F \ll - V_i + j, - W^{\, \intercal}_j + i\rr \rr 
}{
\displaystyle \prod_{i,j = 1}^\infty 
\ll 1 + F \ll j, i \rr \rr
} 
= 
\prod_{\bsquare \in V} \ll 1 + F \ll - A_{\bsquare, V},    - L_{\bsquare, W}      \rr \rr
\prod_{\wsquare \in W} \ll 1 + F \ll   A^{++}_{\wsquare, W}, L^{++}_{\wsquare, V} \rr \rr
\\
= 
\prod_{\wsquare \in W} \ll 1 + F \ll - A_{\wsquare, V},    - L_{\wsquare, W}      \rr \rr
\prod_{\bsquare \in V} \ll 1 + F \ll   A^{++}_{\bsquare, W}, L^{++}_{\bsquare, V} \rr \rr
\end{multline}

\subsubsection{Example}
Figure \ref{A.Young.diagram.01} and Figure \ref{A.Young.diagram.02} illustrate the idea of the proof of this proposition for the case where $V = \ll 4, 3, 3, 1, 1, 1 \rr$ and $W = \ll 5,5,3,2 \rr$.

\subsubsection{Proof of the first equality in Equation 
\ref{general.normalized.product.id}}

To simplify the notation, we write the
$\ll 1 + F \ll m, n \rr \rr$, $m, n \in \ZZ$ as 
$\ll m, n \rr$. 
Consider the positive quadrant of the integer lattice 
$\mathbb{Z}^2_+ = \ZZ_{\geq 1} \times \ZZ_{\geq 1}$. 
We associate to each point 
$\ll j, i \rr \in \mathbb{Z}^2_+$, 
the symbol 
$\ll - V_i + j, - W^{\, \intercal}_j + i\rr$. 
In the other words, we consider the map 
$\psi_{W V} : \mathbb{Z}^2_+ \rightarrow \mathbb{Z}^2$,
and identify $\mathbb{Z}^2_+$ with the associated symbols 
and the map $\psi_{W V}$. 
By multiplying all factors that correspond to all symbols in
$\mathbb{Z}^2_+$, we obtain the product 
$\prod_{i, j = 1}^\infty 
\ll 1 + F \ll - V_i + j, - W^{\, \intercal}_j + i \rr \rr$.

For $V = W =  \emptyset$, Equation \ref{general.normalized.product.id} is trivially true, and $\mathbb{Z}^2_+$ is 
populated by symbols $\ll j,i \rr$ at each point $\ll j,i \rr \in \mathbb{Z}^2_+$. 
We extend this to $\mathbb{Z}^2$ by associating $\ll j, i \rr$ at all points 
$\ll j,i \rr \in \mathbb{Z}^2$, and take $\psi_{\emptyset\emptyset}$ to be the restriction 
of the identity map to the sub-lattice $\mathbb{Z}^2_+$.

For arbitrary Young diagrams $\ll W, V \rr$, 
we construct $\psi_{WV}$ from 
$\psi_{\emptyset\emptyset}$ as follows.
\1 Shift the first coordinate of each symbol $\ll m, n \rr$ in the $i$-th row 
of $\mathbb{Z}^2$ in the positive $j$-direction by $V_i$. 
For example $\ll m, n \rr$ becomes $\ll m - V_i, n \rr$ after the shift, 
\2 Shift the second coordinate of each symbol $\ll m, n \rr$ in the $j$-th column 
of $\mathbb{Z}^2$ in the positive $i$-direction by $W^{\, \intercal}_i$. 
For example $\ll m, n \rr$ becomes $\ll m, n - W^{\, \intercal}_i \rr$ after the shift, and 
\3 Restrict to the sub-lattice $\ZZ^2_+ \subset \ZZ^2$.

Consider $V$ to be arbitrary, and $W = \emptyset$. 
From the construction of $\psi_{\emptyset V}$, all symbols in 
$\psi_{\emptyset\emptyset}$ are preserved, but get pushed to the 
right to make room for the embedding of $V \subset \mathbb{Z}^2_+$. 
All the factors corresponding to symbols in $\psi_{\emptyset\emptyset}$ are
eliminated when dividing by 
$\prod_{i,j=1}^\infty\ll 1 + F\ll j,i \rr \rr$, 
and we are left with the product 
$\prod_{(j,i) \in V}\ll 1 + F\ll -V_i+j,i \rr \rr$, 
so the identity is proven in this case. 

Next, consider the general case where both $V$ and $W$ 
are arbitrary. 
From the above construction, the symbols in $V \subset \mathbb{Z}^2_+$ give 
the product, 

$$
\prod_{(j, i) \in V}
\ll 1 + F 
\ll 
- V_i + j, - W^{\, \intercal}_j + i 
\rr 
\rr = 
\prod_{\bsquare \in V}
\ll 1 + F 
\ll - A_{\bsquare,V}, - L_{\bsquare, W} 
\rr 
\rr
$$

It only remains to show that $W$ along with $\psi_{\emptyset\emptyset}$ can be embedded 
in the $\mathbb{Z}^2_+\setminus V$ part of $\psi_{WV}$. 
That is, the embedding of $W$ must include the symbols 
that represent the factors in the product,

$$
\prod_{\wsquare \in W}
\ll 1 + F
\ll A^{++}_{\wsquare,W}, L^{++}_{\wsquare, V} 
\rr \rr
$$ 

The contributions from
$\psi_{\emptyset\emptyset}$ are eliminated 
when dividing 

$$
\prod_{i, j = 1}^\infty 
\ll 
1 + F
\ll 
- V_i + j, - W^{\, \intercal}_j + i  
\rr 
\rr
$$ 

by $\prod_{i,j}^\infty\ll 1 + F\ll j,i \rr \rr$, 
giving Equation \ref{general.normalized.product.id}.

Now, keep $V$ fixed, and add boxes to $W$, 
completing a row then moving to the next. 
The first box added to $W$ shifts the second 
coordinates of the symbols on the first 
column of $\mathbb{Z}^2$ down by $1$. 
This is equivalent to pushing all symbols above 
the $V^{\, \intercal}_1$-row in the first column 
up by $1$, and refilling the empty spot at 
$\ll 1, V^{\, \intercal}_1+1 \rr$ by 
a new symbol $\ll 1, V^{\, \intercal}_1 \rr$. 
Therefore, all original symbols from $\psi_{\emptyset\emptyset}$ are preserved.

The new symbol $\ll 1, V^{\, \intercal}_1 \rr$ cannot 
be part of $\psi_{\emptyset\emptyset}$ because we have 
just concluded that $\psi_{\, \emptyset \, \emptyset}$ 
gets pushed away with all its symbols preserved. 
This gives an embedding of $W$ into $\psi_{WV}$ 
as the new symbol 
$\ll 1,V^{\, \intercal}_1 \rr$, and $\psi_{\emptyset\emptyset}$ into $\psi_{WV}$ as 
$\mathbb{Z}^2_+ \setminus 
\ll V \union \{\ll 1,V^{\, \intercal}_1+1 \rr\} 
\rr$.

Consider that $W$ consists of a single row of length 
$\ll n - 1 \rr$ and can be embedded in $\psi_{WV}$. 
Adding a box into the row of $W$, so that $W$ is single 
row of length $n$, amounts to shifting the second 
coordinates of the symbols in the $n$-th column of 
$\mathbb{Z}^2$ down by 1. 
Take $V^{\, \intercal}_0 = \infty$, and let,

\begin{equation}
\cN_n = 
\ll 
j \in \ll 1, \cdots, n \rr 
\, | \, 
V^{\, \intercal}_{j-1} - V^{\, \intercal}_j > 0 
\rr
\end{equation}

For each $j \in \cN_n$, all symbols above the $V^{\, \intercal}_j$-th row but below 
the $V^{\, \intercal}_{j-1}$-th row are shifted up by $1$ and for each 
$j \in \cN_n\setminus \{1\}$, the symbol $\ll n - j + 1, V^{\, \intercal}_{j-1} \rr$ 
on the $V^{\, \intercal}_{j-1}$-th row and $n$-th column is pushed to replace 
$\ll n - j+1, V^{\, \intercal}_{j-1} \rr$ at 
$\ll n - \ll j - V_{V^{\, \intercal}_{j-1} + 1} \rr, V^{\, \intercal}_{j-1} + 1\rr$. 
The symbol being replaced was a new symbol created when we added a 
$\ll n - \ll j - V_{V^{\, \intercal}_{j-1} + 1}\rr\rr$-th box to the row of $W$.
The new symbols introduced are $\ll n - j + 1, V^{\, \intercal}_{j} \rr$ for all 
$j \in \cN_n$. 
The remaining symbols needed for the embedding of $W$ are 
$\ll n - j + 1, V^{\, \intercal}_j \rr$ for $j \notin\cN_n$ or 
$V^{\, \intercal}_{j-1} = V^{\, \intercal}_j$. These symbols are in $\psi_{WV}$ 
before we added our last box and it cannot get replaced because, as noted above, 
only $\ll n-j+1, V^{\, \intercal}_{j-1} \rr$ for $j \in \cN_{n}\setminus \{1\}$ 
are replaced. 

The number of symbols introduced is equal to $| \, \cN_n \, |$, which is always one more 
than the number of symbols replaced. Therefore, all symbols that belong to $W$ 
before we added the box but not after, must be replaced. 
Since all the original symbols from $\psi_{\emptyset\emptyset}$ are preserved, we 
have an embedding of a single row of any size $W$ and 
$\psi_{\emptyset\emptyset}$ into $\psi_{WV}$ as required.

Since we work with Young diagrams, shifting the second coordinates of symbols in 
the $n$-th column of $\mathbb{Z}^2$ down by 1 can only be done if we have done it 
for columns $1, \cdots, n-1$. 
Moreover, the operation preserves all but those symbols it created itself in columns 
$1, \cdots, n-1$. 
When we add boxes to the next rows of $W$, the embeddings of all previous rows of 
$W$ along with $\psi_{\emptyset\emptyset}$ are preserved. 
Adding the $i$-th row of $W$ is exactly the same as adding the first row, except 
that all the second coordinates of symbols in every columns that concern us have 
already been shifted down $(i-1)$-times. 
So, wherever we introduce a new symbol $\ll W_1-j+1,V^{\, \intercal}_j\rr$, 
we introduce instead $\ll W_i - j + 1, V^{\, \intercal}_j-i+1 \rr$. 
We have an embedding of $\psi_{\emptyset\emptyset}$ and an arbitrary $W$ into 
$\psi_{WV}$, which concludes the proof of the first equality in Equation 
\ref{general.normalized.product.id}.

\subsubsection{Proof of the second equality of Equation 
\ref{general.normalized.product.id}}
Applying the first equality with $\tilde{F}\ll m, n \rr = F\ll n, m \rr$ and diagrams
$\tilde{V} = W^{\, \intercal}, \tilde{W} = V^{\, \intercal}$, 

\begin{multline}
\frac{
\displaystyle \prod_{i, j = 1}^\infty 
\ll 1 + \tilde{F} \ll - \tilde{V}_i + j, - \tilde{W}^{\, \intercal}_j + i \rr \rr
}{
\displaystyle \prod_{i,j=1}^\infty \ll 1 + \tilde{F}\ll j,i \rr \rr}
\\
= 
\prod_{(j,i) \in \tilde{V}} 
\ll 
1 + \tilde{F} 
\ll 
- \tilde{V}_i           + j,   
- \tilde{W}^{\, \intercal}_j + i
\rr 
\rr
\prod_{(j,i) \in \tilde{W}}
\ll 1 + \tilde{F} 
\ll   
\tilde{W}_i - j + 1, 
\tilde{V}^{\, \intercal}_j - i + 1 
\rr 
\rr
\end{multline}

where, for clarity, the right hand side is given
in detail. Re-writing this in terms of 
$F \ll m, n \rr$ and the Young diagrams $V$ and $W$ gives

\begin{multline}
\frac{
\displaystyle \prod_{i,j = 1}^\infty
\ll 1 + F 
\ll 
- V_i + j, 
- W^{\, \intercal}_j + i 
\rr 
\rr
}{
\displaystyle \prod_{i,j=1}^\infty 
\ll 1 + F
\ll 
j, i 
\rr \rr}
\\ 
= 
\prod_{(j,i) \in W^{\, \intercal}} 
\ll 1 + F 
\ll 
- V_i + j,    
- W^{\, \intercal}_j + i     
\rr 
\rr
\prod_{(j,i) \in V^{\, \intercal}} 
\ll 1 + F 
\ll   
W_i - j + 1, 
V^{\, \intercal}_j - i + 1 
\rr \rr
\\
= 
\prod_{(i,j) \in W} 
\ll 1 + F 
\ll 
- V_i + j, 
- W^{\, \intercal}_j + i       
\rr \rr
\prod_{(i,j) \in V} 
\ll 1 + F 
\ll   
W_i - j + 1, 
V^{\, \intercal}_j - i + 1
\rr \rr
\end{multline}

This is the second equality in Equation \ref{general.normalized.product.id} and the Proposition 
is proven.

\subsubsection{Proof of identity \ref{R.0.normalized.product.formula}}

Let 

\begin{equation}
F\ll j, i \rr = 
\begin{cases}
- Q \, x^{j} y^{i-1}, & i + j - 1 = c \bmod{n} 
\\
  0,                & i + j - 1 \neq c \bmod{n}
\end{cases}
\end{equation}

Then Equation \ref{general.normalized.product.id} gives Equation \ref{R.0.normalized.product.formula}. 

\section{The $N$-strip partition function. A proof by induction}
\label{appendix.B}
\textit{We present a proof by induction of Equation 
\ref{5d.N.strip.partition.function} for the $n$-coloured 5D strip partition function.}

\subsubsection{The unevaluated strip partition function}

Our goal is to obtain Equation  \ref{5d.N.strip.partition.function} from 
Equation \ref{uneval.s.N}. However, it is hard to perform a proof by induction 
directly on $\cS^{\, 5D}_{\, \bV  \, \bW  \, \bDelta} \ll x, y, R \rr$, 
and we find that we need to consider a more general function $\tilde{\cS}^{\, (N)}$.

Let $A^I = \ll A^I_1, A^I_2, A^I_3, \cdots \rr$ be a set of variables 
indexed by $I$ and $\bA = \ll A^1, \cdots, A^N \rr$, and similarly for 
$\bB, \bC, \bD$. 
Let $\bQ' = \ll Q'_1, \cdots, Q'_N \rr$ and 
$\bQ'' = \ll Q''_1, \cdots, Q''_{N-1}\rr$. 
We define, 

\begin{multline}\label{defn.general.function}
\tilde{\cS}^{\, (N)} \ll \bA,\bB,\bC,\bD \, | \, \bQ', \bQ'' \rr = 
\sum_{\xi_1', \cdots, \xi_N', \xi_1'', \cdots, \xi_{N-1}''}
\prod_{I=1}^N
\ll 
(-Q'_I)^{| \, \xi'_I \, |}(-Q''_I)^{|\, \xi''_I \, |} 
\rr
\\ 
\prod_{I=1}^N
\ll 
\sum_{\eta'_I}  
\, s_{\xi''_{I-1}/\eta'_I} \ll A^I \rr 
\, s_{\xi'_I/\eta'_I} \ll B^I \rr 
\sum_{\eta''_I} 
\, s_{\xi^{''\intercal}_I/\eta''_I} \ll C^I \rr 
\, s_{\xi^{'\intercal}_I/\eta''_I} \ll D^I \rr 
\rr
\end{multline}

where $\xi''_0 = \xi''_N = \emptyset$. In the rest of the proof, any Young diagram $\xi$ 
that appears in the product but not being summed over is assumed to be null 
$\xi = \emptyset$.
$\cS^{5D}_{\bV\bW\bDelta}\ll x,y,R \rr$ in Equation \ref{uneval.s.N} 
is expressed in terms of $\tilde{\cS}^{\, (N)}$ as,  

\begin{multline}
\label{s.N.in.term.of.general.func}
\cS^{\, n}_{\, \bV \, \bW \, \bDelta}
\ll x, y, R \rr
= \prod_{I=1}^N
\ll Z^{\, n}_{\, V_I}
\ll x, y \rr 
Z^{\, n}_{\, W^{\, \intercal}_I}
\ll y, x \rr 
\rr
\\
\prod_{c_H = 0}^{\, p - 1}
\tilde{\cS}^{\, (N)}
\ll 
[x^{-\bV}                y^{ \bj - 1}]_{      \,   c_H - c_{\, \bV}},
[x^{ \bi}                y^{-\bV^{\, \intercal}}]_{\, - c_H + c_{\, \bV}},
[y^{-\bW^{\, \intercal}} x^{\, \bi-1}]_{\, - c_H + c_{\bW} - 1},
[y^{\, \bj}                 x^{-\bW}]_{\, c_H - c_{\, \bW} + 1} 
\, | \, 
\bQ, \bQ_M \rr
\end{multline}

where

\begin{multline}
[q^{-\bV}t^{\, \bj-1}]_{c_H-c_{\bV}} = \ll 
[q^{-V_1}t^{\, \bj-1}]_{c_H-c_{V_1}}, 
\cdots, 
[q^{-V_N}t^{\, \bj-1}]_{c_H-c_{V_N}} \rr,
\\ 
[q^{\, \bi} t^{- \bV^{\, \intercal}}]_{ - c_H + c_{\bV}} = 
\ll 
[q^{\, \bi} t^{- V^{\, \intercal}_1}]_{ - c_H + c_{V_1}},
\cdots, 
[q^{\, \bi} t^{- V^{\, \intercal}_N}]_{ - c_H + c_{V_N}} 
\rr,
\\ 
[t^{-\bW^{\, \intercal}}q^{\, \bi-1}]_{-c_H+c_{\bW}-1} = 
\ll 
[t^{-W^{\, \intercal}_1} q^{\i-1}]_{- c_H + c_{W_1} - 1}, 
\cdots, 
[t^{-W^{\, \intercal}_N} q^{\i-1}]_{- c_H + c_{W_N} - 1} 
\rr,
\\ 
[t^{\, \bj} q^{-\bW}]_{c_H - c_{\bW} + 1} = 
\ll 
[t^{\, \bj} q^{-W_1}]_{c_H - c_{W_1} + 1},
\cdots, 
[t^{\, \bj} q^{-W_N}]_{c_H - c_{W_N} + 1} 
\rr,
\\
\bQ   = \ll Q_1,     \cdots, Q_N     \rr, 
\bQ_M = \ll Q_{M_1}, \cdots, Q_{M_N} \rr.
\end{multline}

Therefore, once we prove the following general result, 

\begin{multline}
\label{general.function.N}
\tilde{\cS}^{\, (N)} 
\ll \bA, \bB, \bC, \bD \, | \, \bQ', \bQ''\rr =
\\ 
\prod_{J=1}^N
\prod_{I=1}^{J}
\prod_{i,j}
\ll 
1 - \prod_{K=I}^{J-1} Q''_K
\prod_{K=I}^{J} Q'_K D^{J}_j B^I_i 
\rr 
\prod_{J=1}^{N}
\prod_{I=1}^{J-1}
\prod_{i,j}
\ll 1 - \prod_{K=I}^{J-1} Q''_K
\prod_{K=I+1}^{J-1} Q'_K A^{J}_j C^I_i 
\rr
\\
\prod_{J=1}^{N}
\prod_{I=1}^{J-1}
\prod_{i,j}
\ll 1 - 
\prod_{K=I}^{J-1}Q''_K
\prod_{K=I}^{J-1} Q'_KA^{J}_j B^I_i 
\rr^{-1}
\prod_{J=1}^{N}
\prod_{I=1}^{J-1}
\prod_{i,j}
\ll 1 - 
\prod_{K=I}^{J-1} Q''_K
\prod_{K=I+1}^{J} Q'_K D^{J}_j C^I_i \rr^{-1}, 
\end{multline}

for any $\bA,\bB,\bC,\bD,\bQ',\bQ''$, for $N \geq 1$, 
Equation \ref{5d.N.strip.partition.function} follows 
from Equation \ref{s.N.in.term.of.general.func} and 
our proof is complete. 
We prove Equation \ref{general.function.N} by induction. 

\subsubsection{Step 1. The base case}
From Equation \ref{defn.general.function}, the $N=1$ 
base case is,

\begin{multline}
\tilde{\cS}^{(1)}
\ll \bA,\bB,\bC,\bD\, | \, \bQ',\bQ'' \rr = 
\\
\sum_{\xi'_1}\ll \ll -Q_1' \rr^{\, | \, \xi'_1\, | \, } 
\sum_{\eta'_1}  
    \, s_{\emptyset/\eta'_1} \ll A^1 \rr 
    \, s_{\xi'_1/\eta'_1} \ll B^1 \rr 
\sum_{\eta''_1} 
    \, s_{\emptyset/\eta''_1} \ll C^1 \rr 
    \, s_{\xi^{'\intercal}_1/\eta''_1} \ll D^1 \rr \rr\\
= 
\sum_{\xi'_1} \ll -Q_1'\rr^{\, | \, \xi'_1\, | \, } 
s_{\xi'_1} \ll B^1 \rr 
s_{\xi^{'\intercal}_1} 
\ll D^1 \rr 
= 
\sum_{\xi'_1}s_{\xi'_1} \ll - Q_1'B^1 \rr s_{\xi^{'\intercal}_1} 
\ll D^1 \rr
= 
\prod_{i,j} \ll 1 - Q_1' D^1_j B_i^1 \rr
\end{multline}

which agrees with Equation \ref{general.function.N}.

\subsubsection{Step 2. The $\ll N - 1 \rr$ case}
We assume that Equation \ref{defn.general.function} holds for 
$\ll N - 1 \rr$. 
From Equation \ref{defn.general.function}, by switching the summation 
for $\{\xi_I', \xi_I''\}$ and $\{\eta_I', \eta_I''\}$ and defining 
$\eta'_1 = \eta''_N = \emptyset$, we obtain

\begin{multline}
\tilde{\cS}^{\, (N)} \ll \bA, \bB, \bC, \bD \, | \, \bQ', \bQ'' \rr 
= 
\sum_{\eta'_2,  \cdots, \eta'_N}
\sum_{\eta''_1, \cdots, \eta''_{N-1}}
\prod_{I=1}^N
\ll 
\sum_{\xi'_I} \ll-Q'_I\rr^{\, | \, \xi'_I\, | \, }
\, s_{\xi'_I/\eta'_I} \ll B^I \rr 
\, s_{\xi^{'\intercal}_I/\eta''_I}\ll D^I \rr 
\rr 
\\ 
\prod_{I=1}^{N-1}
\ll 
\sum_{\xi''_I}\ll -Q''_I\rr^{\, | \, \xi''_I\, | \, }
\, s_{\xi''_I/\eta'_{I+1}}(A^{I+1})
\, s_{\xi^{''\intercal}_I/\eta''_I}\ll C^I \rr  
\rr
\end{multline}

Then, using Equation \ref{schur.id.4},

\begin{multline}
\tilde{\cS}^{\, (N)} \ll \bA, \bB, \bC, \bD \, | \, \bQ', \bQ'' \rr = 
\sum_{\eta'_2,  \cdots, \eta'_N}
\sum_{\eta''_1, \cdots, \eta''_{N-1}}
\prod_{I=1}^N
\ll 
\ll - Q'_I \rr^{\, | \, \eta'_I\, | \, } 
\ll - Q''_I\rr^{\, | \, \eta''_I\, | \, } 
\rr
\\
\prod_{I=1}^N
\ll 
\sum_{\xi'_I}
\, s_{\xi'_I/\eta'_I}                \ll - Q'_I B^I \rr 
\, s_{\xi^{'\intercal}_I / \eta''_I} \ll        D^I \rr  
\rr 
\prod_{I=1}^{N-1}
\ll 
\sum_{\xi''_I}
\, s_{\xi''_I/\eta'_{I+1}}          \ll A^{I+1}     \rr 
\, s_{\xi^{''\intercal}_I/\eta''_I} \ll - Q''_I C^I \rr  
\rr
\end{multline}

Then, using Equation \ref{schur.id.2},

\begin{multline}
\tilde{\cS}^{\, (N)} \ll \bA, \bB, \bC, \bD \, | \, \bQ', \bQ'' \rr =
\\ 
\prod_{I=1}^N
\prod_{i,j} 
\ll 1 - Q'_I B^I_i D^I_j \rr 
\prod_{I=1}^{N-1}
\prod_{i,j}
\ll 1 - Q''_I A^{I+1}_ iC^I_j \rr
\sum_{\eta'_2, \cdots, \eta'_N}
\sum_{\eta''_1, \cdots, \eta''_{N-1}}
\prod_{I=1}^N
\ll \ll -Q'_I\rr^{|\eta'_I\, | \, '}\ll -Q''_I\rr^{\, | \, \eta''_I\, | \, } \rr
\\
\prod_{I=1}^N
\ll \sum_{\tau'_I} 
\, s_{\eta^{ '\intercal}_I / \tau^{'\intercal}_I} \ll        D^I \rr 
\, s_{\eta^{''\intercal}_I / \tau'_I}             \ll - Q'_I B^I \rr 
\rr
\prod_{I=1}^{N-1}
\ll 
\sum_{\tau''_I}
\, s_{\eta^{'\intercal}_{I+1}/\tau^{''\intercal}_I} \ll -s_I'' C^I \rr 
\, s_{\eta^{''\intercal}_I / \tau''_I}              \ll A^{I+1}    \rr 
\rr
\end{multline}

Changing the summation for 
$\{\eta'_I, \eta''_I\}$ and $\{\tau_I', \tau_I''\}$ and defining 
$\tau_1' = \tau_N' = \emptyset$, 

\begin{multline}
\tilde{\cS}^{\, (N)} \ll \bA, \bB, \bC, \bD \, | \, \bQ', \bQ'' \rr = 
\\ 
\prod_{I=1}^N
\prod_{i,j} \ll 1 - Q'_I B^I_i D^I_j \rr 
\prod_{I=1}^{N-1}
\prod_{i,j} \ll 1 - Q''_I A^{I+1}_i C^I_j \rr
\\
\sum_{\tau'_2,  \cdots, \tau'_{N-1}}
\sum_{\tau''_1, \cdots, \tau''_{N-1}}
\prod_{I=2}^N \ll
\sum_{\eta'_I}\ll -Q'_I\rr^{\, | \, \eta'_I\, | \, }
\, s_{\eta'_I/\tau^{'\intercal}_I} \ll D^I \rr 
\, s_{\eta'_I/\tau^{''\intercal}_{I-1}} \ll - Q''_{I-1} C^{I-1} \rr 
\rr
\\
\prod_{I=1}^{N-1} 
\ll 
\sum_{\eta''_I}\ll -Q''_I\rr^{\, | \, \eta''_I\, | \, }
\, s_{\eta^{''\intercal}_I/\tau'_I}  \ll -Q'_I B^I \rr 
\, s_{\eta^{''\intercal}_I/\tau''_I} \ll  A^{I+1}  \rr  
\rr
\end{multline}
Applying Equation \ref{schur.id.1} then gives

\begin{multline}
\tilde{\cS}^{\, (N)} \ll \bA, \bB, \bC, \bD \, | \, \bQ', \bQ'' \rr =
\\ 
\prod_{I=1}^N
\prod_{i,j} \ll 1 - Q'_I  B^I_i     D^I_j \rr 
\prod_{I=1}^{N-1}
\prod_{i,j} \ll 1 - Q''_I A^{I+1}_i C^I_j \rr
\\ 
\prod_{I=2}^N
\prod_{i,j} \ll 1 - Q'_IQ''_{I-1} D^I C^{I-1} \rr^{-1}
\prod_{I=1}^{N-1}
\prod_{i,j} \ll 1 - Q'_IQ''_IB^IA^{I+1} \rr^{-1}
\\ 
\sum_{\tau'_2,  \cdots, \tau'_{N-1} }
\sum_{\tau''_1, \cdots, \tau''_{N-1}}
\prod_{I=1}^N 
\ll 
\ll 
-Q_I'\rr^{\, | \, \tau'_I\, | \, } 
\ll 
-Q_I'' \rr^{\, | \,  \tau''_I \, | \, } 
\rr
\\
\prod_{I=2}^N 
\ll
\sum_{\alpha'_I}
\, s_{\tau^{'  \intercal}_I     / \alpha'_I} \ll - Q''_{I-1} C^{I-1} \rr 
\, s_{\tau^{'' \intercal}_{I-1} / \alpha'_I} \ll - Q'_I      D^I     \rr 
\rr
\prod_{I=1}^{N-1}
\ll 
\sum_{\alpha''_I}
\, s_{ \tau'_I / \alpha''_I} \ll - Q''_I A^{I+1} \rr 
\, s_{\tau''_I / \alpha''_I} \ll - Q'_I  B^I     \rr 
\rr
\end{multline}

On changing the summation indices to 
$\xi'_1 = \tau''_1, \cdots,\xi'_{N-1} = 
\tau''_{N-1},\xi''_1 = \tau'_2, \cdots,\xi''_{N-2} = 
\tau'_{N-1}$ and 
$\eta'_2 = \alpha''_2, \cdots,\eta'_{N-1} = 
\alpha''_{N-1}, \eta''_1 = \alpha'_2, \cdots,\eta''_{N-2} = 
\alpha'_{N-1}$ and $\xi''_0 = \xi''_{N-1} = \emptyset$, this becomes, 

\begin{multline}
\tilde{\cS}^{\, (N)} 
\ll \bA, \bB, \bC, \bD \, | \, \bQ',\bQ'' \rr = 
\\ 
\prod_{I=1}^N 
\prod_{i,j} 
\ll 
1 - Q'_I B^I_i D^I_j 
\rr 
\prod_{I=1}^{N-1}
\prod_{i,j}
\ll 
1 - Q''_IA^{I+1}_iC^I_j
\rr
\\ 
\prod_{I=2}^N
\prod_{i,j}
\ll 
1 - Q'_IQ''_{I-1}D^I_jC^{I-1}_i 
\rr^{-1}
\prod_{I=1}^{N-1}
\prod_{i,j}
\ll 
1 - Q'_I Q''_I B^I_j A^{I+1}_i 
\rr^{-1}
\\
\sum_{\xi'_1, \cdots,\xi'_{N-1}, \xi''_1, \cdots, \xi''_{N-2}}
\prod_{I=1}^{N-1}
\ll 
(-Q_I'')^{\, | \, \xi'_I\, | \, }(-Q_{I+1}')^{\, | \, \xi''_I\, | \, } 
\rr
\\
\prod_{I=1}^{N-1}
\ll
\sum_{\eta'_I}
\, s_{\xi''_{I-1}/\eta'_I} \ll -Q''_I A^{I+1} \rr 
\, s_{\xi'_I/\eta'_I} \ll -Q'_I B^I \rr 
\sum_{\eta''_I}
\, s_{\xi^{''\intercal}_I/\eta''_I} \ll -Q''_I C^I \rr 
\, s_{\xi^{'\intercal}_I/\eta''_I}  \ll -Q'_{I+1} D^{I+1} \rr  
\rr
\end{multline}

Using Equation \ref{defn.general.function} then gives

\begin{multline}
\tilde{\cS}^{\, (N)} 
\ll \bA, \bB, \bC, \bD \, | \, \bQ',\bQ'' \rr = 
\\ 
\prod_{I=1}^N
\prod_{i,j} 
\ll 1 - Q'_I B^I_i D^I_j \rr 
\prod_{I=1}^{N-1}
\prod_{i,j}
\ll 1 - Q''_I A^{I+1}_i C^I_j
\rr
\\ 
\prod_{I=2}^N
\prod_{i,j}
\ll 1 - Q'_I Q''_{I-1} D^I_j C^{I-1}_i 
\rr^{-1}
\prod_{I=1}^{N-1}
\prod_{i,j}
\ll 1 - Q'_I Q''_I B^I_j A^{I+1}_i 
\rr^{-1}
\\
\tilde{\cS}^{(N-1)} 
\ll - \bQ'' \bA^+, - \bQ' \bB, -\bQ'' \bC, - \bs^{'+} \bD^+ \, | \, \bQ'',\bQ^{'+} \rr 
\end{multline}

where, for any given set of variables 
$x = \ll x_1, x_2, x_3, \cdots \rr, 
 y = \ll y_1, y_2, Y, \cdots \rr$, we define,

\begin{equation}
x^+ := \ll  x_2,x_3, \cdots \rr, \quad 
xy  := \ll x_1 y_1, x_2 y_2, x_3 Y, \cdots \rr.
\end{equation}

\subsubsection{Step 3. The induction}
By the induction hypothesis, 

\begin{multline}
\tilde{\cS}^{(N-1)} 
\ll 
- \bQ'' \bA^+, - \bQ' \bB, - \bQ'' \bC, - \bQ^{'+} \bD^+
\, | \, 
\bQ'',\bQ^{'+}
\rr 
= 
\\
\prod_{J=1}^{N-1}
\prod_{I=1}^{J}
\prod_{i,j}
\ll 1 - 
\prod_{K=I}^{J-1} Q'_{K+1}
\prod_{K=I}^{J} Q''_K Q'_{J+1} Q'_ID^{J+1}_j B^I_i 
\rr
\\
\prod_{J=1}^{N-1}
\prod_{I=1}^{J-1}
\prod_{i,j}
\ll 1 - 
\prod_{K=I}^{J-1} Q'_{K+1}
\prod_{K=I+1}^{J-1} Q''_K Q''_{J} Q''_IA^{J+1}_j C^I_i 
\rr
\\
\prod_{J=1}^{N-1}
\prod_{I=1}^{J-1}
\prod_{i,j}
\ll 1 - 
\prod_{K=I}^{J-1} Q'_{K+1}
\prod_{K=I}^{J-1} Q''_K Q''_{J} Q'_I A^{J+1}_j B^I_i \rr^{-1}
\\
\prod_{J=1}^{N-1}
\prod_{I=1}^{J-1}
\prod_{i,j}
\ll 1 - 
\prod_{K=I}^{J-1}Q'_{K+1}
\prod_{K=I+1}^{J} Q''_K Q'_{J+1} Q''_I D^{J+1}_j C^I_i 
\rr^{-1}
\end{multline}

from Equation \ref{general.function.N} for $\ll N - 1 \rr$. Therefore,

\begin{multline}
\tilde{\cS}^{\, (N)} \ll \bA, \bB, \bC, \bD \, | \, \bQ', \bQ'' \rr =
\\
\prod_{I=1}^N
\prod_{i,j} \ll 1 - Q'_ID^I_j B^I_i \rr
\prod_{I=1}^{N-1}
\prod_{i,j} \ll 1 - Q''_IA^{I+1}_j C^I_i \rr
\\
\prod_{I=1}^{N-1}
\prod_{i,j} \ll 1 - Q'_{I+1}Q''_ID^{I+1}_j C^I_i \rr^{-1}
\prod_{I=1}^{N-1}
\prod_{i,j} \ll 1 - Q'_IQ''_IA^{I+1}_j B^I_i \rr^{-1}
\\
\prod_{J=1}^{N-1}
\prod_{I=1}^{J}
\prod_{i,j}
\ll 1 - 
\prod_{K=I}^{J} Q''_K
\prod_{K=I}^{J+1} Q'_{K}
D^{J+1}_j B^I_i 
\rr
\prod_{J=1}^{N-1}
\prod_{I=1}^{J-1}
\prod_{i,j}
\ll 1 - 
\prod_{K=I}^{J} Q''_K
\prod_{K=I+1}^{J} Q'_{K}
A^{J+1}_j C^I_i 
\rr
\\
\prod_{J=1}^{N-1}
\prod_{I=1}^{J-1}
\prod_{i,j}
\ll 1 - 
\prod_{K=I}^{J} Q''_K
\prod_{K=I}^{J} Q'_{K}
A^{J+1}_j B^I_i \rr^{-1}
\prod_{J=1}^{N-1}
\prod_{I=1}^{J-1}
\prod_{i,j}
\ll 1 - 
\prod_{K=I}^{J} Q''_K
\prod_{K=I+1}^{J+1}Q'_{K}
D^{J+1}_j C^I_i 
\rr^{-1}
\\
=
\prod_{I=1}^N
\prod_{i,j} \ll 1 - Q'_ID^I_j B^I_i \rr
\prod_{I=1}^{N-1}
\prod_{i,j} \ll 1 - Q''_IA^{I+1}_j C^I_i \rr
\\
\prod_{I=1}^{N-1}
\prod_{i,j} \ll 1 - Q''_IQ'_IA^{I+1}_j B^I_i \rr^{-1}
\prod_{I=1}^{N-1}
\prod_{i,j} \ll 1 - Q''_IQ'_{I+1}D^{I+1}_j C^I_i \rr^{-1}
\\
\prod_{J=1}^{N}
\prod_{I=1}^{J-1}
\prod_{i,j}
\ll 1 - 
\prod_{K=I}^{J-1} Q''_K
\prod_{K=I}^{J} Q'_{K}
D^{J}_j B^I_i 
\rr
\prod_{J=1}^{N}
\prod_{I=1}^{J-2}
\prod_{i,j}
\ll 1 - 
\prod_{K=I}^{J-1} Q''_K
\prod_{K=I+1}^{J-1} Q'_{K}
A^{J}_j C^I_i 
\rr
\\
\prod_{J=1}^{N}
\prod_{I=1}^{J-2}
\prod_{i,j}
\ll 1 - 
\prod_{K=I}^{J-1} Q''_K
\prod_{K=I}^{J-1} Q'_{K}
A^{J}_j B^I_i \rr^{-1}
\prod_{J=1}^{N}
\prod_{I=1}^{J-2}
\prod_{i,j}
\ll 1 - 
\prod_{K=I}^{J-1} Q''_K
\prod_{K=I+1}^{J}Q'_{K}
D^{J}_j C^I_i 
\rr^{-1}
\\
=
\prod_{J=1}^N
\prod_{I=1}^{J}
\prod_{i,j}
\ll 
1 - \prod_{K=I}^{J-1} Q''_K
\prod_{K=I}^{J} Q'_K D^{J}_j B^I_i 
\rr 
\prod_{J=1}^{N}
\prod_{I=1}^{J-1}
\prod_{i,j}
\ll 1 - \prod_{K=I}^{J-1} Q''_K
\prod_{K=I+1}^{J-1} Q'_K A^{J}_j C^I_i 
\rr
\\
\prod_{J=1}^{N}
\prod_{I=1}^{J-1}
\prod_{i,j}
\ll 1 - 
\prod_{K=I}^{J-1}Q''_K
\prod_{K=I}^{J-1} Q'_KA^{J}_j B^I_i 
\rr^{-1}
\prod_{J=1}^{N}
\prod_{I=1}^{J-1}
\prod_{i,j}
\ll 1 - 
\prod_{K=I}^{J-1} Q''_K
\prod_{K=I+1}^{J} Q'_K D^{J}_j C^I_i \rr^{-1}
\end{multline}

This concludes the proof of Equation \ref{general.function.N}.

\end{document}